\documentclass[a4paper,11pt]{article}
\pdfoutput=1
\usepackage{jheppub}
\usepackage{amsmath}
\usepackage{amssymb}
\usepackage{color}
\usepackage{csquotes}
\usepackage{graphicx}
\usepackage{hyperref}
\usepackage[right]{lineno}
\usepackage{xspace,slashed}
\usepackage[utf8]{inputenc}
\usepackage{multirow}
\usepackage{algorithm}
\usepackage{algorithmic}
\usepackage{stackengine}
\usepackage{enumitem}
\usepackage{etoolbox}
\usepackage{siunitx}
\usepackage{tensor}

\newtoggle{draft}

\togglefalse{draft}

\iftoggle{draft} {
  \usepackage{changes}
}{%
  \usepackage[final]{changes}
}

\definechangesauthor[name=pn, color=blue]{pn}
\definechangesauthor[name=km, color=magenta]{km}

\usepackage{booktabs}

\makeatletter
\g@addto@macro\bfseries{\boldmath}
\makeatother
\newcommand{\ch}{{\rm ch}}
\newcommand{\sh}{{\rm sh}}

\newcommand{\ep}{\epsilon}

\iftoggle{draft} {
  \newcommand{\tobedeleted}[1]{\textcolor{azure}{#1}}
  
}{%
  \newcommand{\tobedeleted}[1]{}
  
}
\newcommand{\be}{\begin{equation}}
\newcommand{\ee}{\end{equation}}

\newcommand{\tka}{{\tilde k}_a}

\newcommand{\tkb}{{\tilde k}_b}

\usepackage{soul}
\allowdisplaybreaks

\definecolor{azure}{rgb}{0.0, 0.9, 1.0}

\newcommand{\TTPaff}{Institute for Theoretical Particle Physics,
  KIT, 76128 Karlsruhe, Germany}
\newcommand{\IAPaff}{Institute for Astroparticle Physics, KIT, 76344 Eggenstein-Leopoldshafen, Germany}

\preprint{
  \begin {flushright}
    TTP25-002, P3H-25-011 
  \end{flushright}
}
\title{
Power corrections to the production of a color-singlet final state  in hadron collisions in the   $N$-jettiness slicing scheme at NLO QCD
}
\author[a]{Prem Agarwal,}
\author[a]{Kirill Melnikov,}
\author[a,b]{Ivan Pedron,}
\author[a]{and Philip Pfohl}

\affiliation[a]{\TTPaff}
\affiliation[b]{\IAPaff}
\abstract{ We compute  next-to-leading power corrections in the zero-jettiness variable for the production of colorless final states at hadron colliders at next-to-leading order in QCD. To assess if  the process-independence of leading power contributions   can be extended, we attempt to construct  generic expansions of phase spaces and matrix elements squared through  next-to-leading power in the zero-jettiness. We highlight  challenges associated with the collinear limit, where universality no longer holds  at the subleading power, making the result process-dependent.
We show that 
quantities that need to be calculated in the collinear limit can be obtained  using  Berends-Giele currents,   enabling computation of power corrections  to high-multiplicity final states. As a concrete example, we apply our method to compute power corrections  in the zero-jettiness for lepton pair 
as well as 
multi-photon production in  $q \bar q$ collisions.
}

\begin{document}

\maketitle 
%==

\section{Introduction}

Early perturbative computations in QED (see, e.g. Ref.~\cite{Brown:1952eu}) were performed using methods that, currently, would be classified as ``slicing''. 
The idea of such a   method is to split the real emission contribution into singular and regular parts by introducing a parameter that  distinguishes between  unresolved (soft and collinear)  and  resolved (hard) radiation.  Although recognizing 
the difference between two types of emissions  is helpful, in practice  slicing methods suffer from large cancellations when resolved and unresolved parts of the calculation are 
combined. This drawback led to a rejection of the slicing methods as a suitable tool for 
computing higher-order perturbative (mostly QCD) corrections to complex  collider processes.   Since 
worthy computational alternatives  in the form of  subtraction schemes \cite{Catani:1996vz,Frixione:1995ms} appeared,  development of  slicing methods was put on a back burner for a while. 

Slicing methods made a remarkable comeback with the advent of next-to-next-to-leading order (NNLO) computations, starting with the proposal to use the transverse momentum of a color-singlet final state as a slicing parameter \cite{Catani:2007vq}.
Since this method was only suitable for processes without 
final-state jets at leading order, it was later suggested 
to use the so-called $N$-jettiness as the slicing parameter for generic NNLO computations~\cite{Stewart:2010tn, Boughezal:2015dva, Gaunt:2015pea}. 
We note in passing that other slicing variables have recently been proposed for processes with final-state jets \cite{Buonocore:2023rdw,Fu:2024fgj}.

Nevertheless, the use of slicing methods is still hindered by very large numerical cancellations between the different contributions to physical cross sections. These cancellations are caused by the need to take  the slicing parameter  to be  very small, 
to ensure the independence of the final result on its value. 
Thus, efficient numerical implementations remain  a challenge for modern slicing schemes, especially when applied to  complex processes.  To overcome this challenge, 
one needs to compute the \emph{unresolved} contribution
more accurately; to achieve this, a description of  
 real-emission amplitudes and cross sections  beyond leading soft and collinear limits is  required. 

Such power-suppressed terms were studied in a number of publications in recent years, 
focusing mostly on computations at next-to-leading order~\cite{Cieri:2019tfv,Boughezal:2016zws, DelDuca:2017twk, Boughezal:2018mvf, Moult:2016fqy, Moult:2017jsg, Ebert:2018lzn, Boughezal:2019ggi, vanBeekveld:2019prq,Oleari:2020wvt,Vita:2024ypr,Pal:2023vec,Pal:2024eyr,Beneke:2018gvs,Beneke:2019oqx,Bonocore:2015esa,Bonocore:2016awd,Broggio:2021fnr,Broggio:2023pbu,Ebert:2018gsn,Laenen:2020nrt}.  However,  these calculations typically address relatively simple processes and it  is unclear how to generalize them to \emph{arbitrary} collider processes and higher perturbative orders. 

The goal of this paper is to make a step in this direction and to  explore  
power corrections  that arise when 
a process,  where  an \emph{arbitrary} color-singlet final state 
is produced in the 
collision of a $q \bar q$ pair,  is studied in the context of the $N$-jettiness slicing scheme at NLO QCD.
Calculation of 
 power corrections in the $N$-jettiness variable for  such processes  requires us to 
understand the expansion of 
two building blocks --  the phase space and the matrix element squared  
-- around the limit of 
the vanishing $N$-jettiness
for the radiative process $q \bar q \to X+g$.

Since these building blocks appear to be process-dependent, it is crucial to investigate to what extent  a  process-independent calculation of the first subleading $N$-jettiness power correction is  possible. 
In this respect, the so-called Low-Burnett-Kroll (LBK) theorem~\cite{Low:1958sn, Burnett:1967km, DelDuca:1990gz, Engel:2021ccn}, that allows 
one to compute the next-to-soft 
corrections by calculating derivatives of the Born process 
with respect to momenta of external particles, serves as an 
inspiration. 
Similarly, 
next-to-collinear terms in the 
expansion of a generic matrix element can be related to matrix elements of simpler   processes~\cite{Ebert:2018lzn, Czakon:2023tld} although this case is more complex than the next-to-soft one. 

Another source of these power-suppressed terms is kinematic dependences of various observables or dependence of fiducial cross sections on selection cuts.  For simplicity, in this paper we will only consider observables that have smooth dependence on kinematics
although it is known that this is not always the case \cite{Ebert:2019zkb, Salam:2021tbm, Ebert:2020dfc}. As this paper focuses on providing general framework for  computing  power corrections to arbitrary collider processes, we do not examine such observables in what follows.

In what follows, we discuss the next-to-leading power corrections in the $N$-jettiness variable by considering 
the process $q \bar q \to X+g$ where $X$ 
is an arbitrary colorless state. In Section~\ref{sect2} we begin with the discussion of a toy example -- the power corrections to  vector boson production.  In  
Section~\ref{sect4} we turn to the study of an arbitrary process,  we explain how to use momenta transformations
to enable the expansion of the phase space and the matrix elements in the limit of small 
$N$-jettiness beyond  leading power. 
 In Section~\ref{sect3a} we  combine 
 the soft and collinear contributions 
 obtained  in the previous section, 
 and derive the formula for power corrections.  
  In  Section~\ref{sec:currents} we  explain how various quantities that appear in 
  the final formula can be computed by relating them to  generalizations of 
  Berends-Giele currents \cite{Berends:1988zn}. 
  In Section~\ref{sect:5} we first apply the general formula to the
  processes $q \bar q \to \gamma^* \to e^+e^-$ and 
  $q \bar q \to \gamma \gamma$,
  for which we derive expressions for   power corrections analytically, and then we compute the power correction to $q \bar q \to 4 \gamma$ numerically, further showcasing the general nature of the derived formula. 
  We conclude in Section~\ref{sect:conc}.
  Some technical aspects of the calculation are
discussed in  appendices. 
\\  

%generalize our ideas to consider the production of colorless final-states in a process-independent way. In particular, we consider the case where the kinematics of the colorless final-state particles are restricted by an observable, and we show the effect the observable has in the power corrections.

\section{A simple  example: vector boson production}
\label{sect2}

We start by considering the production of a vector boson in 
$q \bar q$ collisions, studying  power corrections that arise from both the phase space and the matrix element of this process. We note that studying this process has advantages and disadvantages. On the one hand, since this process is very simple,  many computational steps can be performed explicitly.   On the other hand,   because of its $2 \to 1$ nature, the phase space of this process is singular which precludes us from using the same computational methods that we develop in the rest of the paper to address power corrections to a general partonic process. 

 We consider the tree-level process
\be
q(p_1) + \bar q(p_2) \to V,
\label{eq2.0}
\ee
and define the cross section convoluted with a smooth function of the partonic center-of-mass energy squared
\be
\Sigma_0 = \int {\rm } {\rm d} s \; {\cal L}(s) \sigma_0(s).
\label{eq2.1}
\ee
We can think about this function as a product 
of parton distribution function integrated over the ratio of two Bjorken variables keeping their product fixed.  We note that the need to introduce this function into a computation is related to the fact that the phase space of 
process in Eq.~(\ref{eq2.0}) is overconstrained. Precisely for this reason, this step is not necessary when power corrections for more complex processes are studied.  

The cross section $\sigma_0$ for the process in Eq.~(\ref{eq2.0}) evaluates to
\be
\sigma_0 = \frac{2\pi \delta(s-m_V^2) }{8s N_c^2} \; |{\cal M}_0(p_1,p_2)|^2.
\ee
Under the assumption that the  vector boson only has vector coupling to quarks, $-i g_V \gamma^\mu$, the matrix element squared reads
\be
|{\cal M}_0(p_1,p_2)|^2 = g_V^2 N_c \; 4 s (1-\ep).
\ee
 We then find the following result for the convoluted cross section
\be
\Sigma_0 = \frac{\pi g_V^2}{N_c} (1 - \ep)  \; {\cal L}(m_V^2) . \; 
\ee

We are interested in developing a method for a systematic computation of 
power corrections in the zero-jettiness variable. 
To find them, we have to consider the real emission process 
\be
q(p_1) + \bar q(p_2) \to V + g(k), 
\label{eq2.6}
\ee
and work at fixed zero-jettiness  of the final state.
We therefore write
\be
\frac{1}{\Sigma_0} \frac{ {\rm d} \Sigma }{ {\rm d} \tau}
= \frac{1}{\Sigma_0}  \int {\rm d} s\;  {\cal L}(s) \frac{{\rm d} \sigma (s) }{{\rm d} \tau}.
\label{eq2.7c}
\ee

To compute the partonic cross section, 
we   work in the center-of-mass frame.
We write
\be
\frac{{\rm d} \sigma }{{\rm d} \tau} =  
\frac{1}{8 s N_c^2}
\int [{\rm d} k] [{\rm d} p_V] (2 \pi)^d \delta(p_1+p_2 - p_V - k) \delta(\tau - T_0) |M(p_1,p_2,k)|^2,
\ee
where $T_0$ is the zero-jettiness function defined as 
\be
T_0 = {\rm min}\left ( \frac{s_{1k}}{Q_1},\frac{s_{2k}}{Q_2} \right ),
\ee
with $s_{ik} = 2 p_i \cdot k$.
The two quantities $Q_{1,2}$ are arbitrary normalization constants. We will use  $Q_1 = Q_2 = Q$,  
throughout the paper, unless stated otherwise. 
Finally, the matrix element 
squared for the process in 
Eq.~(\ref{eq2.6}) reads 
\be
|{M}(p_1,p_2,k)|^2 = 8 C_F g_V^2 g_s^2 N_c \; (1-\ep) \left[ \frac{2 \ s \; m_V^2}{s_{1k} s_{2k}} + (1-\ep) \left( \frac{s_{1k}}{s_{2k}} + \frac{s_{2k}}{s_{1k}} \right) - 2 \ep \right],
\ee
with $s = 2 p_1  \cdot p_2$. 
\\

To proceed further, we integrate over the momentum of the vector boson $p_V$; the phase space becomes
\be
\begin{split} 
    [{\rm d} p_V] (2 \pi)^d \delta(p_1+p_2 - p_V - k)
   & = 2\pi \; {\rm d}^dp_V \; \delta(p_V^2 - m_V^2) \delta(p_1+p_2 - p_V - k)
\\
   & = 2 \pi \; \delta(s - m_V^2 - 2 \sqrt{s} \omega_k).
\end{split} 
    \ee
We remove the zero-jettiness $\delta$-function by integrating over gluon energy $\omega_k$, finding  that the energy of the gluon is given by 
\be
\omega^*_k = \frac{\tau Q}{\sqrt{s} \psi_k},
\ee
with 
\be
\psi_k = {\rm min} \left ( \rho_{1k}, \rho_{2k} \right ),\;\;\;\ \rho_{ik} = 1 - \cos \theta_{ik}.
\ee
We use this result in the expression for the cross section and obtain  
\be
\frac{{\rm d} \sigma }{{\rm d} \tau}
= \frac{ 2 \pi   [\alpha_s]}{8 s N_c^2} \int [{\rm d} \Omega_k]
\delta \left (s - m_V^2 - \frac{2 \tau Q}{\psi_k} \right )
\left (\frac{\tau Q}{\sqrt{s} \psi_k}  \right )^{1-2\ep} \frac{Q}{\sqrt{s} \psi_k}
\;  \frac{ |{\cal M}(p_1,p_2,k^*)|^2}{g_s^2},
\label{eq2.14}
\ee
where we introduced the following notations
\be
[\alpha_s] = \frac{g_s^2 \Omega^{(d-2)}}{2(2\pi)^{d-1}},
\;\;\;
[{\rm d} \Omega_k]
= \frac{{\rm d} \Omega_k^{(d-1)}}{\Omega^{(d-2)}} .
%= \frac{(4 \pi)^\ep}{8 \pi^2 %\Gamma(1-\ep)} 2 (2 \pi)^{d-%1}{\rm d} \Omega_k$.
\label{eq.2.15alphas}
\ee

Integrating over $s$, we find
\be
\frac{1}{\Sigma_0} \frac{ {\rm d} \Sigma }{{\rm d} \tau}
= \frac{ C_F  [\alpha_s]}{ \tau^{1+2\ep} }
\int [{\rm d} \Omega_k] \;  \frac{ {\cal L}\left (  s^*\right ) }{
  {\cal L}(m_V^2)} \;
\left (\frac{ Q}{\sqrt{s^*} \psi_k}  \right )^{2-2\ep} 
\;  \frac{ \tau^2 |{\cal M}(p_1,p_2,k^*)|^2}{C_F N_c g_s^2 g_V^2(1-\ep) 4 s^*},
\label{eq1.14} 
\ee
where
\be
s^* = m_V^2 + \frac{2 \tau Q}{\psi_k},
\ee
and we should use this value to evaluate energies associated with  momenta $p_{1,2}$.
\\

We have to construct an expansion of ${\rm d}\Sigma/{\rm d} \tau$ in Eq.~(\ref{eq1.14}) in powers of 
$\tau$. The expansion is controlled by the value of the emission angle $\theta_k$
of the gluon which is parametrized by 
$\Omega_k$. 
If  $\theta_k$ is ${\cal O}(1)$, the energy of the emitted gluon is ${\cal O}(\tau)$ and the
required expansion is the soft expansion.  
If, on the
other hand, $\theta_k$ or $\pi - \theta_k$ is ${\cal O}(\tau/m_V)$,  we have to expand 
in $\theta_k$ or $\pi - \theta_k$. This case corresponds to  two collinear contributions, since the  gluon can be emitted either along
the direction of $p_1$ or $p_2$. Two distinct integration regions -- the soft one and the collinear one --  are associated with two ``branches'' of the cross section with respect to  $\tau$: the soft 
contribution is proportional to $\tau^{-1-2\ep}$, whereas the collinear ones are proportional to  $\tau^{-1-\ep}$. Hence, schematically, we can write 
\begin{equation}
\frac{ d \Sigma}{{\rm d} \tau } \sim \tau^{-1-2\ep} f_s(\tau) 
+ \tau^{-1-\ep} f_c(\tau),
\end{equation}
and we compute each of the two  contributions by performing an 
appropriate Taylor expansion 
of the \emph{integrand} and then integrating the resulting expression over the angle $\theta_k$.
\\

We compute the  \emph{soft} contribution by  Taylor-expanding in $\tau$. We obtain 
\be
\begin{split} 
\frac{1}{\Sigma_0} \frac{ {\rm d} \Sigma^{(s)}  }{{\rm d} \tau}
& = \frac{ 4 C_F  [\alpha_s]}{ \tau^{1+2\ep} } \left(\frac{Q }{m_V  }\right)^{-2 \ep}
\int [{\rm d} \Omega_k]
\frac{  \psi_k^{2\ep}  }{\rho_{1k} \rho_{2k}}  
\\
& \times \left[ 1
- \frac{Q\tau}{m_V^2 \psi_k}
  \left (2 - 2 \ep  -  2  m_V^2 {\cal L}_1(m_V^2)  \right ) 
  \right ],
\end{split} 
\label{eq2.19}
\ee
where ${\cal L}_1(m_V^2) = {\cal L}^{-1}(m_V^2) \; {\rm d} {\cal L}(m_V^2)/{\rm d} m_V^2 $.
We note that ${\cal O}(\tau)$ correction contains 
$1/\psi_k$ factor which makes  the integration over angle 
$\theta_k$ \emph{power-divergent.} 
This is a feature that will also appear when zero-jettiness corrections to a general process are  discussed in the next section.  Due to the simplicity of the process and the knowledge of the matrix element, 
explicit integration over  $\theta_k$ is straightforward. It yields
\be
\frac{1}{\Sigma_0} \frac{ {\rm d} \Sigma^{(s)}  }{{\rm d} \tau}
 = \frac{ 4 C_F  [\alpha_s]}{ \ep \tau^{1+2\ep} } \left(\frac{Q }{m_V  }\right)^{-2 \ep}
\left[
 1
   - \frac{Q\tau}{m_V^2} \frac{(1-2\ep)}{(1-\ep)} \left ( 1-\ep -
  m_V^2 {\cal L}_1(m_V^2)  \right ) 
  \right ].
\ee

To construct a Laurent expansion in $\ep$, 
we write 
\be
\tau^{-1-n \ep} = -\frac{\delta(\tau)}{n \ep} + L_0 (\tau) - n \ep \ L_1 (\tau) + \mathcal{O}(\ep^2),
\label{eq1.19}
\ee
with
\be
L_n \left( x \right) = \left[ \frac{\theta(x) \log^n(x)}{x} \right]_{+}.
\label{eq2.22}
\ee
Using it, we obtain the  leading- and subleading-power soft contributions
\be
\begin{split}
\frac{1}{\Sigma_0} \frac{ {\rm d} \Sigma^{(s), LP}  }{{\rm d} \tau}
 = 4 C_F  [\alpha_s] \bigg[ & - \frac{\delta(\tau)}{2 \ep^2} + \frac{L_0 (\tau) + \delta(\tau) \ln \frac{Q}{m_V}}{\ep} - 2 L_1 (\tau) \\
 &-2 L_0 (\tau) \ln \frac{Q}{m_V} - \delta (\tau) \ln^2 \frac{Q}{m_V}  \bigg],
\end{split} 
\label{eq1.21}
\ee
\be
\begin{split}
\frac{1}{\Sigma_0} \frac{ {\rm d} \Sigma^{(s), NLP}  }{{\rm d} \tau}
 = \frac{4 C_F [\alpha_s] Q}{m_V^2} \bigg\{ & \frac{m_V^2 {\cal L}_1(m_V^2)-1}{\ep} \\
 &- \left[ m_V^2 {\cal L}_1(m_V^2) - 1 \right] \left( 1 + 2 \ln \frac{Q \tau}{m_V}\right) + 1 \bigg\}.\\
 \label{eq1.22}
\end{split}
\ee
\\
We continue with the construction of the \emph{collinear} expansion. We consider the case where the gluon is emitted along the direction 
of the incoming quark with the momentum 
$p_1$; the other case, where the gluon is emitted along the direction of the anti-quark with momentum $p_2$ is analogous. To construct the collinear
expansion, we go back to Eq.~(\ref{eq1.14}) and consider the case of the small emission angle , $\theta_{1k} \ll 1$. It follows that the
function $\psi_k$ in this case is given by $\rho_{1k}$.  We  need to perform the expansion of the integrand
assuming that the emission angle is small, and extend the integration region over the angle (or related variable)
to \emph{infinity}.  We write
\be
s^* \to m_V^2 + \frac{ 2 \tau Q}{\rho},
\ee
where $\rho = 1 - \cos \theta_{1k}$ and  assume that  $\rho \sim \tau Q/m_V^2$. The integration region over $\theta_k$ extends from $\rho = 0$ to $\rho = \infty$.  
We then change the integration variable $\rho \to z$, 
\be
\rho = \frac{2 \tau Q}{m_V^2} \frac{z}{1-z},
\ee
with $ 0 < z < 1$, expand the resulting formula in $\tau$ and obtain 
\be
\frac{1}{\Sigma_0} \frac{ {\rm d} \Sigma^{(c1)}  }{{\rm d} \tau}
= \frac{ C_F  [\alpha_s]}{ \tau^{1+2\ep} } \left ( \frac{\tau}{Q} \right )^{\ep} 
\int \limits_{0}^{1} \;
     \frac{{\rm d} z}{z} \frac{ {\cal L} \left ( m_V^2/z\right )}{{\cal L}(m_V^2)} \; (1-z)^{-\ep}
\left[
 \tilde P_{qq}(z)
 +
  \frac{\tau Q}{m_V^2} z \tilde P_{qq}^{(1)}(z)
  \right ],
\ee
where
\be
\tilde P_{qq}(z) = \frac{1+z^2}{1-z} - \ep(1-z),
\;\;\;\;
\tilde P_{qq}^{(1)}(z) = \frac{(4-z)z - 1 + 2\ep z - \ep^2 (1-z)^2 }{(1-z)^2}.
\ee

We note that the function $\tilde P_{qq}^{(1)}(z)$  
has a power singularity at $z= 1$. Again, we will see the appearance of these power singularities in the general case discussed in the next section.  In the current  case, since both the matrix element and the phase space are  simple,   we can integrate by parts obtaining expressions  that can be expanded in $\ep$
\be
\begin{split} 
& \int \limits_{0}^{1} \;
     \frac{{\rm d} z}{z} \; {\cal L} \left ( \frac{ m_V^2}{z} \right ) (1-z)^{-\ep}
    z \tilde P_{qq}^{(1)}(z) = \int \limits_{0}^{1} \;
           {\rm d} z \; (1-z)^{-1-\ep}
           \\
      & \times  \Bigg \{
         2 \left[1 + \frac{(1+\ep^2) (1-z)}{(1+\ep)} \right] {\cal L}\left ( \frac{m_V^2}{z} \right )
         - \frac{2 m_V^2}{z^2} \ {\cal L}^{'} \left (\frac{m_V^2}{z}  \right ) \left[ z -  
         \frac{ (1+\ep^2) (1-z)^2}{2 (1+\ep)} \right]
         \Bigg \}.
       \end{split} 
       \ee
It is straightforward to extract the $1/\ep$ divergences from the above expression  by applying Eq.~\eqref{eq1.19} to construct the $\ep$-expansion of 
 $\tau^{-1-\ep}$ and $(1-z)^{-1-\ep}$.  This leads to the leading-power contribution 
 \be
\begin{split}
\frac{1}{\Sigma_0} \frac{ {\rm d} \Sigma^{(c1), LP}  }{{\rm d} \tau}
= & C_F [\alpha_s] \int \limits_{0}^{1} \; \frac{{\rm d} z}{z} \frac{ {\cal L} \left ( m_V^2/z\right )}{{\cal L}(m_V^2)} \; \bigg\{ \frac{2 \delta(\tau) \delta(1-z)}{\ep^2} - \frac{\delta(\tau)(1+z^2)L_0(1-z)}{\ep} \\
& - 2 \frac{\delta(1-z) \left[L_0(\tau)+\delta(\tau) \ln Q\right]}{\ep} + \delta(\tau) \delta(1-z) \ln^2 Q \\
& + \delta(\tau) \left[ 1-z + (1+z^2) L_1 (1-z)+(1+z^2)L_0(1-z)\ln Q\right] \\
& + 2 \delta(1-z) \left[ L_1(\tau) + L_0(\tau) \ln Q\right] + (1+z^2) L_0 (\tau) L_0 (1-z)\bigg\},
\end{split}
\label{eq1.28}
\ee
and the next-to-leading power contribution 
\be
\begin{split}
\frac{1}{\Sigma_0} \frac{ {\rm d} \Sigma^{(c1), NLP}  }{{\rm d} \tau}
= -& C_F [\alpha_s] \frac{Q}{m_V^2} \int \limits_{0}^{1} \; {\rm d} z \frac{ {\cal L} \left ( m_V^2/z\right )}{{\cal L}(m_V^2)} \; \bigg\{ \frac{2 \delta(1-z) \left[ m_V^2 \ {\cal L}_1 \left ( m_V^2\right ) - 1 \right] }{\ep} \\
& + L_0 (1-z) \left[ 4 - 2 z + \frac{ m_V^2 \ {\cal L}_1 \left ( m_V^2/z\right )  \left( 1 + (z-4)z \right) }{z^2} \right] \\
& + 2 \delta(1-z) \ln \left( Q \tau \right) \left[1 - m_V^2 \ {\cal L}_1 \left ( m_V^2\right) \right] \bigg\}.
\end{split}
\label{eq1.29}
\ee
As we already mentioned, the above results describe the collinear emissions off the 
quark with momentum $p_1$. 
The contribution where an anti-quark with momentum $p_2$ emits  collinear gluons,  doubles the above result.

By comparing Eqs.~\eqref{eq1.22} and~\eqref{eq1.29},  we observe that 
at next-to-leading power all $\ep$-divergences cancel between the soft and collinear contributions.  This is the main result of our analysis.  However, for completeness, we also comment on the leading-power contribution which is well-understood. 

Indeed, to get the cancellation of $\ep$-poles  at leading power,  the virtual contributions and the collinear renormalization  contributions need to be added to the real emission one.
For $q \bar q \to V$ annihilation,  they are given by the following expressions 
\be
\frac{1}{\Sigma_0} \frac{ {\rm d} \Sigma^{(v)}  }{{\rm d} \tau}
= C_F  [\alpha_s] (m_V)^{2 \ep} \left( -\frac{2}{\ep^2}-\frac{3}{\ep} + \frac{2 \pi^2}{3}\right) \delta(\tau),
\label{eq2.32}
\ee
and
\be
\frac{1}{\Sigma_0} \frac{ {\rm d} \Sigma^{(cs1)}  }{{\rm d} \tau}
= \frac{C_F  [\alpha_s] (m_V)^{2 \ep}}{\ep} \int \limits_{0}^{1} \; \frac{{\rm d} z}{z} \frac{ {\cal L} \left ( m_V^2/z\right )}{{\cal L}(m_V^2)}
\left[ (1+z^2) L_0(1-z) + \frac{3}{2} \delta(1-z)\right] \delta(\tau).
\label{eq2.33}
\ee
Upon combining Eqs.~(\ref{eq2.32}, \ref{eq2.33},
\ref{eq1.28}, \ref{eq1.21})
and accounting for the second collinear region, we observe the cancellation of all $1/\ep$ poles in the leading-power contribution. 
\\

The next-to-leading power correction to the zero-jettiness 
regulated cross section for $q \bar q \to V$ is obtained upon combining  results from Eqs.~(\ref{eq1.22}, \ref{eq1.29}).  It reads 
\begin{equation}
\begin{split}
\frac{1}{\Sigma_0} 
\frac{{\rm d} \Sigma^{NLP}}{{\rm d} \tau} 
& = \frac{4 C_F [\alpha_s] Q}{m_V^2} \bigg\{ 1+\left[ 1 - m_V^2 \ {\cal L}_1 \left ( m_V^2\right ) \right] \left( 1 + \ln \frac{Q \tau}{m_V^2}\right) \\
& - \int \limits_{0}^{1} {\rm d} z \frac{ {\cal L} \left ( m_V^2/z\right )}{{\cal L}(m_V^2)} L_0 (1-z) \left[ 2 - z + \frac{ m_V^2 {\cal L}_1 \left ( m_V^2/z\right )  \left( 1 + (z-4)z \right) }{2 \ z^2} \right] \bigg\},
\end{split}
\label{eq2.34}
\end{equation}
where ${\cal L}_1$ is defined right after Eq.~(\ref{eq2.19}) and $L_0$ in Eq.~(\ref{eq2.22}).

To check this result, we 
make use of the fact that, when working at finite $\tau$, the real emission contribution in \eqref{eq1.14}  is $\ep$-finite and can be computed by numerical integration over the gluon emission angle $\theta_k$.  We then write a small-$\tau$ expansion
\be
\frac{1}{\Sigma_0}
\frac{{\rm d} \Sigma}{{\rm d} \tau } 
 = \tau^{-1} 
 \left ( A_{1} 
 \ln \tau 
 + A_2 \right ) 
 + A_3 \ln \tau + A_4 + 
 \tau \left ( A_5 \ln \tau   + A_6 \right )  + A_7 \tau^2 + \dots.
\ee
The coefficients $A_{1,..,7}$
are functions of $m_V, Q$ and 
the function ${\cal L}$, for which  we take a Gaussian distribution 
\begin{equation}
{\cal L}(s) = \frac{2}{\sqrt{2 \pi} \sigma_s} \exp\left( -\frac{s^2}{2 \sigma_s^2} \right).
\label{eq2.36}
\end{equation}
For numerical calculations, we
 choose 
$\sigma_s^2 =  2.47 ~ m_V^2$, and $m_V = 90~{\rm GeV}$.

\begin{figure}
    \centering
    \includegraphics[scale=0.8]{fig/vector_production_X.pdf}
    \caption{The parameter $X(\tau)$ expressed as a function of $\xi$, which relates the small-jettiness expansions of the vector boson production cross section and a numerical integration of \eqref{eq1.14} at $Q = 100~{\rm GeV}$ and $m_V = 90~{\rm GeV}$ and with the Gaussian luminosity distribution given in Eq.~(\ref{eq2.36}). The parameter approaches unity for lower values of $\xi$ before the numerical integration becomes unreliable. A fit of the missing linear terms in $\tau$ is presented in blue.}
    \label{fig1}
\end{figure}

Both the leading-power coefficients $A_1,A_2$, and 
the next-to-leading power 
corrections $A_3, A_4$ are obtained 
from Eqs.~(\ref{eq1.21}, \ref{eq1.22}, \ref{eq1.28}, \ref{eq1.29}).
We then write 
\begin{equation}
X(\tau) = \frac{1}{A_4} 
\left [  \frac{1}{\Sigma_0}
\frac{{\rm d} \Sigma}{{\rm d} \tau } 
 - A_{1} \frac{\ln \tau}{\tau}
 - \frac{A_2}{\tau} - A_3 \ln \tau
\right ],
\end{equation}
and note that at small $\tau$ this quantity should approach one, 
\be
\lim_{\tau \to 0}^{} X(\tau)  
 = 1.
\ee
Since $\tau$ is a dimensionful variable, the expansion proceeds in  powers of the parameter $\xi = Q \tau/m_V^2$. Hence, choosing 
$\tau < 0.1~m_V$ 
and $Q = 100~{\rm GeV}$, 
we expect that the deviation  of $X(\tau)$ from unity  is about a percent at  $\xi \sim 0.1$, and that it decreases  
at  smaller values of $\tau$. Numerical results for $X(\tau)$ are shown in Figure \ref{fig1}, where $X$ is plotted as a function of $\xi$. We observe  that $X(\tau)$ 
approaches the limiting value $X(\tau) = 1$ at  $\xi > 10^{-8}$ and that for smaller values of $\xi$,   it starts to deviate from its  limiting value  due to the loss of accuracy in the numerical integration. A fit of the next-to-next-leading-power terms is also shown, demonstrating they can explain the (tiny)  difference between $X(\tau)$ and its asymptotic value.  
\\

The example presented in this section showcases the main aspects of the calculation of the zero-jettiness power corrections to hadron collisions processes.  In the next section, we will discuss how to generalize this computation to arbitrary color-singlet final state. However, before doing that, we would like to 
reiterate the following points that emphasize the differences between vector boson production and a general case. 

\begin{itemize}
    \item In vector boson production, the final-state phase space is  simple, so that one can  easily parametrize it and perform all integrations explicitly.  However,  for  processes with higher final-state multiplicities this is not  possible. Thus, new approaches are needed to enable a systematic way to expand an arbitrary phase space  in the zero-jettiness variable.

    \item The  matrix element that describes vector boson production is simple and can be written down explicitly. Hence, expanding it through subleading terms in relevant limits  is straightforward.  However,  for  high-multiplicity final states it may not be possible to compute the relevant matrix element analytically.   Hence, one needs to understand how to compute subleading terms in the  soft and collinear expansions  in a process-independent manner. 
    
    \item In the case of vector boson production the phase space is overconstrained,  i.e. it is proportional to $\delta(s - M_V^2)$ at leading order.  This feature is particular to $2 \to 1$ processes, and it makes the  zero-jettiness expansion  of  \emph{partonic} processes in such cases  more complex,  than for  processes with more final-state particles.  To ameliorate this problem,     the zero-jettiness expansion of \emph{hadronic} cross sections for $2 \to 1$  processes is usually studied.     In the toy example discussed in this section, we introduced a luminosity function and computed an analog of a hadronic cross section.  
    In the result for subleading-power corrections shown in Eq.~(\ref{eq2.34}), the derivative of the luminosity function $\mathcal{L}$  appears, which is analogous to the derivatives of parton distribution functions found in the previous studies of the Drell-Yan and Higgs production processes~\cite{Boughezal:2016zws, Moult:2016fqy, Moult:2017jsg, Boughezal:2018mvf, Ebert:2018lzn}.
    \\
\end{itemize}

\section{Power corrections: general considerations}
\label{sect4}

We consider the following leading order process 
\be
f_a(p_a) + f_b (p_b) \to X(P_X),
\label{eq4.1}
\ee
where $f_a$ and $f_b$ are the  initial-state  partons,  which we take to be a quark and an anti-quark, and  $X$  denotes a generic colorless final state with the momentum $P_X$ composed of $m$ massless particles. 
To discuss  the next-to-leading power corrections in the  zero-jettiness variable, we  add a gluon with the momentum $k$ to the process in Eq.~(\ref{eq4.1})
 and 
write the differential cross section as 
\be
\begin{split} 
\frac{ {\rm d}  \sigma}{{\rm d} \tau}
& = {\cal N}
\int [{\rm d} \tilde P_X]_m [{\rm d} k]
(2\pi)^d \delta( p_a + p_b - \tilde P_X - k  )
\\
& \times \delta(\tau - T_0(p_a,p_b,k) ) \; {\cal O}(\tilde P_X) \; \sum \limits_{\rm col, pol}^{} |{\cal M}|^2(p_b,p_a,k,\tilde P_X).
\end{split}
\label{eq4.2}
\ee
In Eq.~(\ref{eq4.2}) we used  $[{\rm d} k]={\rm d}^{d-1}
\vec k/(2 \; (2\pi)^{d-1} k^0)$ and 
$[{\rm d} \tilde P_X]_m = \prod \limits_{i=1}^{m} [{\rm d} \tilde p_i]$. The 
zero-jettiness function  is defined as follows
\be
T_0(p_a,p_b,k) = {\rm min}\left[ \frac{2 p_a \cdot k}{Q},\frac{2 p_b \cdot k}{Q} \right ],
\ee
with $Q$ being an arbitrary normalization factor
of mass dimension one. Furthermore, 
${\cal O}$ is an observable that depends on the momenta of  colorless particles 
comprising the 
final state $X$, and 
${\cal N}$ is the 
cross-section normalization  that contains  the flux factor, color- and spin-averaging 
terms, etc. We note that 
we have used a new notation for the momentum of 
the colorless final state  
by writing it with a tilde, 
$P_X \to \tilde P_X$. The reason for doing this will become clear later. 

Using a reference frame where the collision 
axis is the $z$-axis,  and 
writing the zero-jettiness variable 
in terms of the energy and the polar angle  of the emitted gluon,  it is easy 
to see that the 
constraint $\tau = T_0(p_a,p_b,k)$ implies that  either the gluon energy or  its transverse momentum squared is  ${\cal O}(\tau)$. The expansions around these distinct limits  can be performed independently of each other, as we show below.   We will start with the construction 
of the soft expansion.

\subsection{The soft contribution}
\label{sec:soft}

A gluon with momentum $k$ is considered to be soft if  $k \sim \tau$. Since we  are interested in the relative  ${\cal O}(\tau/\sqrt{s})$ correction, where 
$s = 2 p_a \cdot p_b$,  we only need 
to expand the integrand in 
Eq.~(\ref{eq4.2}) 
to the \emph{first} subleading order 
in $k$.
 To facilitate  this expansion,  we use the momentum mapping that absorbs  $k$ into the momentum of the  colorless final state~\cite{DelDuca:2019ctm}, and  
 write 
\begin{equation}
\label{eq2.3}
P^\mu_{ab} = \lambda^{-1} \left [ \Lambda_s \right ]\indices{^\mu_\nu} (P^\nu_{ab}-k^\nu).
\end{equation}
In Eq.~(\ref{eq2.3})  $P_{ab} = p_a + p_b$,
$\Lambda_s^{\mu \nu}$ is 
the matrix of a Lorentz boost that we specify below, and  $\lambda$ is a constant that  is defined from the condition 
\be
\lambda^2 P_{ab}^2 = (P_{ab} - k)^2.
\ee
It follows from the above equation that 
\begin{equation}
\lambda = \sqrt{ 1 - \frac{2 P_{ab} \cdot k}{P_{ab}^2}} 
\approx 
1- \frac{P_{ab} \cdot  k}{P_{ab}^2} + {\cal O}(k^2).
\end{equation}
We then write 
\begin{equation}
\begin{split}
{\rm d} \Phi_m(p_a,p_b,\tilde P_X,k) & = [{\rm d} \tilde P_X]_m [{\rm d} k]
(2\pi)^d \delta^{(d)}( p_a + p_b  - \tilde P_X - k )
\\
&  = 
 [{\rm d} \tilde P_X]_m [{\rm d} k]
(2\pi)^d \delta^{(d)}\left ( 
\lambda \Lambda_s^{-1} P_{ab}
- \tilde P_X \right )
\\
& = [{\rm d} \tilde P_X]_m [{\rm d} k]
(2\pi)^d \delta^{(d)}\left ( 
\lambda \Lambda_s^{-1} \left ( P_{ab}
- \lambda^{-1} \Lambda_s\tilde P_X \right ) \right )
\\
 & = 
 [{\rm d} \tilde P_X]_m [{\rm d} k] \; 
\lambda^{-d} \; (2\pi)^d \delta^{(d)}\left ( 
 P_{ab}
- \lambda^{-1} \Lambda_s\tilde P_X  \right ).
\end{split}
\end{equation}
To further simplify this expression, we use the fact that $\tilde P_X = \sum \limits_{i=1}^{m} \tilde p_i$, so that 
\begin{equation}
[{\rm d} \tilde P_X]_m 
= \prod \limits_{i=1}^{m} 
\frac{{\rm d}^d \tilde p_i}{(2\pi)^{d-1}} \; \delta_+({\tilde p}^2_i  ). 
\end{equation}
 We then write 
\be
\tilde p_i = \lambda \Lambda_s^{-1} p_i,
\ee
and  since $\Lambda_s$ is a Lorentz transformation, we find 
\begin{equation}
[{\rm d} \tilde P_X]_m 
= \lambda^{m(d -2)} \prod \limits_{i=1}^{m} 
\frac{{\rm d}^d p_i}{(2\pi)^{d-1}} \delta({ p}^2_i  ) 
 = \lambda^{m(d -2)} [{\rm d}  P_X]_m.
\end{equation}
Hence, we obtain 
\begin{equation}
\begin{split}
{\rm d} \Phi_m(p_a,p_b,\tilde P_X,k) & = {\rm d} \Phi_m(p_a,p_b,P_X) \; [{\rm d} k]
\; \lambda^{m(d-2) - d} 
\\
& \approx {\rm d} \Phi_m(p_a,p_b,P_X) \; [{\rm d} k] \; 
\left ( 1 - \kappa_m \; \frac{P_{ab} \cdot k}{P_{ab}^2}
\right ),
\end{split}
\end{equation}
where 
\be
{\rm d} \Phi_m(p_a,p_b,P_X) 
= [{\rm d} P_X]_m\; (2\pi)^d \delta(p_a
+p_b - P_X),
\ee
is the phase space of the Born process 
$q \bar q \to X$, 
and 
\be
\kappa_m = m(d-2) - d.
\label{eq4.12}
\ee
Putting everything together, we 
find 
\begin{equation}
\begin{split} 
\frac{{\rm d} \sigma }{
{\rm d} \tau
} & = 
{\cal N} 
\int {\rm d} \Phi_m(p_a,p_b,P_X)
  \;  \int [{\rm d} k]
\left ( 1 - \kappa_m \frac{P_{ab} \cdot k}{P_{ab}^2}
\right )
\delta(\tau - T_0(p_a,p_b,k) 
)
\\
& \times \; {\cal O}( \lambda \Lambda_s^{-1} P_X) 
\sum \limits_{\rm col,pol}
|{\cal M}|^2(p_b,p_a,k,\lambda \Lambda_s^{-1}  P_X).
\end{split}
\label{eq4.15}
\end{equation}

Since we are interested in 
${\cal O}(\tau)$ corrections, 
we  need the matrix element squared to the first subleading order in the expansion in $k$. 
The matrix element itself scales as $1/k$, so that we need to find ${\cal O}(1)$ terms in the expansion. The required terms can be obtained  from the Low-Burnett-Kroll theorem \cite{Low:1958sn,Burnett:1967km}, as we  explain shortly.  

Before discussing the expansion of the matrix element, we derive the formula  for the Lorentz boost $\Lambda_s^{-1}$. We start with a 
general formula for the boost 
$\Lambda_{\rm gen}(Q_f,Q_i)$, that transforms a vector $Q_i$ to a vector $Q_f$; this formula can be found   in Eq.~(\ref{eqa.2}). 
%In the next  step, we need to %construct the Lorentz %transformation explicitly. This %is straightforward since  \be
%\Lambda_{\rm gen}(\tilde Q, %Q)^{\mu}_{\nu} = g^{\mu}_{\nu} -
%\frac{2( \tilde Q +  Q)^\mu ( %\tilde Q + Q)_{\nu}}{(Q+\tilde %Q)^2} + \frac{ 2 \tilde Q^\mu %Q_\nu }{Q^2},
%\label{eq4.5}
%\ee
%is a Lorentz transformation %matrix 
%that boosts a vector $Q$ to  
% $\tilde Q$, i.e. 
% \begin{equation}
%\tilde Q^\mu = \Lambda_{\rm gen}%(\tilde Q, Q )^{\mu}_{\nu} Q^\nu.
% \end{equation}
The Lorentz transformation that we need (c.f. Eq.~(\ref{eq2.3})) 
reads 
 \be
\Lambda_s^{-1} = \Lambda_{\rm gen}(P_{ab}-k,\lambda P_{ab} ).
 \ee
  Since in the soft limit  $k \sim \tau$, $\Lambda_s^{-1}$ is nearly the identity matrix; to determine 
  ${\cal O}(\tau)$ corrections 
  to the cross section, 
we need the Lorentz boost to the first order in $k$.  Simplifying the expression for $\Lambda_s^{-1}$, we find 
 \begin{equation}
\left [ \Lambda^{-1}_s \right ] _{\mu \nu}= 
g_{\mu \nu}  - B_{\mu \nu} +{\cal O}(k^2),
\end{equation}
where 
\be
B^{\mu \nu} = 
\frac{k^\mu P_{ab}^\nu - P_{ab}^\mu k^\nu}{P_{ab}^2}.
\end{equation}
\\

We turn to the discussion of the expansion 
of the matrix element in the soft limit. 
We ignore the color charges since for the process we consider it is trivial to restore them at the end of the calculation. Separating emissions off the external legs and the structure-dependent radiation, we write  
\begin{equation}
\begin{split}
{\cal M}(p_b,p_a,k,\tilde P_X) & = -g_s \epsilon^*_\mu 
\bar v_b \Bigg [ 
 N(p_b, p_a-k,\tilde P_X) 
\left ( J_a^\mu+ S_a^\mu \right ) 
\\
& + \left ( 
-J_b^\mu + S_b^\mu 
\right ) N(p_b-k,p_a,\tilde P_X)
+ N^\mu_{\rm str}(p_b,p_a,k,\tilde P_X) 
\Bigg  ] 
u_a,
\end{split}
\end{equation}
where 
\begin{equation}
\begin{split}
& J_a^\mu = \frac{2 p^\mu_a - k^\mu }{d_a}, \;\;\;
J_b^\mu = \frac{2 p_b^\mu - k^\mu}{d_b},
\;\;\; S_a^\mu = \frac{\sigma^{\mu \nu} k_\nu}{d_a},
\;\;\;
S_b^\mu = \frac{\sigma^{\mu \nu} k_\nu}{d_b},
\end{split}
\end{equation}
with $d_a = (p_a-k)^2$, $d_b = (p_b-k)^2$ and $\sigma^{\mu \nu} = [\gamma^\mu, \gamma^\nu]/2$.
\\

The structure-dependent contribution to the amplitude can be restored by requiring 
that the Ward identity is  fulfilled, namely that the amplitude 
vanishes if  the gluon  polarization vector $\ep^\mu$ is replaced with its momentum $k^\mu$.
This implies that in the soft limit
\begin{equation}
\begin{split}
{\cal M}   = -g_s \epsilon^*_\mu \bar v_b 
\Bigg  [ 
 N(p_a-k) &
\left (  J_a^\mu+ S_a^\mu \right ) 
 + \left ( 
-J_b^\mu + S_b^\mu 
\right ) N(p_b-k)
\\
& - \left [ \frac{\partial}{\partial p_{a,\mu}}
- \frac{\partial }{ 
\partial p_{b,\mu}}
\right  ] N
\Bigg ]
u_a,
\end{split}
\end{equation}
where we only show the $k$-dependent momenta in the arguments of the function $N$.
Since the currents 
$J_{a,b}^\mu$ scale as $1/k$, we need to expand the $k$-dependent 
functions $N$ in powers of  the gluon momentum.
We then obtain 
\begin{equation}
\begin{split}
{\cal M} &  = -g_s \epsilon^*_\mu \bar v_b 
\Bigg  [ 
J^\mu N
 - ( L^\mu N )
+ 
\left (N S_a^\mu + S_b^\mu N \right )
\Bigg ]
u_a ,
\end{split}
\end{equation}
where the function $N$ is now $k$-independent and 
\begin{equation}
J^\mu = J^\mu_a -J^\mu_b ,
\;\;\;\;\;\;
L^\mu = L_{a}^{\mu} - L_{b}^{\mu},
\end{equation}
with
\be
L_a^\mu = J_a^\mu 
k^\nu \frac{\partial}{\partial p_a^\nu}
+ \frac{\partial}{\partial p_{a,\mu}}
,
\;\;\;\;\;\;\;
L_b^\mu = J_b^\mu 
k^\nu \frac{\partial}{\partial p_b^\nu}
+ \frac{\partial}{\partial p_{b,\mu}}
.
\ee
Upon squaring the soft amplitude and summing over polarizations of 
all external particles, we find 
\begin{equation}
g_s^{-2} |{\cal M}(p_b,p_a,k,\tilde P_X)|^2
\approx  - J_\mu J^\mu 
|{\cal M}|^2(p_b,p_a,\tilde P_X)
+ J^\mu L_\mu
|{\cal M}|^2(p_b,p_a,\tilde P_X) 
+ \dots,
\label{eq4.25}
\end{equation}
where the ellipses denote terms that are finite in the $k \to 0$ limit. 
We note that the first term on the right-hand side in Eq.~(\ref{eq4.25}) 
provides the leading contribution that scales as $1/k^2$. Hence, we need to account for the momenta redefinitions in that term. Momenta redefinitions impact  
particles  that comprise the color-singlet system $X$. Working 
through first subleading order in $k$, we obtain
\begin{equation}
\begin{split}
g_s^{-2} |{\cal M}(p_b,p_a,k,\tilde P_X)|^2
\approx \Bigg [ & - J_\mu J^\mu 
\left ( 1 - \sum \limits_{i=1}^{m}
\left [ 
\frac{P_{ab} \cdot k}{P_{ab}^2} p_i^\rho+ 
B^{\rho \sigma} p_{i, \sigma}
\right ]
\frac{\partial }{\partial p_i^\rho}
\right ) 
\\
& + J^\mu L_\mu
\Bigg ]
|{\cal M}|^2(p_b,p_a, P_X).
\end{split} 
\end{equation}
Combining  this expression with 
Eq.~(\ref{eq4.15}), 
we observe that the integration over $k$ 
can be performed  
in a process-independent way  
and that for computing  the 
soft contribution to the
zero-jettiness cross section   through next-to-leading power,   
we need  
to calculate  two distinct integrals 
\begin{equation}
\begin{split}
I_1 & = g_s^2 \int [{\rm d} k] \;
\delta(\tau - T_0(p_a,p_b,k) 
) \; \frac{ 2p_a \cdot p_b}{ (p_a \cdot k) (p_b \cdot k)}, \; 
\\
I_2^{\mu} & = g_s^2 \int [{\rm d} k] \;
\delta(\tau - T_0(p_a,p_b,k) 
) \; \frac{k^\mu }{ (p_a \cdot k) (p_b \cdot k)}.
\end{split}
\end{equation}
The second integral can be written as 
\begin{equation}
I_2^\mu = I_2 \; \frac{ P_{ab}^\mu}{s},
\end{equation}
so that 
\begin{equation}
I_2 = 2 I_2^\mu p_{b,\mu} 
 = g_s^2 \int [{\rm d} k]
\delta(\tau - T_0(p_a,p_b,k) 
) \; \frac{2}{ (p_a \cdot k)}.
\end{equation}

Using these definitions, we find the following results
for the integrals that are needed to compute ${\rm d} \sigma/{\rm d} \tau$ in the soft limit
\begin{equation}
\begin{split}
    & g_s^2 \int [{\rm d} k] \; 
\delta \left(\tau - T_0(p_a,p_b,k) \right)
\left ( 1 - \kappa_m \frac{P_{ab} \cdot k}{P_{ab}^2}
\right ) \; (-J_\mu J^\mu )
 = I_1   - \kappa_m  I_2, 
 \\
 &  g_s^2 \int [{\rm d} k] \; 
\delta \left(\tau - T_0(p_a,p_b,k) \right)
 \; (J_\mu J^\mu )
\frac{P_{ab} \cdot k}{P_{ab}^2} p_i^\rho
  = -I_2 p_i^\rho,
  \\
   &  g_s^2 \int [{\rm d} k] \; 
\delta \left(\tau - T_0(p_a,p_b,k) \right)
 \; (J_\mu J^\mu )
B^{\rho \sigma}
  = 0,
  \\
   &  g_s^2 \int [{\rm d} k] \; 
\delta \left(\tau - T_0(p_a,p_b,k) \right)
 \; J^\mu L_\mu
  = - I_2
  \left ( 
p_a^\mu \frac{\partial}{\partial p_a^\mu}
+ p_b^\mu \frac{\partial}{\partial p_b^\mu}
  \right ).
\end{split}
\end{equation}
Putting everything together, and accounting for the fact that the observable 
${\cal O}$ also depends on the boosted momenta, we obtain 
\begin{equation}
\begin{split} 
\frac{{\rm d} \sigma^{(s)} }{
{\rm d} \tau
} & = 
{\cal N}
\int [ {\rm d} \Phi_m(p_a,p_b,P_X) ]
\;   \Bigg \{  {\cal O}(P_X) \;
\Bigg [ 
I_1 - \kappa_m I_2 
\\
&  - I_2 \sum \limits_{i \in L_f }^{}
p_i^\mu \frac{\partial }{\partial p_i^\mu }
\Bigg  ]
|{\cal M}|^2(p_b,p_a,P_X)
- I_2 \; |{\cal M}|^2(p_b,p_a,P_X)
\sum \limits_{i =1}^{m}
p_i^\mu \frac{\partial }{\partial p_i^\mu }
{\cal O}(P_X)
\Bigg \}
,
\end{split}
\label{eq5.29}
\end{equation}
where $L_f$ is the list that includes all particles in the Born process Eq.~(\ref{eq4.1}).
We note that in the last term 
in Eq.~(\ref{eq5.29})
the sum can be extended to include initial partons if an observable does not depend on them and the corresponding derivatives vanish.

To finalize the calculation, we need to compute  integrals $I_{1,2}$.  To do this, we integrate 
over the energy of the gluon with momentum $k$,
removing  the zero-jettiness $\delta$-function. 
Then, using the following expressions for the angular integrals 
\begin{equation}
    \int \left[ {\rm d} \Omega^{(d-1)}_{\vec k} \right]  \;   \frac{\psi_k^{2\ep}}{\rho_{ak} \rho_{bk}}
  = \frac{1}{\ep},
  \;\;\;\;\int \left[ {\rm d} \Omega^{(d-1)}_{\vec k} \right]  \;  
  \;  \frac{\psi_k^{2\ep-1}}{\rho_{ik}}  =
\frac{1}{2 \ep} -\frac{1}{2}  -\frac{\ep}{2} + {\cal O}(\ep^2),\;\;\; i = a,b, 
\label{eq3.31}
\end{equation}
where $\left[ {\rm d} \Omega_{\vec k}^{(d-1)} \right] = {\rm d} \Omega_{\vec k}^{(d-1)}/\Omega^{(d-2)}
$, 
$\psi_k = {\rm min}(\rho_{ak},\rho_{bk})$ 
and $\rho_{ik} = 1 - \cos \theta_{ik}$, $i = a,b$,
we find 
\begin{equation}
I_1= 
  [\alpha_s] 
\left ( \frac{Q}{\sqrt{s} }\right )^{-2\ep} \frac{4}{\ep \; \tau^{1+2\ep}},
\;\;\;
I_2 = 
[\alpha_s] 
\left ( \frac{Q \tau}  {\sqrt{s}} \right )^{-2\ep} 
\frac{4 Q}{s}
\left ( 
\frac{1}{2 \ep}
-\frac{1}{2} 
-\frac{\ep}{2} 
+{\cal O}(\ep^2)
\right ),
\label{eq2.32a}
\end{equation}

where $[\alpha_s]$ is defined in Eq. (\ref{eq.2.15alphas}). It is straightforward to use 
Eq.~(\ref{eq5.29}) together with the 
results for the two integrals $I_{1,2}$ 
to determine both leading and subleading 
 zero-jettiness 
contributions to the cross section 
of a process  in Eq.~(\ref{eq4.1}), that 
originate from the emission of a soft gluon.
Since the 
physical result requires including the contributions of the collinear emissions, 
we refrain from presenting the expansion 
of Eq.~(\ref{eq5.29}) in powers of $\ep$. Nevertheless, 
for illustration purposes, 
we show the  $1/\ep$-divergence of the subleading soft contribution which can be  easily obtained from Eq.~\eqref{eq5.29}. This contribution comes entirely from the divergent part of the integral $I_2$. After restoring the appropriate color factor, we obtain
\be
\frac{{\rm d} \sigma^{s,\rm div} }{
{\rm d} \tau
 } = -\frac{2 C_F [\alpha_s] \; {\cal N}  }{\ep} 
 \frac{(Q \tau)^{-2\ep}}{s^{-\ep} } \frac{Q}{s}
\left ( \kappa_m + \sum \limits_{i \in L_f} 
p_i^\mu \frac{ \partial}{\partial p_i^\mu} 
\right ) 
|{\cal M}|^2 \;  {\cal O}(P_X)
.
\ee
%Th complete result for the subleading soft contribution is also easy to obtain. We use it in 
%Section~\ref{sect3a}, where  
%soft and colliear contributions 
%are combined, and the complete 
%result for next-to-leading power %correction   is obtained. 
\\

\subsection{The first collinear contribution: $\vec k || \vec p_a$ }
\label{sec:cola}

As the next step, we need to construct  expansions in the  zero-jettiness variable around  the collinear limits.  We will start with the case where the gluon is emitted along the direction of the incoming quark with momentum $p_a$. The case where the gluon is emitted along the direction of the incoming anti-quark is completely analogous; we discuss it in the next subsection. 

Similarly to the case of the soft emission considered earlier, we perform a momenta mapping \cite{DelDuca:2019ctm} that allows us to construct the collinear expansion. To do this,  we start by re-writing  the gluon momentum $k$ as follows
\be
k  = \frac{k \cdot P_{ab} }{p_a \cdot p_b} p_a  + \tka = (1-x) p_a  + \tka.
\label{eq4.34}
\ee
The momentum conservation condition\footnote{At variance with the previous section, here we  denote the momentum of the colorless final state $X$ as $\tilde Q_X$. We do this because 
we need  several redefinitions of this momentum,  before we reach the final formula 
in Sec.~\ref{sect3a}. There,  we will 
return to the notation $P_X$ for the momentum of the final state $X$.
} 
\be
p_a + p_b = k + \tilde Q_X,
\ee
becomes
\be
x p_a + p_b - Q_X = 0,
\ee
where 
\be
Q_X =  \tilde Q_X + \tka.
\ee
It is easy to show that $Q_X^2$ 
and $\tilde Q_X^2$ are the same
\be
\begin{split} 
 Q_X^2 & = (\tilde Q_X + \tka)^2 = (p_a + p_b - k + \tka )^2
= (P_{ab} - \frac{k \cdot P_{ab} }{p_a \cdot p_b} p_a)^2
\\
& = P_{ab}^2 - 2 k \cdot P_{ab} = (P_{ab} - k)^2 = {\tilde Q}_X^2.
\end{split} 
\ee
Since $\tilde Q_X^2 = Q_X^2$, we can obtain one
of these momenta by  Lorentz-boosting the other.
We therefore write
\be
\begin{split}
  & [{\rm d} {\tilde Q}_X]_m [{\rm d} k] (2 \pi)^d \delta(P_{ab} -{\tilde Q}_X - k)
   = [{\rm d} {\tilde Q}_X]_m [{\rm d} k] (2 \pi)^d \delta( x p_a+p_b - Q_X)
\\
&    = [{\rm d} {\tilde Q}_X]_m [{\rm d} k] (2 \pi)^d \delta( x p_a+p_b - \Lambda_a( Q_X, \tilde Q_X)  \tilde Q_X),
\end{split} 
\ee
where the Lorentz boost  $\Lambda_a( Q_X, \tilde Q_X) $ is defined as follows
\be
Q_X = \Lambda_a( Q_X, \tilde Q_X) \;  \tilde Q_X.
\ee
Since ${\tilde Q}_X = \sum \limits_{i=1}^{m} \tilde p_i$, we perform the required boost for each final-state particle
$\tilde p_i = \Lambda_a^{-1}(Q_X, \tilde Q_X) p_i$
and obtain 
\be
\begin{split} 
   &  \prod \limits_{i=1}^{m}  [{\rm d}  p_i]
   [{\rm d} k] (2 \pi)^d \delta\left( x p_a+p_b - \sum \limits_{i=1}^{m} p_i \right)
\\
   & =
   \int {\rm d} \xi \; \prod \limits_{i=1}^{m}  [{\rm d} p_i] (2 \pi)^d \delta\left( \xi p_a+p_b - \sum \limits_{i=1}^{m} p_i \right)
   \; [{\rm d} k] \; \delta( x - \xi).
   \end{split} 
\ee
The Lorentz transformation 
$\Lambda_a(Q_X, \tilde Q_X)$ can be found  in Eq.~\eqref{eqa.2}, 
where one should identify 
$Q_i = \tilde Q_X$, $Q_f= Q_X$.

Since 
we will have to apply this transformation to all final-state particles and then expand the result  around the collinear limit, we need to simplify $\Lambda_a$. 
To do this, we introduce the notation  
\be
Q_a = x  p_a  + p_b,\;\;\;\;
\label{eq2.41}
\ee
so that 
\be
Q_f = Q_a,\;\;\; 
Q_i =  Q_a - \tka.
\ee
We use Eq.~(\ref{eq4.34}) 
to write  
$\tka$  
as 
\be
\tka^\mu = k^\mu - \frac{k \cdot P_{ab}}{p_a \cdot p_b} \;  p_a^\mu.
\label{eq3.38a}
\ee
To simplify the expression for $\tilde k_a$
further,  we perform the Sudakov decomposition of the vector $k$ and write  
\be
k^\mu = \alpha p_a^\mu + \beta p_b^\mu + k^\mu_\perp,
\ee
where the transverse momentum  $k_\perp$ satisfies $p_{a,b} \cdot k_\perp = 0$.
We then compute the coefficients $\alpha$ and $\beta$, and  find 
\be
\tka^\mu  = \frac{\omega_k}{\sqrt{s}}(2- \rho_{ak}) p_a^\mu + \frac{\omega_k}{\sqrt{s}} \rho_{ak} p_b^\mu + k_\perp^\mu,
\label{eq3.37a}
\ee
where $\omega_k$ is 
the gluon's energy and 
$\rho_{ak} = 1 - \cos \theta_{ak}$ was introduced 
earlier. 
%with $\theta_{ak}$ being the angle between the %three momenta $\vec p_a$ 
%and $\vec k$. 
Furthermore, 
 we also made use of the fact  that vectors $p_{a,b}$ are back-to-back,  and that $2 p_a \cdot p_b =s$.

The absolute value of the vector $k_\perp$ is determined from the on-shell condition $k^2=0$. We derive
\be
k_\perp^2 = -\omega_k^2\rho_{ak}(2 - \rho_{ak}).
\ee
Using Eq.~(\ref{eq3.37a}), we write $\tka^\mu$ in Eq.~(\ref{eq3.38a})
as follows
\be
\tka^\mu = \frac{\omega_k}{\sqrt{s}} \rho_{ak} (p_b - p_a)^\mu + \omega_k \sqrt{\rho_{ak} (2-\rho_{ak} )} n_\perp^\mu
= \frac{2k p_a}{s} (p_b-p_a)^\mu + k_\perp^\mu.
\label{eq4.46}
\ee
The important point is that $\tka^\mu$ vanishes in the soft $\omega_k \to 0$ and  in the
collinear $\rho_{ak} \to 0$ limits, which allows us to construct the expansion of the Lorentz-boost matrix $\Lambda_a$, 
which becomes the identity matrix in both 
of these limits. 

The boost operator depends on $Q_a$ and $\tka$; 
the collinear expansion is the expansion in small $\tka$. 
Since $\tka
\sim \sqrt{\rho_{ak}}
$ and we need to account for 
${\cal O}(\rho_{ak})$ terms, 
 we must  expand $\Lambda_a$
 to  \emph{second order} in $\tka$.
 The expansion can be simplified
if we notice that
\be
2 \tka \cdot Q_a = \tka^2.
\ee
The above equation follows from the equality $Q_f^2 = Q_a^2 = Q_i^2 = (Q_a - \tka)^2$.  Hence, in this case 
\be
(Q_f + Q_i)^2  = 
4 Q_a^2 -\tka^2.
\ee
Using this result, we easily arrive at the expressions for the boost operator $\Lambda_a$ and its inverse, 
shown in Eqs~(\ref{eqa.4}, \ref{eqa.5}).

We continue with the simplification of the starting expression for the cross section  in 
Eq.~(\ref{eq4.2}) 
in the collinear $\vec k || \vec p_a$ limit. The first point is that 
the jettiness constraint is simplified in this limit, since $\psi_k = \rho_{ak}$. 
It is important to emphasize 
that the above formula is valid not only in the strict $\vec k || \vec p_a$ limit, but also in its 
neighborhood. Because of this, 
we do not expand the 
zero-jettiness function 
around the collinear limit 
below. 
Hence, in the first step we write 
\be
\begin{split} 
\frac{ {\rm d}  \sigma^{ca}}{{\rm d} \tau}
& = {\cal N}
\int [{\rm d} \tilde Q_X]_m [{\rm d} k]
(2\pi)^d \delta( p_a + p_b - k - \tilde Q_X )
\\
& \times 
\delta \left ( \tau -  \frac{\sqrt{s} \omega_k \rho_{ak} }{Q} \right ) \; 
\; {\cal O}(\tilde Q_X) \; \sum \limits_{\rm col, pol}^{} |{\cal M}|^2(p_b,p_a,k,\tilde Q_X).
\end{split}
\ee

Following the above discussion, we perform the momenta transformation and obtain
\be
\begin{split} 
 \frac{{\rm d} \sigma^{ca} }{{\rm d} \tau}
& = {\cal N}
\int \limits_0^1 {\rm d} x \;
     [{\rm d} Q_X]_m (2\pi)^d \delta( x p_a + p_b - Q_X)
     \; \int [{\rm d} k]
     \delta \left ( 1 - \frac{2\omega_k}{\sqrt{ s}}  - x \right )
     \\
&   \times    \delta \left ( \tau -  \frac{\sqrt{s} \omega_k \rho_{ak} }{Q} \right )
            {\cal O}( \Lambda_a^{-1}  Q_X) \;
            \sum \limits_{\rm pol, col}^{}  |{\cal M}(p_b,p_a,k,\Lambda^{-1}_a Q_X) |^2.
     \end{split} 
     \label{eq2.51}
\ee
The matrix of the inverse Lorentz transformation $\Lambda_a^{-1}$ is  
given in Eq.~(\ref{eqa.5}). 

The product of the 
gluon phase-space element $[{\rm d}k]$ and two 
$\delta$-functions in Eq.~(\ref{eq2.51}) can be simplified, 
since these delta-functions fix the gluon energy and its emission angle relative to the direction of the quark with momentum $p_a$. We find
\be
\begin{split} 
   & [{\rm d} k] \;
 \delta \left ( 1 - \frac{2\omega_k}{\sqrt{ s}}  - x \right )
 \delta \left ( \tau -  \frac{\sqrt{s} \omega_k \rho_{ak} }{Q} \right )
 = \frac{\Omega^{(d-2)} }{2 (2\pi)^{d-1} } \; 
  {\rm d} \omega_k \; 
  \omega_k^{1-2\ep} \; 
\\
& \; \times {\rm d} \rho_{ak} \; \rho_{ak}^{-\ep}  \; (2-\rho_{ak})^{-\ep}
\frac{ {\rm d} \Omega^{(d-2)}}{\Omega^{(d-2)}}
  \delta \left ( 1 - \frac{2\omega_k}{\sqrt{ s}}  - x \right )
  \delta \left ( \tau -  \frac{\sqrt{s} \omega_k \rho_{ak} }{Q} \right )
  \\
  & =  \frac{\Omega^{(d-2)} }{2 (2\pi)^{d-1} } \; \frac{Q^{1-\ep} \tau^{-\ep}}{2} (1-x)^{-\ep}
    \left (1 + \frac{ \ep \rho_{ak}^*}{2} \right ) [{\rm d} \Omega^{(d-2)} ],
   \end{split} 
\ee
where
\be
\omega_k = \frac{ \sqrt{s}}{2} (1-x),
\;\;\;\; \rho^*_{ak} = \frac{2 Q \tau}{s(1-x)}, 
\;\;\;\;
[{\rm d} \Omega^{(d-2)} ]
= \frac{ {\rm d} \Omega^{(d-2)}}{\Omega^{(d-2)}}.
\ee

We now put everything together and write the collinear contribution as follows 
\be
\begin{split} 
 \frac{{\rm d} \sigma^{ca} }{{\rm d} \tau}
& = \frac{C_F [\alpha_s] Q^{1-\ep} }{2 \tau^{1+\ep}} 
{\cal N} \;
\int \limits_0^1 {\rm d} x \; {\rm d} \Phi_m^{xa}
\;  \left[ {\rm d} \Omega_k^{(d-2)} \right]  \;   (1-x)^{-\ep}  \left ( 1 + \frac{ \ep \rho^*_{ak}}{2}  \right )
\\
& \times 
       \; {\cal O}( \Lambda_a^{-1} Q_X)
       \; \sum \limits_{\rm pol, col}^{}  C_F^{-1} g_s^{-2} \tau \ |{\cal M}(p_b,p_a,k,\Lambda_a^{-1}  Q_X) |^2,
\end{split}
\label{eq4.56}
\ee
where 
\begin{equation}
{\rm d} \Phi_{m}^{xa}
 = [{\rm d} Q_X]_m
 (2\pi)^d \delta(xp_a 
 + p_b - Q_X).
 \label{eq2.56}
\end{equation}

To determine the subleading contributions to the  cross section, we use 
Eq.~(\ref{eq4.56}) 
as a starting point  
and expand the 
 matrix element squared
and the observable 
around the collinear limit 
using explicit expressions for the 
Lorentz boost $\Lambda_a$. We  
 integrate the result of the expansion over the azimuthal angle  of the emitted gluon,
 leading to the final expression which depends on  the scalar products of $p_{a,b}$ and $p_{i}$, as well as on the derivatives of the \emph{observable} with respect to the momenta $p_i$.  This expression will  have to be combined with the contribution 
of the other collinear limit
($\vec k || \vec p_b$) and the contribution of the soft limit, to arrive
at the next-to-leading power correction to the differential cross section 
subject to the zero-jettiness constraint.
\\

To proceed further, we  need to construct the collinear expansion of the matrix element squared.  We write 
\begin{equation}
g_s^2 C_F F_a = \sum \limits_{\rm pol} 
|{\cal M}|^2(p_b,p_a,k, \Lambda_a^{-1}  Q_X) 
=  \sum \limits_{\rm pol} 
|{\cal M}|^2(
\Lambda_a p_b,
 \Lambda_a p_a, \Lambda_a k, Q_X), 
\end{equation}
where we have used the Lorentz invariance of the matrix element squared to move the action of the  Lorentz boost  to $p_a$, $p_b$ and $k$.
The advantage of the above formula is that the action of the boost is now ``localized''; we need to consider changes in the momenta of the initial partons and the momentum of the gluon $k$, but the momenta of the final-state color-neutral particles do not need to be changed.  

We require the expression of the matrix element that is suitable for the study of the collinear limit.  Furthermore, the collinear limit should be written in such a way that the   soft singularity  in the corresponding expressions can be isolated. Finally, it is convenient to first expand the matrix element squared
around the collinear limit, and apply the boost later; this approach keeps the expressions more compact since the boost will have to be applied to fewer terms. 

We write the matrix element as 
\be
{\cal M} 
= - g_s T^a \epsilon^*_\nu 
\bar v_b 
\left [ N(p_b,p_a-k,Q_X) \frac{(\hat p_a - \hat k) \gamma^\nu}{(-2 p_a \cdot k)} + 
 N_{{\rm fin},a}^\nu 
\right ] u_a. 
\label{eq4.69}
\ee
The collinear $\vec k || \vec p_a$ singularity is present in the first term of the above expression, whereas the second term is subleading in the collinear limit. However, that term still has a soft singularity. For this reason, it is convenient to write it as
follows 
\begin{equation}
\label{eq4.70}
N_{{\rm fin},a}^\nu 
= R_{\rm fin}^{\nu}
(p_b,p_a,k,Q_X)
+ \gamma^\nu \frac{(\hat p_b - \hat k)}{2p_b \cdot k}\; 
 N(p_b-k,p_a,Q_X).
\end{equation}
We note that    $R_{\rm fin}^\nu$
arises from diagrams where the gluon is emitted from off-shell lines and, therefore, it contains neither collinear nor soft singularities.  In what follows 
we will denote $N(p_b,p_a-k,Q_X)$ as  $N_a (p_b,p_a-k,Q_X)$ and $N(p_b-k,p_a,Q_X)$ as $ N_b(p_b-k,p_a,Q_X)$, and we will not  write their arguments explicitly unless it is   needed.   
\\

In the collinear limit $\vec k || \vec p_a$, we write the matrix element 
squared as a sum of three terms 
\be
F_a  =  
F_{aa} + F_{ar} + F_{rr,a},
\label{eq2.59}
\ee
where the first term 
on the right-hand side refers
to the square of the diagram where the gluon is emitted off the parton $a$, the second term refers to the  interference of this diagram with the remaining ones,  and the 
third one to the contribution of the remaining diagrams   squared.

We begin by considering the first term on the right-hand 
side of Eq.~(\ref{eq2.59}) 
and write it as 
\be
F_{aa} = \frac{1}{(2 p_a k)^2} {\rm Tr} 
\left [  N_a \; \left( \hat p_a - \hat k \right) \gamma^\mu \hat p_a \gamma^\nu 
\left( \hat p_a - \hat k \right) \; N_a^+ \hat p_b
\right ] \; \rho^{(a)}_{\mu \nu},
\ee
where 
\be
\label{eq2.61}
\rho^{(a)}_{\mu \nu} =
-g_{\mu \nu} + \frac{k_\mu p_{b,\nu} + p_{b,\mu} k_\nu}{k \cdot p_b}, 
\ee
and the arguments of $N_a$ are not shown. 
The superscript $a$ in the gluon density 
matrix  in Eq.~(\ref{eq2.61})  indicates that  
 a gauge choice for the gluon polarization vector is made to simplify the expansion 
 around  the  $\vec k || \vec p_a$  collinear limit. 
 We will need to expand $F_{aa}$ through 
terms that scale as ${\cal O}( (k p_a)^0)$ and, once the expansion is constructed, apply the Lorentz boost 
$\Lambda_a$ to momenta $p_a$, $p_b$ and $k$ in the resulting formula. 

We begin with  the simplification of  $F_{aa}$. A straightforward algebra gives 
\be
\begin{split} 
( \hat p_a - \hat k) \gamma^\mu \hat p_a \gamma^\nu 
( \hat p_a - \hat k) 
\; \rho^{(a)}_{\mu \nu} 
& = (2 k \cdot p_a) 
\Big  [ (d-2) \hat k 
\\
& + \frac{1}{k \cdot p_b} 
\left ( 
\hat p_a \hat p_b ( \hat p_a - k) 
+ \left( \hat p_a - \hat k \right) \hat p_b \hat p_a
\right )
\Big  ].
\end{split} 
\ee
Since 
\be
\hat p_a \hat p_b ( \hat p_a - k) 
+ ( \hat p_a - \hat k) \hat p_b \hat p_a 
= 2 (p_a - k) \cdot p_b \; \hat p_a 
+ 2 p_a \cdot k \; \hat p_b 
+ 2 p_a \cdot p_b \; ( \hat p_a  - \hat k),
\ee
we obtain 
\begin{equation}
F_{aa} = \frac{1}{2 p_a \cdot  k}
{\rm Tr} 
\left [  N_a
\left ( 
-2(\hat p_a - \hat k) - 2\ep \hat k 
+ \frac{2 p_a \cdot p_b}{p_b \cdot k}
\left ( 2 \hat p_a - \hat k \right ) 
+ \frac{2 p_a \cdot k}{p_b \cdot k} 
\hat p_b
\right )
 N_a^+ \hat p_b\right ].
\end{equation}
We can further simplify the above equation by substituting $k = (1-x) p_a + \tilde k_a$ and neglecting all terms that  contribute to  the expansion around the collinear limit beyond the next-to-leading power.   
We then find 
\begin{equation}
\begin{split} 
F_{aa}  = \frac{1}{2 p_a \cdot k}
{\rm Tr} 
\Big  [ N_a 
\Big  ( 2 \hat p_a P_{qq}(x)
& -2 \left ( \frac{ x}{1-x} +\ep \right ) \hat { \tilde k}_a 
\\
& + \frac{2 p_a \cdot k}{(1-x) p_a \cdot p_b}
\left [ \hat p_b + \frac{1+x}{1-x} \hat p_a \right ]
\Big ) N_a^+ \hat p_b
\Big  ],
\label{eq4.75}
\end{split} 
\end{equation}
where
\be
P_{qq}(x) = \frac{1+x^2}{1-x} - \ep (1-x),
\ee
is the standard collinear splitting function. We note that the last term
in Eq.~(\ref{eq4.75}) is already of  the right order in the collinear expansion. For this reason it does not require further manipulations, i.e. the 
Lorentz boost does not need to be applied to it.  Furthermore, it is convenient to express ${\tilde k}_a$ through $k_\perp$ using Eq.~(\ref{eq4.46}).  We obtain 
\begin{equation}
\begin{split}
F_{aa} & = \frac{2 P_{qq}(x)} {2 p_a \cdot k}
{\rm Tr} 
\left [ N_a 
\; \hat p_a \;
 N_a^+ \hat p_b 
\right ]
-\frac{2}{2 p_a \cdot k} 
\left ( \frac{ x}{1-x} +\ep \right ) 
{\rm Tr} 
\left [  N_a
\hat { k}_\perp
 N_a^+ \hat p_b
\right ]
\\
& +\frac{2}{s}
\left ( \frac{(1+2x -x^2)}{(1-x)^2} 
+ \ep 
\right ) {\rm Tr} 
\left [  N_a \hat p_a N^{+}_a \hat p_b 
\right ]
+ 
\frac{2(1-\ep)}{s}
{\rm Tr} 
\left [ 
 N_a \hat p_b 
 N_a^{+} \hat p_b 
\right ].
\label{eq4.79}
\end{split}
\end{equation}
The last two terms do not require further manipulations, they are already in the right form. The first term is the leading collinear  contribution; 
it must  be expanded to account for subleading collinear terms. 

The second term on the right-hand side in Eq.~(\ref{eq4.79}) requires  discussion.  As we mentioned earlier, 
we will have to boost the momenta $p_a$, $p_b$, $k$ to compute the matrix element squared.  Since the second term in Eq.~(\ref{eq4.79}) is proportional to $k_\perp$, anything that arises from $N_{a,b}$ or $\hat p_{a,b}$ after the boost can only contain $k_\perp$ since 
all other terms contribute beyond next-to-leading order  in the zero-jettiness expansion. However, boosting $k_\perp$ will generate 
a term that is proportional to $2 p_a k$, which will then contribute to $F_{aa}$ 
at the right order.  
Since 
\be
{\Lambda_a}^{\mu}_{~\nu} \; k^\nu_\perp = 
k^\mu_\perp + \frac{ Q_a^\mu}{ Q_a^2} (1-x) (2k \cdot p_a),
\ee
with $Q_a = x p_a + p_b$,
it is convenient to introduce a new vector 
\be
\kappa_a^\mu = 
k_\perp^\mu - \frac{\tilde Q_a^\mu}{\tilde Q_a^2} (1-x) ( 2k \cdot p_a),
\label{eq4.73}
\ee
which  
after the boost becomes $k_\perp$, 
\be
\Lambda^\mu_{a,\nu} \;  \kappa_a^\nu 
 = k_\perp^\mu + {\cal O}(k_\perp^3). 
\ee
Thus, if we express the before-the-boost result  through 
$\kappa_a$, computing the boost 
for $\kappa_a$-dependent  terms 
becomes straightforward.

Hence,  we rewrite the vector $k_\perp$ through 
$\kappa_a$ using Eq.~(\ref{eq4.79}), and find  
\be
{\rm Tr} 
\left [  N_a
\hat { k}_\perp 
 N_a^+ \hat p_b
\right ]
= 
{\rm Tr} 
\left [  N_a
\hat { \kappa_a } 
 N_a^+ \hat p_b
\right ]
+ 2k \cdot p_a \frac{ (1-x) }{sx}
{\rm Tr} 
\left [  N_a
\left (x \hat p_a + \hat p_b  \right ) 
 N_a^+ \hat p_b
\right ].
\label{eq2.72}
\ee
Combining the last term in Eq.~(\ref{eq2.72})  with the third and the fourth term
in Eq.~(\ref{eq4.79}), we obtain
\begin{equation}
\begin{split}
F_{aa} & = \frac{2 P_{qq}(x)} {2 p_a \cdot k}
{\rm Tr} 
\left [ N_a 
\; \hat p_a \;
 N_a^+ \hat p_b 
\right ]
-\frac{2}{2 p_a \cdot k} 
\left ( \frac{ x}{1-x} +\ep \right ) 
{\rm Tr} 
\left [  N_a
\hat { \kappa_a}  
 N_a^+ \hat p_b
\right ]
\\
& +\frac{2}{s}
\left ( \frac{(1+x +x^2 -x^3)}{(1-x)^2} 
+ \ep x 
\right ) {\rm Tr} 
\left [ N_a \hat p_a  N^{+}_a \hat p_b 
\right ]
-
\frac{2\ep}{sx}
{\rm Tr} 
\left [ 
 N_a \hat p_b 
 N_a^{+} \hat p_b 
\right ].
\end{split}
\label{eq4.84}
\end{equation}
As we already mentioned this form is convenient because after the boost,  $\kappa_a$ will become $k_\perp$; this implies that only $k_\perp$ terms from other momenta will be needed. Since, after averaging 
\be
k_\perp^\mu k_\perp^\nu 
\to 
\frac{ k_\perp^2 }{2(1-\ep)} g_\perp^{\mu \nu}
=
-\frac{ 2p_a k (1-x) }{2(1-\ep)} g_\perp^{\mu \nu},
\ee
where 
\be
g_\perp^{\mu \nu} 
= g^{\mu \nu} - \frac{p_a^\mu p_b^\nu + p_b^\mu p_a^\nu}{p_a \cdot p_b},
\ee
such terms do not lead to soft and collinear singularities.  
Hence, the soft singularities are only present in the {\it first} and  {\it third} terms 
on the r.h.s. of  Eq.~(\ref{eq4.84}).
\\

We continue with the discussion of the interference contribution in Eq.~(\ref{eq2.59}). We write it as 
\begin{equation}
F_{ar}
={\rm Tr} 
\left [  
\frac{ N_a (\hat p_a - \hat k) \gamma^\mu \hat p_a N_{{\rm fin},a}^{+,\nu} \hat p_b}{(-2 p_a \cdot k)}
\right ] \; \rho^{(a)}_{\mu \nu}
+ 
{\rm c.c.}
\label{eq4.86}
\end{equation}
We now split $\rho^{(a)}_{\mu \nu}$ into two terms
\be
\begin{split} 
\rho^{(a)}_{\mu \nu} 
= \rho^{(a,1)}_{\mu \nu} 
+ \rho^{(a,2)}_{\mu \nu},
\end{split}
\ee
where 
\begin{equation}
\rho^{(a,1)}_{\mu \nu} 
= -g_{\mu \nu} 
+ \frac{p_{b\mu} k_{\nu}}{k \cdot p_b },
\;\;\; 
\rho^{(a,2)}_{\mu \nu} 
= \frac{k_\mu p_{b\nu}}{k \cdot p_b}.
\end{equation}
We will first compute 
$F_{ar}^{(2)}$ which 
we obtain by replacing 
$\rho^{(a)}_{\mu \nu}$ in 
Eq.~(\ref{eq4.86}) 
with $\rho^{(a,2)}_{\mu \nu}$.  Using 
\begin{equation}
(\hat p_a - k) \gamma^\mu \hat p_a \rho^{(a,2)}_{\mu \nu} 
=
\hat p_a \gamma^\mu 
 \hat p_a 
  \rho^{(a,2)}_{\mu \nu} 
=\frac{2 kp_a}{k \cdot p_b}
\hat p_a \; p_{b\nu},
\end{equation}
we obtain 
\be
F_{ar}^{(2)}
 = -\frac{2 p_{b,\nu}}{s(1-x)}
 {\rm Tr} 
 \left [ 
 N_a \hat p_a  N_{{\rm fin},a}^{+,\nu} \hat p_b 
 \right ]
 +{\rm c.c.},
 \label{eq4.90}
\ee
where we already took the 
collinear limit.  

It is important to understand how the soft limit can be extracted from this expression especially since there is a term in $N_{{\rm fin},a}^{\nu,+}$
that contains the soft singularity.  We use 
Eq.~(\ref{eq4.70}) and write 
\begin{equation}
N_{{\rm fin},a}^{\nu,+} 
= R_{\rm fin}^{\nu,+}
+  N_b^+  \frac{(\hat p_b - \hat k)}{2p_b \cdot k} \gamma^\nu
.
\end{equation}
Using it  in Eq.~(\ref{eq4.90}), we find 
\be
F_{ar}^{(2)}
 = -\frac{2 p_{b,\nu} }{s(1-x)}
 {\rm Tr} 
 \left [ 
 N_a \hat p_a  R_{\rm fin}^{+,\nu} \hat p_b 
 \right ]
 +{\rm c.c.},
 \label{eq4.92}
\ee
and the soft singularity is now explicit.

The other  contribution $F_{ar}^{(1)}$ is 
obtained 
by replacing 
$\rho^{(a)}_{\mu \nu}$ in 
Eq.~(\ref{eq4.86}) 
with $\rho^{(a,1)}_{\mu \nu}$. To simplify the result 
in this case, 
we write 
\be
\left( \hat p_a - \hat k \right) 
\gamma^\mu \hat p_a \rho^{(a,1)}_{\mu \nu} 
= \left( \hat p_a - \hat k \right) 
2 p_a^\mu \rho^{(a,1)}_{\mu \nu} 
+ \hat k \hat p_a \gamma^\mu \rho^{(a,1)}_{\mu \nu}. 
\ee
Replacing $\hat k \hat p_a$ 
in the second term with 
$\hat {\tilde k}_a \hat p_a$ and 
 writing there
$\hat p_a \gamma^\mu = 2 p_a^\mu 
- \gamma^\mu \hat p_a$, 
we find 
\begin{equation}
\left( \hat p_a - \hat k \right) 
\gamma^\mu \hat p_a \rho^{(a,1)}_{\mu \nu} 
= 
\left( \hat p_a - \hat k + \hat {\tilde k}_a \right) 
2 p_a^\mu \rho^{(a,1)}_{\mu \nu} 
- \hat {\tilde k}_a  \gamma^\mu \hat p_a  \rho^{(a,1)}_{\mu \nu}. 
\end{equation}

It is easy to show that,  through the right order in the  zero-jettiness expansion, the following equation holds
\be
p_a^\mu  \rho^{(a,1)}_{\mu \nu} = \frac{\tilde k_{a,\nu}}{1-x}  + \frac{2 k \cdot p_a
}{(1-x) s}\;  p_{a ,\nu}.
\ee
We also find 
\be
\hat {\tilde k}_a  \gamma^\mu \hat p_a \rho^{(a,1)}_{\mu \nu}
= 
-\hat {\tilde k}_a \gamma_\nu  \hat p_a 
+ \frac{1}{k \cdot p_b} \hat {\tilde k}_a \hat p_b \hat p_a  k_\nu.
\ee
The Ward identity implies 
\be
\hat p_a   N_{{\rm fin},a}^{+,\nu} 
\hat p_b \;  k_\nu = \hat p_a N_a^+ \hat p_b.
\ee
Putting the above results together, we obtain 
\be
\begin{split} 
F_{ar}^{(a,1)} 
& = 2 {\rm Tr} 
\left [ 
\frac{ N_a  x \hat p_a   N_{{\rm fin},a,\nu}^{+} \hat p_b }{(-2 p_a \cdot k) }
\right ]
\left ( 
\frac{\tka^\nu}{1-x} + \frac{2 k \cdot p_a}{(1-x) s} p_a^\nu
\right )
+
{\rm Tr} 
\left [ 
\frac{ N_a \hat { \tilde k}_a 
\gamma_\nu \hat p_a   N_{{\rm fin},a}^{+,\nu} \hat p_b }{(-2 p_a \cdot k) }
\right ]
\\
& -
\frac{1}{k \cdot p_b}
{\rm Tr} 
\left [ 
\frac{ N_a \hat { \tilde k}_a \hat p_b \hat p_a
N^+_a \hat p_b }{(-2 p_a \cdot k) }
\right ] + 
{\rm c.c.} .
\end{split}
\label{eq4.98}
\ee
We observe that 
the interference terms are proportional to $\tilde k_a/(-2 p_a \cdot k)$. Hence,  
to obtain the  final result, the functions $N$ and $ N_{\rm fin}^{+,\nu}$
have to be expanded to first order in $\tilde k_a$.
We also note that 
in the last term in 
Eq.~(\ref{eq4.98})
we can replace the  $1/(k \cdot p_b)$ factor with $2/(s(1-x))$, without compromising the accuracy of  the collinear expansion. 

We continue with the analysis of the individual terms in Eq.~(\ref{eq4.98}), aiming  at isolating those that 
have a soft singularity. 
We begin  with the first term  in Eq.~(\ref{eq4.98}) and write 
\begin{equation}
\begin{split}
& 2 {\rm Tr} 
\left [ 
\frac{N_a  x \hat p_a  N_{{\rm fin},a,\nu}^{+} \hat p_b }{(-2 p_a \cdot k) }
\right ]
\left ( 
\frac{\tilde k_a^\nu}{1-x} + \frac{2 k \cdot p_a}{(1-x) s} \;p_a^\nu
\right ) 
\\
& = -\frac{2}{s(1-x)} 
{\rm Tr} \left [ N_a x \hat p_a 
\left (R_{\rm fin}^{\nu,+} p_{b,\nu} 
+ N_b^+ 
\right ) \hat p_b \right]
\\
& +
\frac{2}{(1-x)(-2 p_a \cdot  k)}
{\rm Tr} 
\left [ 
N_a x \hat p_a 
\left ( R_{\rm fin}^{\nu,+} k_{\perp,\nu} 
\hat p_b 
- N_b^+ \frac{\hat p_a  \hat k_\perp}{s}
\hat p_b
\right )
\right ].
\end{split}
\label{eq2.90a}
\end{equation}
It is convenient to 
express the last term in 
Eq.~(\ref{eq2.90a}) 
through a vector $\tilde \kappa_a$, following the  discussion above.
We  find 
\be
\begin{split}
& 2 {\rm Tr} 
\left [ 
\frac{N_a  x \hat p_a  N_{{\rm fin},a,\nu}^{+} \hat p_b }{(-2 p_a \cdot k) }
\right ]
\left ( 
\frac{\tilde k_a^\nu}{1-x} + \frac{2 k \cdot p_a}{(1-x) s} p_a^\nu
\right ) 
= 
\\
& -\frac{2}{s(1-x)} 
{\rm Tr} \left [ N_a x \hat p_a 
\left (R_{\rm fin, \nu}^+ \frac{ p_b^\nu }{x}
+ N_b^+ 
\right ) \hat p_b \right]
-
\frac{2}{s}
{\rm Tr}
\left [ 
  N_a x \hat p_a 
 R_{\rm fin}^{\nu, +} 
 p_{a,\nu}  
\hat p_b 
\right ]
\\
& +
\frac{2 \kappa_{a,\nu} }{(1-x)(-2 p_a \cdot k)}
{\rm Tr} 
\left [ 
N_a x \hat p_a 
\left ( R_{\rm fin}^{\nu,+} 
- N_b^+ \frac{\hat p_a  \gamma^\nu }{s}
\right ) \hat p_b
\right ],
\label{eq4.100}
\end{split}
\end{equation}
and the soft singularity  is only present in the first term on the right-hand side of the above equation, thanks 
to the argument mentioned below 
Eq.~(\ref{eq4.84}).

Next, we need to consider  
the second term in Eq.~(\ref{eq4.98}). 
We write 
\be
\begin{split}
{\rm Tr} 
\left [ 
\frac{N_a \hat { \tilde k}_a \gamma_\nu \hat p_a  N_{{\rm fin},a}^{+,\nu} \hat p_b }{(-2 p_a \cdot k) }
\right ]
 & = 
 {\rm Tr} 
\left [ 
\frac{N_a \hat k_\perp \gamma_\nu \hat p_a  N_{{\rm fin},a}^{+,\nu} \hat p_b }{(-2 p_a \cdot k) }
\right ]
\\
& -
 \frac{1}{s} {\rm Tr} 
\left [ 
N_a (\hat p_b  - \hat p_a) \gamma_\nu \hat p_a  N_{\rm fin}^{+,\nu} \hat p_b 
\right ],
\end{split}
\ee
and then replace  $k_\perp$ with $\kappa_a$ in the first term.  After simplifications, we find 
\be
\begin{split}
& {\rm Tr} 
\left [ 
\frac{N_a \hat { \tilde k}_a \gamma_\nu \hat p_a  N_{{\rm fin},a}^{+,\nu} \hat p_b }{(-2 p_a \cdot k) }
\right ]
 = 
 {\rm Tr} 
\left [ 
\frac{N_a \hat { \kappa}_a  \gamma_\nu \hat p_a  N_{{\rm fin},a}^{+,\nu} 
\hat p_b }{(-2 p_a \cdot k) }
\right ]
+
\frac{1}{sx} 
{\rm Tr} 
\left [ 
 N_a \hat p_b \gamma^\nu \hat p_a  N_b^+ \frac{\hat p_a \gamma_\nu \hat p_b}{s}
\right ]
\\
& 
+ \frac{1}{sx}
{\rm Tr} 
\left [
N_a (x^2 \hat p_a - \hat p_b) \gamma_\nu \hat p_a  R^{\nu,+}_{\rm fin} \hat p_b
\right ]
+
\frac{2x}
{s (1-x)}
{\rm Tr} 
\left [  N_a  \hat p_a  N_b^+ \hat p_b \right ].
\end{split} 
\label{eq4.102}
 \ee
The soft divergence resides in the last term on the right-hand side 
of the above equation. 

Finally, we need to analyze the last term in 
Eq.~(\ref{eq4.98}).
It reads 
\be
\begin{split}
 -\frac{1}{k p_b} 
{\rm Tr}
\left [ 
\frac{  N_a \hat {\tilde k}_a \hat p_b \hat p_a  N_a^+ \hat p_b}{(-2 p_a \cdot k)}
\right ]
= 
& -\frac{2}{s(1-x)} 
{\rm Tr}
\left [ 
\frac{ { N_a \hat { \kappa}_a \hat p_b \hat p_a  N_a^+ \hat p_b}}{(-2 p_a \cdot k)}
\right ]
\\
& - \frac{2x}{s(1-x)}
{\rm Tr} 
\left [ 
N_a \hat p_a 
 N_a^+ \hat p_b 
\right ].
\label{eq5.90}
\end{split}
\ee
The soft singularity is in the second term on the right-hand side 
of Eq.~(\ref{eq5.90}). 
\\

The last contribution we need to compute  is  $F_{rr,a}$; it is finite in the collinear $\vec k || \vec p_a$ limit. It is also finite in the soft limit   thanks to
our choice of the gluon density matrix. The result reads 
\begin{equation}
\begin{split}
F_{rr,a}
& = \frac{1}{s}
{\rm Tr}
\left [ 
 R_{\rm fin}^\mu \hat p_a  N_b^+ 
\hat p_a \gamma_\mu \hat p_b 
\right ] + {\rm c.c.} \\
& \ +\frac{2}{s} 
{\rm Tr} 
\left [ N_b \hat p_a 
 N_b^+ \hat p_a 
\right ]
-{\rm Tr}
\left [R_{\rm fin}^\nu \hat p_a R_{\rm fin}^{\mu,+} 
\hat p_b\right ] g_{\perp,\mu \nu}.
\end{split} 
\label{eq2.92}
\end{equation}

We now collect all the terms  that contribute to the function $F_a$ defined in 
Eq.~(\ref{eq2.59}) 
%\be
%F_a = F_{aa} + %F_{ar}+F_{rr,a},
%\ee
through the required order in the collinear expansion.
We pay particular attention to separating  terms that exhibit soft and collinear singularities\footnote{
Such singular terms 
come  from Eqs~(\ref{eq4.84}, \ref{eq4.92})
and Eqs~(\ref{eq4.100}, \ref{eq4.102}, \ref{eq5.90}).}
from the ones that do not. 
 We then write (discarding ${\cal O}(\ep)$ 
contributions in terms that are neither  soft- nor collinear-divergent) 
\be
\begin{split} 
F_a & =  \frac{2 P_{qq}(x)}{2 p_a \cdot k}
{\rm Tr} \left [  N_a \hat p_a 
N_a^+ \hat p_b \right ]
+ 
\frac{2}{s}  \frac{ (1+x+x^2 - x^3)}{(1-x)^2} 
{\rm Tr} 
\left [  N_a \hat p_a N_a^+ \hat p_b \right ]
\\
& -\frac{4 p_b^\nu}{s(1-x)}
{\rm Tr} 
\left [ N_a \hat p_a R_{\rm fin, \nu}^+ \hat p_b  \right ] + {\rm c.c.}
-\frac{2x}{s(1-x)} 
{\rm Tr}
\left [ N_a \hat p_a  N_a^+ \hat p_b \right ]
+{\rm c.c.} 
\\
& 
+
\frac{1}{sx} 
{\rm Tr} 
\left [ 
 N_a \hat p_b \gamma^\nu \hat p_a  N_b^+ \frac{\hat p_a \gamma_\nu \hat p_b}{s}
\right ] + {\rm c.c.} - \frac{1}{sx}
{\rm Tr} 
\left [
N_a \hat p_b \gamma_\nu \hat p_a  R^{\nu,+}_{\rm fin} \hat p_b]
\right ] + {\rm c.c.}
\\
& -\frac{2}{2 p_a \cdot k} \frac{ x}{1-x} 
{\rm Tr} 
\left [  N_a
\hat { \kappa}_a  
 N_a^+ \hat p_b
\right ]
+ {\rm Tr} 
\left [ 
\frac{N_a \hat { \kappa}_a  \gamma_\nu \hat p_a  N_{{\rm fin},a}^{+,\nu} 
\hat p_b }{(-2 p_a \cdot k) }
\right ] + {\rm c.c.}
\\
& +
\frac{2 \kappa_{a,\nu} }{(1-x)(-2 p_a \cdot k)}
{\rm Tr} 
\left [ 
N_a x \hat p_a 
\left ( R_{\rm fin}^{\nu,+} 
- N_b^+ \frac{\hat p_a  \gamma^\nu }{s}
\right ) \hat p_b
\right ] + {\rm c.c.}
\\
& 
-\frac{2}{s(1-x)} 
{\rm Tr}
\left [ 
\frac{  N_a \hat { \kappa}_a \hat p_b \hat p_a  N_a^+ \hat p_b}{(-2 p_a \cdot k)}
\right ] + {\rm c.c.} + F_{rr,a}.
\end{split}
\label{eq2.95}
\ee
We note that the complex conjugation, indicated by  ${\rm c.c.}$ in the above formula, always refers to the term that appears immediately to the left of it. 

The next-to-last term in Eq.~(\ref{eq2.95}) can be simplified if we combine it 
with its conjugate. Then
\be
{\rm Tr}
\left [ 
\frac{ { N_a \hat { \kappa}_a \hat p_b \hat p_a N_a^+ \hat p_b}}{(-2 p_a \cdot k)}
\right ] + {\rm c.c.}
= 
{\rm Tr}
\left [ 
\frac{ { N_a 
\left [ \hat { \kappa}_a \hat p_b \hat p_a 
+ 
\hat p_a \hat p_b  \hat { \kappa}_a 
\right ]
 N_a^+ \hat p_b}}{(-2 p_a \cdot k)}
\right ].
\ee
Since 
\be
\hat { \kappa}_a \hat p_b \hat p_a 
+ 
\hat p_a \hat p_b  \hat { \kappa}_a 
= 2 (\kappa_a \cdot p_b) \hat p_a  
- 2 (\kappa_a \cdot p_a) \hat p_b 
+ s \hat \kappa_a, 
\ee
we obtain 
\be
\begin{split} 
F_a  
&= \frac{2 P_{qq}(x)}{2 p_a \cdot k}
{\rm Tr} \left [  N_a \hat p_a 
N_a^+ \hat p_b \right ]
+ 
\frac{2}{s} \frac{ (1+x+x^2 - x^3)}{(1-x)^2}
{\rm Tr} 
\left [  N_a \hat p_a N_a^+ \hat p_b \right ]
\\
& -\frac{4 p_b^\nu}{s(1-x)}
{\rm Tr} 
\left [ N_a \hat p_a R_{\rm fin, \nu}^+\hat p_b  \right ] + {\rm c.c.}
-\frac{2x}{s(1-x)} 
{\rm Tr}
\left [N_a \hat p_a  N_a^+ \hat p_b \right ]
+{\rm c.c.} 
\\
& 
+
\frac{1}{sx} 
{\rm Tr} 
\left [ 
 N_a \hat p_b \gamma^\nu \hat p_a  N_b^+ \frac{\hat p_a \gamma_\nu \hat p_b}{s}
\right ] + {\rm c.c.} - \frac{1}{sx}
{\rm Tr} 
\left [
 N_a \hat p_b \gamma_\nu \hat p_a  R^{\nu,+}_{\rm fin} \hat p_b
\right ] + {\rm c.c.}
\\
& 
-\frac{2}{s} 
{\rm Tr}
\left [ 
 N_a \hat p_a   N_a^+ \hat p_b
\right ] 
+\frac{2}{sx} 
{\rm Tr}
\left [ 
 N_a  \hat p_b  N_a^+ \hat p_b
\right ] +F_{rr,a}
\\
& +\frac{2}{2 p_a \cdot k} 
{\rm Tr} 
\left [  N_a
\hat { \kappa}_a  
 N_a^+ \hat p_b
\right ]
 + {\rm Tr} 
\left [ 
\frac{N_a \hat { \kappa}_a  \gamma_\nu \hat p_a  N_{{\rm fin},a}^{+,\nu} 
\hat p_b }{(-2 p_a \cdot k) }
\right ] + {\rm c.c.}
\\
& +
\frac{2 \kappa_{a,\nu} }{(1-x)(-2 p_a \cdot k)}
{\rm Tr} 
\left [ 
N_a x \hat p_a 
\left ( R_{\rm fin}^{\nu,+} 
- N_b^+ \frac{\hat p_a  \gamma^\nu }{s}
\right ) \hat p_b
\right ] + {\rm c.c.} .
\end{split}
\label{eq4.109}
\ee
All  terms that appear in the above 
formula should be evaluated for 
the boosted momenta $p_1$, $p_2$, $k$. In practice, this concerns the first term 
and the last three terms 
on the right-hand side  in Eq.~(\ref{eq4.109})  
which, after the boost, 
will have  to be expanded to the required order in $k_\perp$.
The result of this expansion for the last three terms in 
Eq.~(\ref{eq4.109}) 
is free of both 
collinear and soft singularities but it is somewhat messy; we present the corresponding formulas  in  Appendix~\ref{appB}.

\subsection{The second collinear contribution: $\vec k || \vec p_b$ }
\label{sec:colb}

We need to consider the second collinear contribution which  arises 
when the gluon is emitted along the direction of an anti-quark with momentum  $p_b$.
The construction of the Lorentz transformation and the parametrization of the gluon momentum is identical 
to what has been discussed in the 
previous section except that the replacement  $p_a \leftrightarrow p_b$ should be applied. 

The simplification of the matrix element proceeds as in the previous subsection.  It is easy to see 
that in addition to the $p_a \leftrightarrow p_b$ transformation, we also need to perform the  replacement $N_a \leftrightarrow  - N_b^+$.  We find 
\be
\begin{split}
F_b  
&= \frac{2 P_{qq}(x)}{2 p_b \cdot k}
{\rm Tr} \left [  N_b^+ \hat p_b 
N_b \hat p_a \right ]
+ 
\frac{2}{s} \frac{ (1+x+x^2 - x^3)}{(1-x)^2}
{\rm Tr} 
\left [  N_b^+ \hat p_b N_b \hat p_a \right ]
\\
& +\frac{4 p_a^\nu}{s(1-x)}
{\rm Tr} 
\left [ N_b^+ \hat p_b R_{\rm fin,\nu} \hat p_a  \right ] + {\rm c.c.}
-\frac{2x}{s(1-x)} 
{\rm Tr}
\left [N_b^+ \hat p_b  N_b \hat p_a \right ]
+{\rm c.c.} 
\\
& 
+
\frac{1}{sx} 
{\rm Tr} 
\left [ 
 N_b^+ \hat p_a \gamma^\nu \hat p_b  N_a \frac{\hat p_b \gamma_\nu \hat p_a}{s}
\right ] + {\rm c.c.} + \frac{1}{sx}
{\rm Tr} 
\left [
 N_b^+ \hat p_a \gamma_\nu \hat p_b  R^{\nu}_{\rm fin} \hat p_a
\right ] + {\rm c.c.}
\\
& 
-\frac{2}{s} 
{\rm Tr}
\left [ 
 N_b^+ \hat p_b   N_b \hat p_a
\right ] 
+\frac{2}{sx} 
{\rm Tr}
\left [ 
 N_b^+  \hat p_a N_b \hat p_a
\right ] +F_{rr,b}
\\
& +\frac{2}{2 p_b \cdot k} 
{\rm Tr} 
\left [  N_b^+
\hat { \kappa}_b  
 N_b \hat p_a
\right ]
 + {\rm Tr} 
\left [ 
\frac{N_b^+ \hat { \kappa}_b  \gamma_\nu \hat p_b  N_{{\rm fin},b}^{\nu} \; 
\hat p_a }{(2 p_b \cdot k) }
\right ] + {\rm c.c.}
\\
& +
\frac{2 \kappa_{b,\nu} }{(1-x)(2 p_b \cdot k)}
{\rm Tr} 
\left [ 
N_b^+ x \hat p_b 
\left ( R_{\rm fin}^{\nu} 
+ N_a \frac{\hat p_b  \gamma^\nu }{s}
\right ) \hat p_a
\right ] + {\rm c.c.},
\end{split}
\label{eq5.111}
\ee
where 
\be
N_{{\rm fin},b}^\nu = 
R_{\rm fin}^{\nu} - 
\frac{N_a ( \hat p_a - \hat k) \gamma^\nu}{2 p_a \cdot k}.
\ee
Similar to the  collinear case $ \vec k || \vec p_a$ discussed in the preceding section, 
the first two lines contain divergent 
terms and the remaining terms 
are finite in the collinear and soft limits.  The boost that needs to be applied here differs from the boost in the case $\vec k || \vec p_a$. We denote the required  Lorentz boost as  $\Lambda_b$; it is given in Appendix~\ref{appA}. 
\\

\section{Combining soft and collinear contributions}
\label{sect3a}

In this section, we extract  collinear and soft singularities from the different contributions to the differential cross section,  and derive the finite result 
for the next-to-leading  term in the 
zero-jettiness expansion. 

\subsection{The first collinear region: $ \vec k || \vec p_a$}

We begin with the contribution of the first collinear region where the gluon is emitted along the direction of the incoming quark with momentum $p_a$. 
The differential cross section reads 
\be
\begin{split}
 \frac{ {\rm d} \sigma^{ca}}{{\rm d} \tau}
= \frac{[\alpha_s] C_F Q^{1-\ep} }{2 \tau^{1+\ep} } {\cal N} 
\int \limits_{0}^{1} \; {\rm d}x
& \; 
{\rm d} \Phi_m^{xa} \left[ {\rm d} \Omega_k^{(d-2)} \right]
(1-x)^{-\ep}  
\\
\times 
& \left ( 1 + \frac{\ep Q \tau}{s(1-x)}
\right ) \; 
{\cal O}(\tilde Q_X) \;
\tau \; F_a,
\end{split}
\ee
where the Born phase space 
${\rm d} \Phi_m^{xa}$ can be found 
in Eq.~(\ref{eq2.56}), 
$\tilde Q_X = \Lambda_a^{-1} 
Q_X$, 
$F_a$ is given in Eq.~(\ref{eq4.109}), and the momenta $p_a$, $p_b$, $k$ which  appear in that equation  should be boosted.  It is convenient to 
write, with the required  accuracy, 
\be
\begin{split}
  \left ( 1 + \frac{\ep Q \tau}{s(1-x)}
\right ) \;  F_a 
 = & 
\frac{2 P_{qq}(x)}{2 p_a \cdot k}
{\rm Tr} \left [ N_a \hat p_a 
N_a^+ \hat p_b \right ]
+ 
\frac{4(1+\ep) }{s(1-x)^2} 
{\rm Tr} 
\left [ N_a \hat x p_a N_a^+ \hat p_b \right ]
\\
&  -\frac{4 p_b^\nu}{s(1-x)}
{\rm Tr} 
\left [N_a \hat p_a R_{\rm fin, \nu}^+ \hat p_b  \right ] +{\rm c.c.} 
+ F_{a,\rm reg},
\end{split}
\label{eq6.2}
\ee
where the function $F_{a,\rm reg}$ does not have  soft or collinear singularities. Among the four terms that appear on the right-hand side of Eq.~(\ref{eq6.2}), the first one requires the  expansion of the reduced matrix element  in $k_\perp$,  the second term has a \emph{power} divergence  at $x=1$, 
and the third one has a regular soft singularity. 
\\

We begin with the  discussion of  the first term on the right-hand side  of Eq.~(\ref{eq6.2}). 
We note that momenta that appear in that term still have to be  boosted. Hence, we write 
\be
\begin{split}
& \frac{ {\rm d} \sigma^{ca,1}}{{\rm d} \tau}
= \frac{[\alpha_s] C_F Q^{1-\ep} }{2 \tau^{1+\ep} } {\cal N} 
\int \limits_{0}^{1} \; {\rm d}x \; 
{\rm d} \Phi_m^{xa} \left[ {\rm d} \Omega_k^{(d-2)} \right]
(1-x)^{-\ep} {\cal O}(\Lambda_a^{-1} Q_X)  \\
& 
\times \; \;
\tau \; \frac{2 P_{qq}(x)}{(2 p_a \cdot k )\; x} \; 
{\rm Tr} \left [ N_a(p_b,p_a-k,Q_X) \;  x \hat p_a 
N_a^+(p_b,p_a-k,Q_X) \hat p_b \right ]_{\Lambda_a}.
\end{split}
\label{eq6.3}
\ee
The subscript of the trace  
function indicates that momenta 
$p_a,p_b$ and $k$ should be boosted 
with the matrix $\Lambda_a$.
%{\rm Tr} \left [ N_a(\Lambda_a(p_a-%k),\Lambda_a p_b, P_X) ( \Lambda_a  x %\hat p_a)   \hat N_a^+(\Lambda_a(p_a-%k),\Lambda_a p_b, P_X ) 
%( \Lambda_a \hat p_b)  \right ].

To proceed further, we need to  expand the  trace  function 
and the observable 
in Eq.~(\ref{eq6.3}) around the collinear limit, and 
 extract the soft singularity that is present in $P_{qq}(x)$ 
 from all terms in such an expansion.
As the   first step, we  discuss  the (standard) leading collinear contribution which is obtained by 
setting 
$\Lambda_a \to 1$ 
and neglecting the transverse momentum of the gluon $k$. We  
find 
\be
\begin{split}
& \frac{ {\rm d} \sigma^{ca,1,\rm LP}}{{\rm d} \tau}
= \frac{[\alpha_s] C_F Q^{-\ep} }{ \tau^{1+\ep} } {\cal N} 
\int \limits_{0}^{1} \; {\rm d}x \; 
{\rm d} \Phi_m^{xa} \; 
 {\cal O}(Q_X)  
 \; \frac{ P_{qq}(x)}{ x(1-x)^{\ep}} \;  |{\cal M}(p_b,xp_a,Q_X)|^2.
\end{split}
\label{eq3.4}
\ee
In deriving Eq.~(\ref{eq3.4}) 
we have used the equality  $ 2p_a k = \tau Q$, 
and the fact that in the collinear limit 
\be
{\rm Tr} \left [ N_a(p_b,p_a-k,Q_X) \;  x \hat p_a 
N_a^+(p_b,p_a-k,Q_X) \hat p_b \right ]_{\Lambda_a}
\to |{\cal M}(p_b,xp_a,Q_X)|^2,
\ee
where $|{\cal M}(p_b,xp_a,Q_X)|^2$ 
is the spin-summed matrix element 
for the elastic process $q \bar q \to X$ where  the quark $q$ 
and the anti-quark $\bar q$
have 
 momenta $xp_a$ and  $p_b$, 
 respectively.
There is a soft singularity present in the splitting function,  but 
it is straightforward to extract it 
and we do not discuss this point 
further.

The \emph{subleading} terms require more effort. We start with the discussion 
of the trace function and write 
\be
\begin{split} 
& {\rm Tr} \left [ N_a(p_b,p_a-k,Q_X) \;  x\hat p_a \; 
N_a^+(p_b,p_a-k,Q_X) \; \hat p_b \right ]_{\Lambda_a}  
\\
&=  {\rm Tr} \left [ N_a(\Lambda_a p_b, \Lambda_a(p_a-k),Q_X) \; ( \Lambda_a x \hat p_a) \; N_a^+(\Lambda_a p_b,\Lambda_a(p_a-k), Q_X ) \; ( \Lambda_a p_b ) \right ]
\\
& = {\rm Tr} \big [ N_a(p_b + \delta p_{a1} ,xp_a - \delta p_{a1}, Q_X) \; ( x \hat p_a + \delta \hat p_{a2})  \\
& \quad \qquad \times N_a^+(
p_b + \delta p_{a1} ,xp_a - \delta p_{a1} , Q_X ) \; 
\left ( \hat p_b + {\delta \hat p_{a1}} 
\right ) \big ].
\end{split}
\label{eq4.132}
\ee
The momenta shifts shown in  Eq.~(\ref{eq4.132}) are easily obtained 
using the explicit form of the boost operator $\Lambda_a$, c.f. Appendix~\ref{appA}. We find
\be
\label{eq3.7}
\begin{split} 
& \delta p_{a1} = \frac{k_\perp}{2} 
+ \frac{2 p_a k}{s} 
\left ( p_b + (1-x) \pi_{a1}
\right ), 
\\
& \delta p_{a2} = 
\frac{k_\perp}{2} 
- \frac{2 p_a k}{s} \left ( 
p_a -(1-x)  \pi_{a2}
\right ), 
\end{split}
\ee
with 
\begin{equation}
\pi_{a1} = \frac{p_a}{4}  + \frac{3 p_b}{4 x},
\;\;\;
\pi_{a2} = \frac{3 p_a}{4}  + \frac{ p_b}{4x}.
\end{equation}

The non-trivial step, 
required to move forward, 
is the expansion of Eq.~(\ref{eq4.132}) in powers of $k_\perp$. Since the shifts in 
Eq.~(\ref{eq3.7}) are linear in $k_\perp$, the trace in Eq.~(\ref{eq4.132}) needs to be expanded through  \emph{second} order in $k_\perp$. To organize this expansion efficiently, it is convenient to rewrite Eq.~(\ref{eq4.132}) by introducing the following momenta
\be
 {\cal P}_a  
= xp_a - \delta p,\;\;\;\;{\cal P}_b  
= p_b + \delta p,
\;\;\;\;\;\delta p = \frac{k_\perp}{2}
+ \frac{k_\perp^2}{4 s x} (p_b - x p_a).
\ee
These momenta are constructed 
in such a way that 
${\cal P}_a^2 = {\cal P}_b^2 = 0$
with ${\cal O}(k_\perp^2)$ accuracy, and ${\cal P}_a + {\cal P}_b = xp_a + p_b$.
Using these momenta, we find 
\be
xp_a - \delta p_{a1}
 = {\cal P}_a - \frac{2k \cdot p_a}{sx} p_b, 
 \;\;\;
 p_b + \delta p_{a1}
 = {\cal P}_b + 
 \frac{2k \cdot p_a}{sx}p_b.
\ee
 Furthermore, we find 
 \be
x p_a + \delta p_{a2}
 = {\cal P}_a + k_\perp - 
 \frac{2 k \cdot p_a}{s} x p_a.
 \ee
We are now in a position to 
rewrite Eq.~(\ref{eq4.132}) in the following way
\be
\begin{split} 
& {\rm Tr} \left [ N_a(p_b,p_a-k,Q_X) \;  x\hat p_a \; 
N_a^+(p_b,p_a-k,Q_X) \; \hat p_b \right ]_{\Lambda_a}
\\
& = {\rm Tr} \Bigg[ N_a\left(
{\cal P}_b + \frac{2k \cdot p_a}{sx} p_b,
{\cal P}_a - \frac{2k \cdot p_a}{sx} p_b,
 Q_X \right) \; \left( 
\hat {\cal P}_a + \hat k_\perp - \frac{2k \cdot p_a}{s} x \hat p_a
\right)  \\
& \quad \qquad \times 
N^+_a\left( {\cal P}_b + \frac{2k \cdot p_a}{sx} p_b,{\cal P}_a - \frac{2k \cdot p_a}{sx} p_b,
 Q_X \right)
\; 
 \left( \hat {\cal P}_b + 
\frac{2k \cdot p_a}{sx} \hat p_b
 \right) \Bigg].
\\
& = \left ( 1+\frac{2k \cdot p_a}{s} \frac{1-x}{x} \right )
|{\cal M}|^2({\cal P}_b, {\cal P}_a,Q_X)
+ k_{\perp,\mu} 
{\rm Tr}\left [ N_a \gamma^\mu N_a^+ \hat p_b \right ]
\\
& -\frac{2 k \cdot p_a}{sx} p_{b,\mu} \left ( 
{\rm Tr}[N_a^{(1),\mu} x p_a N_a^+ p_b] +{\rm c.c.}
\right ) 
-\frac{k_{\perp,\mu} k_\perp^\nu}{2} D^{ax,b}_\nu 
{\rm Tr}\left [ N_a \gamma^\mu N_a^+ \hat p_b \right ],
\end{split}
\label{eq3.12}
\ee
where $D_\mu^{xa,b}  = x^{-1}\partial/\partial p_{a}^{\mu} - \partial/\partial p_{b}^{\mu}$, and 
the quantity $N_a^{(1),\mu}$ is defined through the expansion of the function $N_a$ as follows 
\be
N_a(q_b-\delta q,q_a + \delta q, Q_X)  
= N_a(q_b,q_a,Q_x) + 
\delta q_\mu N_a^{(1),\mu}(q_b,q_a,Q_X).
\ee

Note that in Eq.~(\ref{eq3.12}) we have written the matrix element as a function 
of two momenta ${\cal P}_a$, ${\cal P}_b$. This is possible because both of these momenta are on-shell and the momentum conservation ${\cal P}_a + 
{\cal P}_b = Q_X$ is assured.  It remains to expand the matrix element squared through the right order in $k_\perp$.
The result reads 
\be
\begin{split}
|{\cal M}|^2({\cal P}_b, {\cal P}_a,Q_X)
=  & \Bigg[ 1
-\frac{k_\perp^\mu}{2} D_\mu^{xa,b}
- \frac{k_\perp^2}{4sx}(p_b^\mu - x p_a^\mu)
D_\mu^{xa,b} 
\\
& + 
\frac{k_\perp^\mu k_\perp^\nu}{8} D_\mu^{xa,b} D_\nu^{xa,b} 
\Bigg] |{\cal M}|^2(p_b,xp_a,Q_X).
\label{eq3.14a}
\end{split}
\ee

As the last step, we need to 
combine Eqs~(\ref{eq3.12}, \ref{eq3.14a}) 
and average over the directions 
of the vector $k_\perp$ in ${\cal O}(k_\perp^2)$  terms.\footnote{Note that we do not discard terms linear in $k_\perp$ because such terms may get combined with the collinear expansion of an observable ${\cal O}$ producing a non-vanishing 
result. However, since 
${\cal O}(k_\perp^2) $ terms that appear from the  expansion of the matrix element squared 
will always be multiplied by  
${\cal O}(k_\perp^0)$ 
terms in  the  observable, we can  average over directions of $k_\perp$ 
in such terms right away.} It is 
convenient to write the result as follows 
\be
\begin{split} 
& {\rm Tr} \left [ N_a(p_b,p_a-k, Q_X) \; x \hat p_a \; N_a^+(p_b,p_a-k, Q_X ) \; \hat p_b ) \right ]_{\Lambda_a}
= \; |{\cal M}|^2(p_b,xp_a,Q_X)   
\\
& - \frac{k_\perp^\mu}{2} \left (  D^{xa,b}_\mu 
|{\cal M}|^2(p_b,xp_a,Q_X) - 2 {\rm Tr} \left [N_a \gamma_{\mu} 
N_a^+ \hat p_b \right ] \right ) +  
\frac{2 k p_a}{s} W_a(x).
\end{split} 
\ee
In the above equation the function $W_a(x)$ is defined as 
\be
\begin{split}
W_{a}(x) =
- p_{b,\mu} 
W_{a1}^\mu(x) 
+ (1-x) W_{a2}(x), 
\end{split}
\label{eq3.10}
\ee
where 
\be
\begin{split}
W_{a1}^\mu (x) =& 
{\rm Tr} 
\left [ 
N_a^{(1),\mu} \; x \hat p_a \; N_a^{+} \hat p_b \right ]
+ {\rm c.c.},\;\;\;
\\
W_{a2}(x) = & 
-\frac{1}{x} p_{b,\mu} W_{a1}^\mu 
+\frac{1}{4x} \left ( 4
+ (p_b^\mu - x p_a^\mu) D_{\mu}^{xa,b} 
\right ) |{\cal M}|^2(p_b,xp_a,Q_X)
\\
& - \frac{s}{16} g^{\mu \nu}_{\perp} \left( D^{xa,b}_\nu D^{xa,b}_\mu \; |{\cal M}|^2(p_b,xp_a,Q_X)  - 4 D^{xa,b}_\mu \; {\rm Tr} \left [N_a \gamma_{\nu} 
N_a^+ \hat p_b \right ] \right).
\end{split}
\ee
\\

Another quantity that we need to expand is the observable ${\cal O}$ because it 
depends on the transformed momentum of the final-state colorless particles
\be
 \tilde Q_X = \Lambda_a^{-1} Q_X.
\ee
We write 
\be 
\left [  \Lambda^{-1}_{a}
\right ]^{\mu \nu}(x) = g^{\mu \nu} 
+ b^{\mu \nu,\alpha}_{a} k_{\perp,\alpha}
+ \frac{2k \cdot p_a}{s} l_a^{\mu \nu}, 
\ee
where 
\be
\begin{split} 
& {b_a}^{\mu \nu}_{\alpha} = 
\frac{Q^\mu_a g^{\nu}_{\alpha} 
-Q^{\nu}_{a} g^{\mu}_{\alpha}
}{sx},\;\;\;  l_{a}^{\mu \nu}(x)
= \omega^{\mu \nu}_{ab}  + 
(1-x)  
\left [ 
\frac{1}{2 x} 
\omega_{ab}^{\mu \nu}
+\frac{Q_a^\mu Q_a^\nu}{2s x^2} 
+ \frac{g_{\perp}^{\mu \nu}}{4 x} 
\right ],
%b^{\mu \nu},
\end{split}
\label{eq3.20}
\ee 
with
\be
\omega_{ab}^{\mu \nu} 
= \frac{p_a^\mu p_b^\nu - p_b^\mu p_a^\nu} {p_a \cdot p_b}.
\ee
We note that in  Eq.~(\ref{eq3.20}), we have 
replaced  $k_\perp^\mu k_\perp^\nu$ with its average value, 
since  there will be  no further dependencies on $k_\perp$ when  this term is combined with  an 
amplitude squared.
Furthermore, we took the four-dimensional limit because such terms do not present soft or collinear singularities.  Finally, 
we note that 
\be
\lim_{x \to 1} l_a^{\mu \nu}(x)  = 
\omega_{ab}^{\mu \nu}.
\ee
Using the above results, we easily  expand the observable ${\cal O}$ 
around  the collinear limit
\be
\begin{split}
{\cal O}(\Lambda_a^{-1} Q_X) 
= & \Bigg [ 1 + 
\left ( k_\perp^\alpha {b_a}^{\mu \nu}_\alpha 
+ \frac{2 k \cdot p_a}{s} l_a^{\mu \nu} (x)   
\right ) L_{\mu \nu}
\\
& -\frac{1}{2} (1-x) k \cdot p_a \;t_a^{\mu \mu_1, \nu \nu_1} 
L_{\mu \mu_1} L_{\nu \nu_1} 
\Bigg ] {\cal O}(Q_X).
\end{split} 
\label{eq3.23}
\ee
In Eq.~(\ref{eq3.23}), 
the differential operator $L^{\mu \nu}$ reads
\be
L^{\mu \nu} = 
\sum \limits_{i = 1}^{m} p_i^\nu \;
\frac{\partial }{\partial p_{i,\mu}},
\label{eq3.24}
\ee
and 
\be
t_a^{\mu \mu_1, \nu \nu_1} = g_{\perp}^{\alpha \beta} \;  {b_a}^{\mu \mu_1}_{\alpha} {b_a}^{\nu \nu_1}_\beta,
\ee
is a rank-four tensor. 
\\

Having computed all the different 
terms  in  Eq.~(\ref{eq6.3}) 
to the required order in the collinear expansion, we are in the position to 
write the different contributions to the cross section as an expansion in $\ep$.
In this respect, 
we note that the only divergence 
in the subleading term comes from the soft singularity of the splitting function, so that it is  straightforward to extract it. 
We write 
\be
P_{qq}(x)\; (1-x)^{-\ep} 
 = -\frac{2}{\ep} \delta(1-x) + 
 \bar P_{qq}(x) + {\cal O}(\ep),
\ee
where 
\be
\bar P_{qq}(x) = 
\frac{2}{(1-x)_+} -(1+x).
\ee
Using this representation, we derive 
the following result 
for the next-to-leading power contribution that originates 
from the first term in Eq.~(\ref{eq6.2})
\be
\begin{split} 
 \frac{ {\rm d} \sigma^{ca,1,\rm NLP}}{{\rm d} \tau}
& = -\frac{2 [\alpha_s] C_F Q^{1-\ep}  {\cal N} }{s \tau^{\ep} \ep } 
  {\rm d} \Phi_m(p_a,p_b,P_X)  \Big [ 
-p_{b,\mu} W_{a1}^\mu(1)
\\
&+ |{\cal M}|^2(p_b,p_a,..)
\; 
\omega_{ab}^{\mu \nu} L_{\mu \nu}
  \Big ] {\cal O}(P_X) 
\\
& + 
\frac{[\alpha_s] C_F Q {\cal N} }{s}  
\int \limits_{0}^{1}  {\rm d}x \;
{\rm d} \Phi_m(xp_a,p_b,P_X) 
 \frac{\bar P_{qq}(x)}{ \; x} \; 
 \Big \{ 
W_a(x) 
\\
&+\frac{s}{4}(1-x) g_\perp^{\rho \alpha} 
  \Big [ D^{xa,b}_\rho \; |{\cal M}|^2(p_b,xp_a, ...) 
  - 2 {\rm Tr} \left [N_a \gamma_{\rho} 
N_a^+ \hat p_b \right ] \Big ] 
 {b_a}^{\mu \nu} _\alpha L_{\mu \nu}
 \\
& +  |{\cal M}|^2(p_b,xp_a,...) 
 \; l_a^{\mu \nu}(x) 
 L_{\mu \nu} 
\\
& -\frac{s \; (1-x)}{4} |{\cal M}|^2(p_b,xp_a,...) 
t_a^{\mu \mu_1, \nu \nu_1} 
 L_{\mu \mu_1} L_{\nu \nu_1} 
 \Big \} \; {\cal O}(P_X).
\end{split}
\label{eq3.22}
\ee
\\

Next we consider the second term in 
Eq.~(\ref{eq6.2}). That  term is already  subleading in the collinear limit  which means that no  Lorentz boost needs to be applied to it. Consequently, we can replace $\tilde Q_X$
with $Q_X$ everywhere. We also note  that the trace in that term gives the squared matrix element  of the leading order  process with $x$-dependent kinematics 
\be
{\rm Tr} 
\left [ N_a(p_b,x p_a,Q_X)  x \hat p_a N_a^+(p_b,x p_a, Q_X) \hat p_b \right ] 
= |{\cal M}|^2(p_b,xp_a,Q_X) .
\ee
However, the complication arises because  
this term is \emph{linearly divergent} in the soft limit; for this reason,  it requires additional manipulations. 
We begin by writing this term explicitly   
\be
 \frac{ {\rm d} \sigma^{ca,2}}{{\rm d} \tau}
= \frac{2(1+\ep) [\alpha_s] C_F  }{s\tau^{\ep} Q^{-1+\ep} } {\cal N} 
\int \limits_{0}^{1} \; {\rm d}x \; 
{\rm d} \Phi_m^{xa}\;
\frac{{\cal O}(Q_X) }{(1-x)^{2+\ep}} \; 
|{\cal M}|^2(p_b,xp_a,Q_X),
\label{eq6.23}
\ee
where  ${\rm d} \Phi_{m}^{xa}$ is 
given in Eq.~(\ref{eq2.56}).

To extract singularities from this expression, we 
would like to remove the $x$-dependence from the phase space. We do this by performing a boost, along the lines of what was done for the discussion of the soft contribution 
in Sec.~\ref{sec:soft}. 
To this  end, we write 
\be
0 = x p_a + p_b - Q_X = 
p_a + p_b -(1-x) p_a - Q_X,
\ee
and  treat  $(1-x) p_a$ as the 
``soft gluon momentum''. Similarly 
to the discussion in 
Sec.~\ref{sec:soft}, 
we  remove it using a Lorentz boost along with the rescaling
\be
p_a + p_b = \lambda^{-1} \Lambda_{ax} \;  (x p_a + p_b).
\ee

Using the fact that boosts do not change 
squares of four-momenta, it is easy to see that $\lambda = \sqrt{x}$. Following the steps 
discussed in the section 
dedicated to soft emissions, we find 
\be
{\rm d} \Phi_m^{xa} = {\rm d} \Phi_m(xp_a,p_b,Q_X)
= {\rm d} \Phi_m(p_a,p_b, P_X) \;
\lambda^{\kappa_m},
\ee
where $\kappa_m$ is  defined in 
Eq.~(\ref{eq4.12})
and the relation between $Q_X$ 
and $P_X$ reads 
\be
Q_X = \lambda \Lambda_{ax}^{-1} 
P_X.
\ee
We then find  
\be
\begin{split}
 \frac{ {\rm d} \sigma^{ca,2}}{{\rm d} \tau}
= \frac{2 (1+\ep) 
[\alpha_s] C_F Q^{1-\ep} }{s\tau^{\ep} } {\cal N} 
\int \limits_{0}^{1} \; {\rm d}x \; 
{\rm d} \Phi_m^{ab} \; \lambda^{\kappa_m} 
(1-x)^{-\ep-2} \; 
\\
\times {\cal O}(
\lambda \Lambda_{ax}^{-1} P_X) \;
|{\cal M}|^2(p_b,xp_a,\lambda \Lambda_{ax}^{-1} P_X).
\end{split}
\ee
To extract singularities 
from this expression 
and to regulate them, 
we require the expansion  of the matrix element squared, the $\lambda^{\kappa_m}$ factor  
and the observable ${\cal O}$  
through first order in $(1-x)$. 
The boost operator, as well as its expansion 
in $(1-x)$ is given in 
Eqs~(\ref{eqa.13}, \ref{eqa.14}). Using those equations and the expansion of $\lambda$ around $x=1$, 
$\lambda = 1 - (1-x)/2+{\cal O}( (1-x)^2 )$, we obtain  
\be
\lambda (\Lambda^{-1}_{ax})^{\mu \nu}  = g^{\mu \nu} 
- \frac{ (1-x) }{2}
\left ( g^{\mu \nu} + \omega_{ab}^{\mu \nu}
\right ) 
+{\cal O}((1-x)^2).
\ee
We then find 
\be
\begin{split} 
& |{\cal M}|^2 \left (p_b,x p_a,  \lambda \Lambda_{ax}^{-1} 
P_X \right )
= 
\\
& \Bigg[ 1- \frac{(1-x)}{2} 
\left ( 
2 p_a^\mu \frac{ 
\partial }{\partial p_{a,\mu} }
+ \left ( g^{\rho \sigma } 
+ \omega_{ab}^{\rho \sigma}
\right ) L_{\rho \sigma}
\right ) 
\Bigg]
|{\cal M}|^2 \left ( p_b, 
p_a, 
P_X \right ) + {\cal O}((1-x)^2),
\end{split}
\label{eq6.31}
\ee
and 
\be
{\cal O}(\lambda \Lambda_{ax}^{-1} P_X) 
= 
\Big  [ 1 -   \frac{ (1-x)}{2} 
\left ( g^{\rho \sigma } 
+ \omega_{ab}^{\rho \sigma}
\right ) L_{\rho \sigma}
 \Big ]
{\cal O}( P_X)
+{\cal O}((1-x)^2).
\label{eq6.32}
\ee

It is now convenient to define a  new function to represent the subtracted expression
\be
\begin{split}
& W^{(a)}_3(x,p_b,p_a,P_X,{\cal O}(P_X)) =  \lambda^{\kappa_m} 
{\cal O}(\lambda \Lambda_{ax}^{-1} P_X) 
|{\cal M}|^2(p_b,x p_a, \lambda \Lambda_{ax}^{-1} P_X)
 -  
 \\
 & \Bigg[ 1- 
 \frac{(1-x)}{2} 
\left ( 
\kappa_m + 
2 p_a^\mu \frac{ 
\partial }{\partial p_a^\mu }
 +\left ( g^{\rho \sigma } 
+ \omega_{ab}^{\rho \sigma}
\right ) L_{\rho \sigma}
\right ) 
\Bigg] |{\cal M}|^2 \left ( p_b,p_a, 
P_X \right ) 
{\cal O}(P_X),
\end{split}
\label{eq6.33}
 \ee
where we have assumed that the observable  ${\cal O}$ is independent
of the momentum $p_a$.
It follows from Eqs~(\ref{eq6.31}, \ref{eq6.32}) 
 that 
in the soft limit  $W^{(a)}_3$
vanishes as ${\cal O}((1-x)^2)$ and, 
therefore, can be integrated with the 
$1/(1-x)^2$  factor which appears in the cross section computation,
c.f. Eq.~(\ref{eq6.23}). 

We can now write the complete result that originates from  
the second term in Eq.~(\ref{eq6.2}) in the following way\footnote{We provide a detailed derivation of this formula in Appendix~\ref{app:eq340}.}
\be
\label{eq3.40}
\begin{split}
 \frac{ {\rm d} \sigma^{ca,2,\rm NLP}}{{\rm d} \tau}
& = \frac{ (1+\ep) [\alpha_s] C_F Q^{1-\ep} }{s \tau^{\ep} \ep } {\cal N} 
\; 
{\rm d}  \Phi_m(p_a,p_b,P_X) \Bigg[ -\frac{2\ep}{1+\ep} + \kappa_m  
\\
& +   
\left ( 
2 p_a^\mu \frac{ 
\partial }{\partial p_a^\mu }
+\left ( g^{\rho \sigma } 
+ \omega_{ab}^{\rho \sigma}
\right ) L_{\rho \sigma}
\right ) 
\Bigg] {\cal O}(
 P_X) \;
|{\cal M}|^2(p_b,p_a, P_X)
\\
& + \frac{2 [\alpha_s] 
C_F Q}{s} {\cal N} 
\int \limits_{0}^{1} {\rm d} x \; 
{\rm d} \Phi_m(p_a,p_b,P_X)  
\frac{W^{(a)}_3(x,p_b,p_a,P_X,{\cal O}(P_X) )}{(1-x)^2} \\
& = \frac{ [\alpha_s] C_F Q^{1-\ep} }{s \tau^{\ep} \ep } {\cal N} 
\; 
{\rm d} \Phi_m(p_a,p_b,P_X)   \Bigg[ -2\ep + \kappa_m 
\\
& +   
\left ( 
2 p_a^\mu \frac{ 
\partial }{\partial p_a^\mu }
+\left ( g^{\rho \sigma } 
+ \omega_{ab}^{\rho \sigma}
\right ) L_{\rho \sigma}
\right ) 
\Bigg] {\cal O}(
 P_X) \;
|{\cal M}|^2(p_b,p_a, P_X)
\\
& - \frac{[\alpha_s] 
C_F Q}{s} {\cal N} 
\int \limits_{0}^{1} {\rm d} x \; 
{\rm d} \Phi_m(x p_a,p_b,P_X) 
 \;
\frac{1}{(1-x)_+} \\
& \times \left ( \kappa_m + 2 p_a^\mu \frac{ 
\partial }{\partial p_a^\mu }
+\left ( g^{\rho \sigma } 
+ \omega_{ab}^{\rho \sigma}
\right ) L_{\rho \sigma} \right ) {\cal O}(P_X) \; |{\cal M}|^2(p_b,x p_a,P_X),
\end{split}
\ee
where in the first term integration over $x$ has been performed and the $W_3^{(a)}$ term was rewritten in terms of the plus  distribution. 
\\

Finally, we need to consider the 
 third  term in Eq.~(\ref{eq6.2}). This term is also subleading in the collinear expansion
which means that no boost is required. 
Its contribution to the cross section reads
\be
\begin{split}
 \frac{ {\rm d} \sigma^{ca,3}}{{\rm d} \tau}
& = -\frac{[\alpha_s] C_F Q^{1-\ep} }{2 \tau^{\ep} } {\cal N} 
\int \limits_{0}^{1} \; {\rm d}x \; 
\frac{ {\rm d}  \Phi_m(xp_a,p_b,P_X) \; 
{\cal O}(P_X)  }{
(1-x)^{1+\ep} } 
\\
& \times \; 
\;
\frac{4 p_b^\nu}{s}
\left ( 
{\rm Tr} 
\left [N_a \hat p_a R_{\rm fin, \nu}^+ \hat p_b  \right ] +{\rm c.c.} 
\right ).
\end{split}
\ee
  We then replace 
$(1-x)^{- 1 - \ep}$ with the plus distribution in the standard way 
and find 
\be
\begin{split}
 \frac{ {\rm d} \sigma^{ca,3}}{{\rm d} \tau}
&=  -\frac{[\alpha_s] C_F Q^{1-\ep} }{2 \tau^{\ep} } {\cal N} 
\int \limits_{0}^{1} \; {\rm d}x \; 
{\rm d} \Phi_m(xp_a,p_b,P_X) 
\left [ -\frac{1}{\ep} \delta(1-x) 
+ \frac{1}{( 1-x)_+}  
\right ]
\\
& \times \; {\cal O}(P_X) \;
\frac{4 p_b^\nu}{s}
\left ( 
{\rm Tr} 
\left [N_a \hat p_a R_{\rm fin, \nu}^+ \hat p_b  \right ] +{\rm c.c.} 
\right ).
\end{split}
\ee
The $1/\ep$ divergent term 
requires $R_{\rm fin, \nu}(p_b,p_a,k,P_X)$, with $k = (1-x) p_a$  
at $x =1$ (i.e., the soft limit).  We can obtain 
it using gauge invariance.
From the transversality of 
the gluon emission amplitude
it follows that 
\be
\bar v_b 
\left ( 
N(p_b-k,p_a,P_X)
- N(p_b,p_a-k,P_X) 
+ k^\mu R_{\rm fin, \mu} (p_b,p_a,k,P_X)
\right ) u_a = 0. 
\label{eq3.43}
\ee
We are interested in the soft 
$k \to 0$ limit. Since $R_{\rm fin, \nu}(p_b,p_a,k,P_X)$ in that equation is multiplied with $k_\nu$, we can replace it with 
$R_{\rm fin, \nu}(p_b,p_a,0,P_X)$.
The difference of the two Green's functions can be computed using the discussion 
in Sec.~\ref{sec:currents},
where it is explained how such an expansion should be  constructed. 
In particular, if we compute it starting from the incoming quark momentum and do not let momentum $k$ flow into the colorless final state, 
then we have to replace 
$p_b$ with $P_X + k - p_a$
in the functions $N$ in Eq.~(\ref{eq3.43}).
It follows that 
\be
N(p_b-k,p_a,P_X)
- N(p_b,p_a-k,P_X) 
= k^\mu N^{(1),\mu}(p_b,p_a,P_X). 
\ee
Employing this result in 
Eq.~(\ref{eq3.43}) and making 
use of the fact that it is valid for small, but otherwise arbitrary   vectors $k^\mu$, 
we find 
\be
p_b^\nu \bar v_b R_{\rm fin, \nu} u_a \Big |_{k \to 0}
= -p_{b,\nu} \bar v_b N_a^{(1),\nu} u_a \Big |_{k \to 0}.
\ee

%We obtain it from the 
%gauge invariance condition 
%\be
%\lim_{x \to 1} \bar u_b %R_\nu^{\rm fin} u_a 
%= - \bar u_b \left ( %D^{ab}_\nu N_a \right ) u_a,
%\ee
%where $D^{ab}_\nu = %\partial/\partial p_{a,\nu} %- \partial/\partial %p_{b,\nu}$. 
Hence, we obtain  
\be
\begin{split}
 \frac{ {\rm d} \sigma^{ca,3, \rm NLP}}{{\rm d} \tau}
& = -\frac{[\alpha_s] C_F Q^{1-\ep} }{2 \tau^{\ep} \ep } {\cal N} 
\; 
 {\rm d} \Phi_m(p_a,p_b,P_X) \; 
{\cal O}(P_X) \;
\\
& \times 
\frac{4 p_b^\nu}{s}
\left ( 
{\rm Tr} 
\left [N_a \hat p_a  N^{(1),+}_\nu  \hat p_b  \right ]_{x=1} +{\rm c.c.} 
\right )
\\
& -\frac{[\alpha_s] C_F Q }{2 } {\cal N} 
\int \limits_{0}^{1} \; \frac{{\rm d}x}{( 1-x)_+}   \; 
{\rm d} \Phi_m(xp_a,p_b,P_X) \; {\cal O}(P_X) \;
\\
& \times \frac{4 p_b^\nu}{s}
\left ( 
{\rm Tr} 
\left [N_a \hat p_a R_{\rm fin, \nu}^+ \hat p_b  \right ] +{\rm c.c.} 
\right ).
\end{split}
\label{eq3.39}
\ee
We note that the 
${\cal O}(1/\ep)$ term in  Eq.~\eqref{eq3.39} 
is exactly 
canceled by the first term in Eq.~\eqref{eq3.22}.

\subsection{The second collinear region: $\vec k || \vec p_b$}

We continue with the contribution of  the collinear region  where 
the gluon is emitted along 
the direction of the incoming 
anti-quark. 
The differential cross section in this case reads 
\be
\begin{split}
 \frac{ {\rm d} \sigma^{cb}}{{\rm d} \tau}
& = \frac{[\alpha_s] C_F Q^{1-\ep} }{2 \tau^{1+\ep} } {\cal N} 
\int \limits_{0}^{1} {\rm d}x 
{\rm d} \Phi_m(p_a,x p_b,\tilde Q_X) \left[ {\rm d} \Omega_k^{(d-2)} \right]
(1-x)^{-\ep} 
\\
&  \times \left ( 1 + \frac{\ep Q \tau}{s(1-x)}
\right ) \; {\cal O}(\tilde Q_X) \;
\tau \; F_b,
\end{split}
\ee
where $\tilde Q_X = \Lambda_b^{-1} Q_X$, $F_b$ is given in Eq.~(\ref{eq5.111}) and the momenta $p_a$, $p_b$, $k$ that appear there should be boosted.  Similar to the 
collinear  case where $\vec k || \vec p_a$, it is convenient to 
write, with the required  accuracy, 
\be
\begin{split}
&  \left ( 1 + \frac{\ep Q \tau}{s(1-x)}
\right ) \;  F_b 
 = 
\frac{2 P_{qq}(x)}{2 p_b \cdot k}
{\rm Tr} \left [ N_b^+ \hat p_b 
N_b \hat p_a \right ]
+ 
\frac{4(1+\ep) }{s(1-x)^2} 
{\rm Tr} 
\left [ N_b^+ \hat x p_b N_b \hat p_a \right ]
\\
&  +\frac{4 p_a^\nu}{s(1-x)}
{\rm Tr} 
\left [N_b^+ \hat p_b R_{\rm fin, \nu} \hat p_a  \right ] +{\rm c.c.} 
+ F_{b,\rm reg},
\end{split}
\label{eq6.41}
\ee
where the function $F_{b,\rm reg}$ is free from both soft and  collinear singularities.

In principle, the discussion 
of the $\vec k || \vec p_b$ collinear limit follows very closely the discussion in the 
previous section. Nevertheless, we decided to repeat it one more time for the sake of clarity. 

We begin with the   first term on the right-hand side  of Eq.~(\ref{eq6.41}). 
We note that momenta that appear in that term still have to be boosted
with the matrix $\Lambda_b$. Hence, we write 
\be
\begin{split}
 \frac{ {\rm d} \sigma^{cb,1}}{{\rm d} \tau}
&= \frac{[\alpha_s] C_F Q^{1-\ep} }{2 \tau^{1+\ep} } {\cal N} 
\int \limits_{0}^{1} \; {\rm d}x \; 
{\rm d} \Phi_m(p_a,xp_b,Q_X) [{\rm d} \Omega_k^{(d-2)}]
(1-x)^{-\ep} {\cal O}(\Lambda_b^{-1} Q_X)  \\
& 
\times \; 
\tau \; \frac{2 P_{qq}(x)}{2 p_b \cdot k \; x} \; 
{\rm Tr} \left [ N_b^+(p_b-k,p_a,Q_X) \;  x \hat p_b 
\; N_b(p_b-k,p_a,Q_X) \; \hat p_a \right ]_{\Lambda_b},
\end{split}
\label{eq4.129}
\ee
where the subscript of the trace  
function indicates that momenta 
$p_a$, $p_b$ and $k$ should be boosted 
with the matrix $\Lambda_b$.

Following the discussion 
of the  $\vec k || \vec p_a$ case, we first show the result for the leading power contribution that  
is obtained by setting 
$\Lambda_b \to 1$. We find 
\be
\begin{split}
 \frac{ {\rm d} \sigma^{cb,1,\rm LP}}{{\rm d} \tau}
&= \frac{[\alpha_s] C_F Q^{-\ep} }{ \tau^{1+\ep} } {\cal N} 
\int \limits_{0}^{1} \; {\rm d}x \; 
{\rm d} 
\Phi_m(p_a,xp_b,Q_X)
 \;
 \; \frac{ P_{qq}(x)}{ x(1-x)^{\ep}} \;
 \\
& \times   {\cal O}(Q_X)  \;
 |{\cal M}(x p_b,p_a,P_X)|^2,
\end{split}
\ee
where we have used $ 2p_b k = \tau Q$, 
and the fact that in the collinear limit 
\be
{\rm Tr} \left [ N_b^+(p_b-k,p_a,Q_X) \;  x \hat p_b 
\; N_b(p_b-k,p_a,Q_X) \; \hat p_a \right ]_{\Lambda_b}
\to |{\cal M}(xp_b,p_a p_b,Q_X)|^2.
\ee
 There is a soft singularity present in the splitting function $P_{qq}$, but 
it is straightforward to extract it. 

Computing the subleading terms in the $\tau$-expansion requires more effort. We start by showing  formulas for the trace 
\be
\begin{split} 
& 
{\rm Tr} \left [ N_b^+(p_b-k,p_a,Q_X) \;  x \hat p_b 
\; N_b(p_b-k,p_a,Q_X) \; \hat p_a \right ]_{\Lambda_b}
\\
& = {\rm Tr} \big [ N_b^+(
xp_b- \delta p_{b1},
p_a + \delta p_{b1}, Q_X) ( x \hat p_b + \delta \hat p_{b2})  
\\
& \qquad \qquad \qquad \times N_b(x p_b - \delta p_{b1} , p_a + \delta p_{b1}, Q_X ) 
( \hat p_a + \hat \delta p_{b1} ) \big ],
\end{split}
\label{eq6.45}
\ee
where 
\be
\begin{split} 
& \delta p_{b1} = \frac{k_\perp}{2} 
+ \frac{2 k \cdot p_b}{s} 
\left ( p_a + (1-x) \pi_{b1}
\right ), 
\\
& \delta p_{b2} = 
\frac{k_\perp}{2} 
- \frac{2 p_b \cdot  k}{s} \left ( 
p_b -(1-x)  \pi_{b2}
\right ), 
\\
\end{split}
\ee
and
\begin{equation}
\pi_{b1} = \frac{p_b}{4}  + \frac{3 p_a}{4 x},
\;\;\;
\pi_{b2} = \frac{3 p_b}{4}  + \frac{ p_a}{4x}.
\end{equation}
We note that, thanks to the 
explicit $(1-x)$ factors in front of  vectors $\pi_{b1,b2}$, they do not contribute to soft singularities. 
Proceeding as in the previous section, we find 
\be
\begin{split} 
& {\rm Tr} \left [ N_b^+(p_b-k,p_a,  Q_X) \; x \hat p_b \; N_b(p_b-k,p_a, Q_X ) \; \hat p_a ) \right ]_{\Lambda_b}
=  |{\cal M}|^2(x p_b,p_a,Q_X)   
\\
& - \frac{k_\perp^\mu}{2} \left( D_\mu^{xb,a} 
\left [ |{\cal M}|^2(x p_b,p_a,Q_X)  \right ] - 2 {\rm Tr} \left[ N_b^+ \gamma_\mu 
N_b \hat p_a \right] \right) +  
\frac{2 k \cdot p_b}{s} W_b(x),
\end{split} 
\ee
where $D^{xb,a}_\mu = x^{-1} \partial/\partial p_{b}^\mu 
- \partial/\partial {p_a}^\mu$ 
and we defined 
\be
\begin{split}
W_b(x) =
- p_{a,\mu} \; 
W_{b1}^\mu(x) 
+ (1-x) W_{b2}(x), 
\end{split} 
\ee
with 
\be
\begin{split}
  W_{b1}^\mu & = 
 {\rm Tr} 
\left [ 
N_b^{(1),\mu,+} \;  x \hat p_b \; N_b \; \hat p_a 
 \right ] + {\rm c.c.}   ,
\\
 W_{b2} & = 
-\frac{1}{x} p_{a,\mu} W_{b1}^\mu + \frac{1}{4x} \left ( 4
+ (p_a^\mu - x p_b^\mu) 
D^{xb,a}_\mu \right) |{\cal M}|^2(xp_b,p_a,Q_X)
\\
& - \frac{s}{16} g^{\mu \nu}_{\perp} \left( D^{xb,a}_\nu D^{xb,a}_\mu \; |{\cal M}|^2(x p_b,p_a ,Q_X)  - 4 D^{xb,a}_\mu \; {\rm Tr} \left [N_b^+ \gamma_{\nu} 
N_b \hat p_a \right ] \right).
\end{split}
\ee
The function $N_b^{(1),\mu}$ in the above 
equation is defined in Appendix~\ref{appB}. 

We also need to expand the observable ${\cal O}$ 
since it 
depends on the transformed momenta $
\tilde Q_X = \Lambda_b^{-1} Q_X$.
We write 
\be 
{\Lambda_b^{-1}}^{\mu \nu}(x) = g^{\mu \nu} 
+ b^{\mu \nu, \alpha}_{b} k_{\perp,\alpha}
+ \frac{2k \cdot p_b}{s} l_b^{\mu \nu}, 
\ee
where 
\be
\begin{split} 
& {b_b}^{\mu \nu}_{\alpha} = 
\frac{ Q^\mu_b g^{\nu}_{\alpha} 
- Q^{\nu}_{b} g^{\mu}_{\alpha}
}{sx},\;\;\; l_b^{\mu \nu}(x)
= \omega^{\mu \nu}_{ba}  + 
(1-x)  
\left [ 
\frac{1}{2 x} 
\omega_{ba}^{\mu \nu}
+\frac{ Q_b^\mu Q_b^\nu}{2s x^2} 
+ \frac{1}{4 x} g_{\perp}^{\mu \nu}
\right ],
%b^{\mu \nu},
\end{split}
\ee 
and we have replaced  $k_\perp^\mu k_\perp^\nu$ with its average value, 
since  there will be  no further dependencies on $k_\perp$ when the contribution of this term is taken 
into account. We also took the four-dimensional limit because such terms do not contribute to soft and collinear singularities.  Finally, 
we note that 
\be
\lim_{x \to 1} l_b^{\mu \nu}(x)  = 
\omega_{ba}^{\mu \nu}.
\ee
Using the above results, we find 
\be
\begin{split}
& {\cal O}(\Lambda_b^{-1} Q_X) 
= \Big [ 1 + 
\left ( k_\perp^\alpha {b_b}^{\mu \nu}_\alpha 
+ \frac{2 k \cdot p_b}{s} l_b^{\mu \nu}(x)   
\right ) L_{\mu \nu}
\\
& -\frac{1}{2} (1-x) k \cdot p_b \;t_b^{\mu \mu_1, \nu \nu_1} 
L_{\mu \mu_1} L_{\nu \nu_1} 
\Big ] {\cal O}(Q_X),
\end{split} 
\ee
where the differential operator $L^{\mu \nu}$  is given in 
Eq.~(\ref{eq3.24}) 
and the tensor $t_b^{\mu \mu_1, \nu \nu_1}$ reads 
\be
t_b^{\mu \mu_1, \nu \nu_1} = g_{\perp}^{\alpha \beta} \;  {b_b}^{\mu \mu_1}_{\alpha} {b_b}^{\nu \nu_1}_\beta .
\ee

Using these results and following the discussion of the $ \vec k || \vec p_a$ case, we obtain  
\be
\begin{split} 
 \frac{ {\rm d} \sigma^{cb,1,\rm NLP}}{{\rm d} \tau}
 & = \frac{2 [\alpha_s] C_F Q^{1-\ep}  {\cal N} }{s \tau^{\ep} \ep } 
  {\rm d} \Phi_m(p_a,p_b,P_X)  \Big [ 
p_{a,\mu} W_{b1}^\mu(1)
\\
& - |{\cal M}|^2(p_b,p_a,...)
\; 
\omega_{ba}^{\mu \nu} L_{\mu \nu}
  \Big ] {\cal O}(P_X) 
\\
& + 
\frac{[\alpha_s] C_F Q {\cal N} }{s }  
\int \limits_{0}^{1}  {\rm d}x \;
{\rm d} \Phi_m(p_a,x p_b,P_X) 
 \frac{\bar P_{qq}(x)}{ \; x} \; 
 \Big \{ 
W_b(x)
\\
& +\frac{s}{4}(1-x) g_\perp^{\rho \alpha} 
 \Big  [ D^{xb,a}_\rho |{\cal M}|^2(xp_b,p_a,..) 
 \\
&  - 2 {\rm Tr} \left[ N_b^+ \gamma_\rho 
N_b \hat p_a \right] \Big ]
 {b_b}_\alpha^{\mu \nu} L_{\mu \nu}
 +  |{\cal M}|^2(xp_b,p_a,..) 
 \; l_b^{\mu \nu}(x) 
 L_{\mu \nu}
 \\
&  -\frac{s(1-x)}{4} |{\cal M}|^2(xp_b,p_a,...) 
t_b^{\mu \mu_1, \nu \nu_1} 
 L_{\mu \mu_1} L_{\nu \nu_1} 
 \Big \} {\cal O}(P_X).
\end{split}
\label{eq3.58}
\ee
\\

Next we consider the second term in 
Eq.~(\ref{eq6.41}). 
The trace evaluates to 
\be
{\rm Tr} 
\left [ N_b^+(x p_b,p_a,Q_X) \; x \hat p_b \; N_b^+( xp_b,p_a,Q_X) \; \hat p_a \right ] 
= |{\cal M}|^2(x p_b,p_a,Q_X),
\ee
and, following steps 
described in the previous 
section, we find 
\be
 \frac{ {\rm d} \sigma^{cb,2}}{{\rm d} \tau}
= \frac{2(1+\ep) [\alpha_s] C_F  }{s\tau^{\ep} Q^{-1+\ep} } {\cal N} 
\int \limits_{0}^{1} \; {\rm d}x \; 
{\rm d} \Phi_m(p_a,xp_b,Q_X)
\frac{{\cal O}(Q_X) }{(1-x)^{2+\ep}} \; 
|{\cal M}|^2(x p_b,p_a,Q_X).
\label{eq6.60}
\ee
Extracting the singularity at $x = 1$,  as discussed in the previous section, we obtain
\be
\begin{split}
 \frac{ {\rm d} \sigma^{cb,2,\rm NLP}}{{\rm d} \tau}
& = \frac{ [\alpha_s] C_F Q^{1-\ep} }{s \tau^{\ep} \ep } {\cal N} 
\; 
{\rm d} \Phi_m(p_a,p_b,P_X)  \Big  [ -2\ep + \kappa_m 
\\
& +   
\left ( 
2 p_b^\mu \frac{ 
\partial }{\partial p_b^\mu }
+\left ( g^{\rho \sigma } 
+ \omega_{ba}^{\rho \sigma}
\right ) L_{\rho \sigma}
\right ) 
\Big ] {\cal O}(
 P_X) \;
|{\cal M}|^2(p_b,p_a, P_X)
\\
& - \frac{[\alpha_s] 
C_F Q}{s} {\cal N} 
\int {\rm d} x \; 
{\rm d} \Phi_m(p_a,xp_b,P_X) \;
\frac{1}{(1-x)_+} \\
& \times \left ( \kappa_m + 2 p_b^\mu \frac{ 
\partial }{\partial p_b^\mu }
+\left ( g^{\rho \sigma } 
+ \omega_{ba}^{\rho \sigma}
\right ) L_{\rho \sigma} \right ) {\cal O}(P_X) \; |{\cal M}|^2(xp_b,p_a, P_X).
\end{split}
\ee

Finally, we need to consider the third  term in Eq.~(\ref{eq6.41}). This term is also subleading in the collinear expansion 
which means that no boost is required. 
The contribution to the cross section reads 
\be
\begin{split}
 \frac{ {\rm d} \sigma^{cb,3}}{{\rm d} \tau}
& = \frac{[\alpha_s] C_F Q^{1-\ep} }{2 \tau^{\ep} } {\cal N} 
\int \limits_{0}^{1} \; {\rm d}x \; 
 \frac{ {\rm d} \Phi_m(p_a,xp_b,P_X) }{
(1-x)^{1+\ep}}  
  {\cal O}(P_X) \;
  \\
& \times \frac{4 p_a^\nu}{s}
\left ( 
{\rm Tr} 
\left [ N_b^+ \hat p_b R_{\rm fin, \nu} \hat p_a  \right ] +{\rm c.c.} 
\right ).
\end{split}
\ee
  We then replace 
$(1-x)^{-\ep - 1}$ with the plus distribution in the standard way 
and find 
\be
\begin{split}
\frac{ {\rm d} \sigma^{cb,3}}{{\rm d} \tau}
& = \frac{[\alpha_s] C_F Q^{1-\ep} }{2 \tau^{\ep} } {\cal N} 
\int \limits_{0}^{1} \; {\rm d}x \; 
{\rm d} \Phi_m(p_a,xp_b,P_X)
\left [ -\frac{1}{\ep} \delta(1-x) 
+ \frac{1}{( 1-x)_+}  
\right ]
\\
& \times \; {\cal O}(P_X) \;
\frac{4 p_a^\nu}{s}
\left ( 
{\rm Tr} 
\left [ N_b^+ \hat p_b R_{\rm fin, \nu} \hat p_a  \right ] +{\rm c.c.} 
\right ).
\end{split}
\ee
We compute the $x =1$ contribution following the discussion in the  previous section and find  
\be
\begin{split}
 \frac{ {\rm d} \sigma^{cb,3,\rm NLP}}{{\rm d} \tau}
& = \frac{[\alpha_s] C_F Q^{1-\ep} }{2 \tau^{\ep} \ep } {\cal N} 
\; 
{\rm d} \Phi_m(p_a, p_b,P_X) \; 
{\cal O}(P_X) \;
\\
& \times \frac{4 p_a^\nu}{s}
\left (  {\rm Tr} 
\left [ N^+ \hat p_b  N^{(1)}_\nu \hat p_a  \right ]_{x=1} +{\rm c.c.} 
\right )
\\
& 
+\frac{[\alpha_s] C_F Q }{2 } {\cal N} 
\int \limits_{0}^{1} \; \frac{{\rm d}x}{( 1-x)_+}   \; 
{\rm d} \Phi_{m}(p_a,xp_b,P_X) \; {\cal O}(P_X) \;
\\
& \times \frac{4 p_a^\nu}{s}
\left ( 
{\rm Tr} 
\left [ N_b^+ \hat p_b R_{\rm fin, \nu} \hat p_a  \right ] +{\rm c.c.} 
\right ).
\end{split}
\ee
The $1/\ep$ pole in the first term of the above equation is canceled by the first term in Eq.~\eqref{eq3.58}.

\subsection{The final result for the next-to-leading  power correction}

In this section, 
we combine all the different contributions,  and derive the 
final formula for the production of an arbitrary colorless final state $X$ in the  $q \bar q \to  X$ process at next-to-leading power in the zero-jettiness expansion at NLO QCD.  We need to account for the soft and  two collinear contributions, presented in Eqs~(\ref{eq5.29}, \ref{eq4.109}, \ref{eq5.111}), using  further simplifications of the last two equations (the collinear contributions) discussed  in Secs.~\ref{sec:cola} and~\ref{sec:colb}, and in Appendix~\ref{appB}.

Using the above results, it is straightforward to check  that, at next-to-leading power,  
all $1/\ep$ poles  cancel after  summing soft and collinear contributions. However,  the result contains a $\ln \tau$-enhanced term, 
which  appears as a consequence of the mismatch of the $\ep$-dependent exponents  of  $\tau$ in the soft and collinear contributions.  We note that the $1/\ep$ poles proportional to the tensor 
$\omega^{\mu \nu}_{a b}$ cancel when taking the sum of both collinear regions. 

We  write the next-to-leading power 
contribution in the expansion of the 
$q \bar q \to X$ cross section in the zero-jettiness  
as the sum of three terms  
\be
\begin{split}
\frac{{\rm d} \sigma^{{\rm NLP}} }{{\rm d} \tau} 
  = & \frac{[\alpha_s] C_F Q }{s} {\cal N} \Bigg\{
    2 \left [ \ln \left( \frac{Q \tau}{s} \right) +1 \right ]  {\rm C}^{{\rm NLP},s} + {\rm C}^{{\rm NLP},a} + {\rm C}^{{\rm NLP},b}
\Bigg\}.
\end{split}
\label{eq3.66}
\ee
 The finite remnant of the soft and soft-collinear contributions read 
\be
{\rm C}^{{\rm NLP},s} = \int 
{\rm d} \Phi_m(p_a,p_b,P_X) \left ( \kappa_m + \sum \limits_{i \in L_f} 
p_i^\mu \frac{ \partial}{\partial p_i^\mu} 
\right ) 
|{\cal M}(p_b,p_a,P_X)|^2 \; {\cal O}(P_X),
\label{eq3.71a}
\ee
where the sum extends over all particles 
in the process.\footnote{We remind the reader that the validity of Eq.~(\ref{eq3.71a})  requires  that  the observable ${\cal O}$ is independent of momenta $p_{a},p_b$. 
}  We note that   
for amplitudes with \emph{massless} particles \emph{only},
the following equation holds
\be
\left ( \kappa_m + \sum \limits_{i \in L_f} 
p_i^\mu \frac{ \partial}{\partial p_i^\mu} 
\right ) 
|{\cal M}(p_b,p_a,P_X)|^2   = 0.
\label{eq3.74a}
\ee
This result follows from the fact that the mass dimension of the amplitude squared with two initial-state
and $m$ final-state particles is $(-\kappa_m)$ and that the derivative
operator in the above equation probes 
the mass dimension of the amplitude squared in the massless 
case.

The expressions for the two
collinear remnants $C^{{\rm NLP},a(b)}$  are more complex. The $\vec k || \vec p_a$ contribution reads
\begin{align}
{\rm C}^{{\rm NLP},a} & = -2 \int 
{\rm d} \Phi_m \; |{\cal M}(p_b,p_a,P_X)|^2 \; {\cal O}(P_X) + \int {\rm d} x \; {\rm d} \Phi_m^{xa} \Bigg \{ \frac{\bar P_{qq}(x)}{ \; x} \; 
 \bigg [ 
W_a(x)
\nonumber
\\
&+\frac{s}{4}(1-x) g_\perp^{\rho \alpha} 
  \left( D^{xa,b}_\rho \; |{\cal M}|^2(p_b,xp_a, ...) - 2 {\rm Tr} \left [N_a \gamma_{\rho} 
N_a^+ \hat p_b \right ] \right) 
 {b_a}^{\mu \nu} _\alpha L_{\mu \nu}
 \nonumber 
 \\
& 
+  |{\cal M}|^2(p_b,xp_a,...) 
 \; l_a^{\mu \nu}(x) 
 L_{\mu \nu} -\frac{s \; (1-x)}{4} |{\cal M}|^2(p_b,xp_a,...) 
t_a^{\mu \mu_1, \nu \nu_1} 
 L_{\mu \mu_1} L_{\nu \nu_1} 
 \bigg ]
 \nonumber 
\\
& - \frac{1}{(1-x)_+} \left ( \kappa_m + 2 p_a^\mu \frac{ 
\partial }{\partial p_a^\mu }
+\left ( g^{\rho \sigma } 
+ \omega_{ab}^{\rho \sigma}
\right ) L_{\rho \sigma} \right ) |{\cal M}|^2(p_b,x p_a,...)
\nonumber 
\\
&- \frac{2 p_b^\nu}{( 1-x)_+} \;
\left ( 
{\rm Tr} 
\left [N_a \hat p_a R_{\rm fin, \nu}^+ \hat p_b  \right ] +{\rm c.c.} 
\right ) +  F_{{\rm fin},a}
\label{eq3.68}
\\
& + \frac{s}{4} (1-x) g_\perp^{\alpha \beta} \Bigg [
- 2 {\rm Tr} \left [  N_a \gamma_\beta N_a^+ \hat p_b \right ] 
\nonumber 
\\
& + {\rm Tr} 
\left [ N_a \gamma_\beta  \gamma_\rho \hat p_a \left( R_{{\rm fin}}^{\rho,+} + \frac{N_b^+ (\hat p_b - (1-x) \hat p_a) \gamma^\rho}{(1-x) s}\right) 
\hat p_b \right ] + {\rm c.c.}
\nonumber 
\\
& +
\frac{2 x}{1-x}
{\rm Tr} \left [ 
N_a \hat p_a \left ( R_{\rm fin, \beta}^+
- \frac{N_b^+ \hat p_a  \gamma_\beta }{s}
\right ) \hat p_b
\right ] + {\rm c.c.} \; \Bigg ] {b_a}^{\mu \nu} _\alpha L_{\mu \nu}
 \Bigg \} \; {\cal O}(P_X),
\nonumber 
\end{align}
where 
\be
{\rm d} \Phi_m = {\rm d} \Phi(p_a,p_b,P_X),
\;\;\;\;
{\rm d} \Phi_m^{xa}
={\rm d} \Phi
(xp_a,p_b,P_X),
\ee
$W_a(x)$ is defined in Eq.~\eqref{eq3.10}, 
$F_{{\rm fin},a}$ 
%the coefficients ${\rm %C}^{ka}_i$ 
can be found in 
Appendix~\ref{appB}, and the functions $N_a$, $N_b$ and $R^\mu_{\rm fin}$ appearing in the above expression should be evaluated with the following arguments
\be
\begin{split}
& N_a = N_a (p_b,x p_a,P_X), \\
& N_b = N_b (p_b-(1-x) p_a, p_a,P_X), \\
& R^\mu_{\rm fin} = R^\mu_{\rm fin}(p_b, p_a, (1-x) p_a, P_X).
\end{split}
\ee
We note that many terms  
in Eq.~(\ref{eq3.68}) involve derivatives of the observable 
${\cal O}$;
these terms are written 
for a generic case and may simplify significantly  
if a definite  observable 
is considered. We will see 
examples of this in what follows.

The second 
collinear contribution with 
$\vec k || \vec p_b$, that we referred to as  
${\rm C}^{{\rm NLP},b}$ above, can be obtained from Eq.~\eqref{eq3.68} by making the following replacements
\be
\begin{split}
p_a &\leftrightarrow p_b,\;\;\;\;
N_a  \leftrightarrow -N_b^+,
\end{split}
\ee
which also implies replacing the following quantities
\be
\begin{split}
W_a(x) \to & W_b(x),\;\; D^{xa,b} \to D^{xb,a},\;\;
b_{a} \to b_{b},\;\; l_{a} \to l_{b}, \;\; t_{a} \to t_{b},\\
&\omega_{ab}^{\rho \sigma} \to \omega_{ba}^{\rho \sigma}, \;\; F_{{\rm fin},a} \to F_{{\rm fin},b},\;\;
{\rm d} \Phi_m^{xa } \to {\rm d} \Phi_m^{xb}.
\end{split}
\ee
Furthermore, the Green's functions that would appear 
in $C^{\rm NLP,b}$  will have to be evaluated for the following arguments 
\be
\begin{split}
& N_a = N_a (p_b,p_a-(1-x)p_b,P_X), \\
& N_b = N_b (xp_b, p_a, P_X), \\
& R^\mu_{\rm fin} = R^\mu_{\rm fin}(p_b, p_a, (1-x) p_b, P_X).
\end{split}
\ee
\\

Several terms in the collinear contributions 
$C^{{\rm NLP},a(b)}$  can  be simplified further although 
we do not try to do this 
systematically. 
As an example, consider the term 
 \be
\omega_{ab}^{\mu \nu} 
L_{\mu \nu} |M|^2(p_b,p_a,P_X).
  \ee
in Eq.~(\ref{eq3.68}).
Since $\omega_{ab}^{\mu \nu}$ is an antisymmetric tensor, we can think of it as part of an infinitesimal Lorentz transformation 
\be
\left [ \Lambda_\delta 
\right ]^{\mu \nu} 
 =  g^{\mu \nu} 
 + \delta \omega_{ab}^{\mu \nu}
 + {\cal O}(\delta^2).
 \label{eq3.73}
\ee
Because the matrix element 
squared is invariant under Lorentz transformations, we can write 
\be
|{\cal M}|^2( p_b, p_a, 
\Lambda_\delta P_X) 
= |{\cal M}|^2( \Lambda_\delta^{-1}  p_b, \Lambda_\delta^{-1} p_a, P_X).
\label{eq3.74}
\ee
The inverse infinitesimal  transformation is obtained by 
replacing $\delta \to -\delta$ in Eq.~(\ref{eq3.73}). Finally, 
expanding Eq.~(\ref{eq3.74}) 
in $\delta$, we find 
\be
\label{eq3.75}
\omega_{ab}^{\mu \nu} 
L_{\mu \nu} 
|{\cal M}|^2( p_b, p_a, 
 P_X)
  = -\left ( p_a^\mu 
  \frac{\partial}{\partial p_a^\mu} - p_b^\mu \frac{\partial }{\partial p_b^\mu} 
  \right ) |{\cal M}|^2( p_b, p_a, 
 P_X), 
\ee
which might be  helpful for calculating this  quantity for 
complex  physical processes. 
\\

\section{How to compute Green's functions  that appear in the formula for power corrections}
\label{sec:currents}

The general formula for subleading zero-jettiness 
corrections, derived in the previous section, 
is complicated because it involves 
 Green's functions whose relation 
to amplitudes is obscure. Thus, for such a  formula  to be useful, one has to understand 
how the relevant Green's functions can be calculated.
It turns out that methods developed for computing high-multiplicity 
QCD  amplitudes more than thirty years ago \cite{Berends:1988zn} are suitable for this purpose.\footnote{The extension of these methods beyond QCD is discussed in  
Ref.~\cite{Gleisberg:2008fv}.}

Although we are certain that the 
discussion in this section can be made fully general, for the sake of definiteness, we consider the case when  the state $X$ consists of $N$ photons. The observable function ${\cal O}(P_X)$
is chosen in such a way that photons are hard and not collinear to the incoming quark and anti-quark; hence,  we treat them as hard particles
throughout the calculation.   

\begin{figure}
    \centering
    \includegraphics{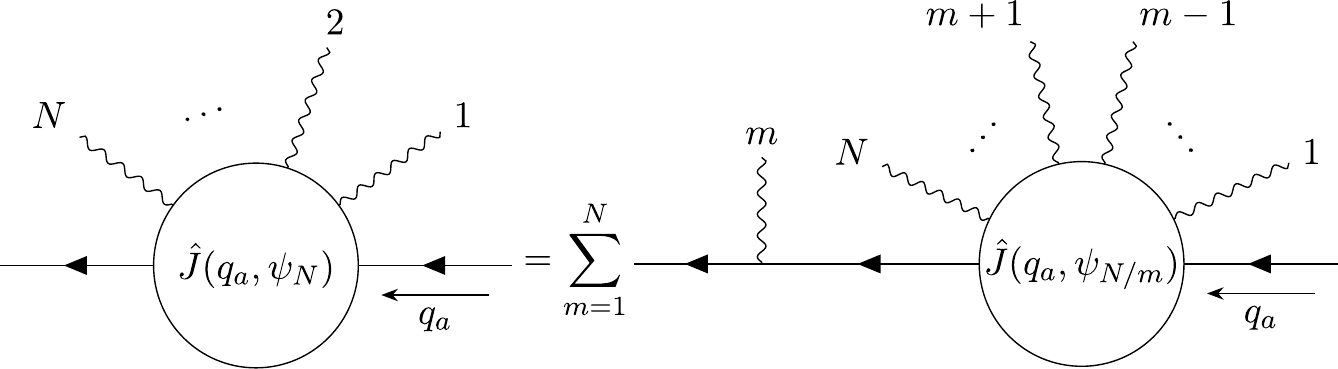}
    \caption{Pictorial representation of Eq.~\eqref{eq4.111}.}
    \label{fig1c}
\end{figure}

We need to understand how to compute the Green's functions $N_{a,b}$,  $N_{a,b}^{(1),\mu}$ etc., 
as well as 
$R_{\rm fin}^{\nu}$ and its expansion to first order in $k_\perp$. We will start with the discussion of the two simplest Green's functions $N_{a,b}$. 
To calculate them,  we introduce the quark 
current $\hat J$ (c.f. Fig.~\ref{fig1c}) which  depends on the momentum of the  incoming quark (that we  denote as $q_a$) 
and the momenta and polarization vectors  of $N$ photons. The momentum of the anti-quark is obtained  from the momentum conservation. The current reads 
$\hat J(q_a,\psi_N)$,
where the set $\psi_N$ is given by 
$\psi_N = \{ (p_1, \epsilon_1), (p_2, \epsilon_2), ..,
(p_N,\;\epsilon_N) \}$, and  $(p_i,\epsilon_i)$ denote  the 
momentum and the polarization vector of the photon $i$. 
The current is a four-by-four matrix that satisfies the following recurrence  relation 
\be
\hat J(q_a,\psi_N)
= \frac{i}{\hat q_a - \hat Q_N} \sum \limits_{m=1}^{N} 
\left (i e_q \hat \epsilon_m \right )
\hat J(q_a,\psi_{N/m}),
\label{eq4.111}
\ee
where $e_q$ is the quark electric charge, 
\be
Q_N = \sum \limits_{m=1}^{N}p_m,
\ee
$\psi_{N/m}$ denotes the original set $\psi_N$ from which the photon $m$ is removed, and the recursion starts 
by identifying $\hat J(q_a,\{\})$ with the identity matrix. A schematic representation of Eq.~\eqref{eq4.111} is shown in Fig.~\ref{fig1c}.

Eq.~(\ref{eq4.111}) is general; it allows us to  compute the current $\hat J$ and 
obtain the Green's functions $N_{a,b}$ from it. This is achieved 
by simply removing the propagator 
$i/(\hat q_a - \hat Q_N)$ from 
Eq.~(\ref{eq4.111}). We then 
find 
\be
N_{a,b} = 
\sum \limits_{m=1}^{N} 
\left ( i e_q \hat \epsilon_m \right )
\hat J \left (q_{a,b},\psi_{N/m} \right ),
\label{eq4.3}
\ee
where the two vectors $q_{a,b}$  
are different for the two cases. For example, many terms in the final 
formula involve functions 
$N_a$ and $N_b$ in the  collinear $ \vec k || \vec p_a$ limit.
In that case
\begin{equation}
q_a = x p_a, \;\;\; q_b = p_a.
\label{eq4.4a}
\end{equation}
For the $\vec k || \vec p_b$ case,  
\begin{equation}
q_a = p_a - (1-x) p_b,
\;\;\;\; q_b = p_a. 
\label{eq4.4b}
\end{equation}

In addition, we require the expansion of these Green's functions for certain deformations of the quark momentum  $q$; 
we will denote such deformations by   $\delta q$. 
The important feature 
of these deformations is that 
they do not affect momenta and polarizations of colorless particles of the final 
state $X$. Thanks to this feature, it becomes straightforward to compute the 
expansion of the functions $N_{a,b}$
with respect to such deformations. 
Writing 
\be
\hat J(q + \delta q,\psi_N)
= \hat J^{(0)}(q ,\psi_N)
+ 
\delta q_{\mu} \hat  J^{(1),\mu}(q, \psi_N) + 
\frac{\delta q_{\nu} \delta q_{\mu}}{2} \hat  J^{(2),\mu \nu}(q, \psi_N) +{\cal O}(\delta q^3),
\ee
we can derive equations that  currents $\hat J^{(0)}$, $\hat J^{(1),\mu}$ and $\hat J^{(2),\mu \nu}$ satisfy. In fact, the equation for  $J^{(0)}$ 
is identical to Eq.~(\ref{eq4.111}). 
The equations for $\hat J^{(1),\mu}$ and $\hat J^{(2),\mu \nu}$ read 
\be
\begin{split}
& \hat J^{(1),\mu}(q_a,\psi_N)
 = -\frac{1}{\hat q_a - \hat Q_N} \gamma^\mu
 J^{(0)}(q_a,\psi_N)
 + \frac{i}{\hat q_a - \hat Q_N} \sum \limits_{m=1}^{N} 
\left (i e_q \hat \epsilon_m \right )
\hat J^{(1),\mu}(q_a,\psi_{N/m}), \\
& \hat J^{(2),\mu \nu}(q_a,\psi_N)
 = -\frac{1}{\hat q_a - \hat Q_N} \left[ \gamma^\mu
 J^{(1),\nu}(q_a,\psi_N) + \gamma^\nu
 J^{(1),\mu}(q_a,\psi_N) \right] \\
& \qquad \qquad \qquad \quad + \frac{i}{\hat q_a - \hat Q_N} \sum \limits_{m=1}^{N} 
\left (i e_q \hat \epsilon_m \right )
\hat J^{(2),\mu \nu}(q_a,\psi_{N/m}).
\end{split}
\label{eq4.7}
\ee
To start the recursion, we use $\hat J^{(0)}(q_a) = \hat 1$, $\hat J^{(1),\mu}(q_a) = 0$ and $\hat J^{(2),\mu \nu}(q_a) = 0$.  
To compute relevant Green's functions, we 
write  their 
expansions as 
\be
N_{a,b}(q+ \delta q,P_X)
 = N_{a,b}(q,P_X) + \delta q_\mu N_{a,b}^{(1),\mu}(q,P_X)
 + \frac{\delta q_\mu \delta q_\nu}{2} N_{a,b}^{(2),\mu \nu}(q,P_X) + \dots ,
\ee
where $q$ is the quark momentum and 
ellipses stand for terms with higher 
powers of $\delta q$. Then, 
using Eqs~(\ref{eq4.3}, \ref{eq4.7}), we find 
\be
\begin{split}
 N_{a,b}^{(1),\mu} &= \sum \limits_{m=1}^{N} 
\left (i e_q \hat \epsilon_m \right )
\hat J^{(1),\mu}  \left (q_{a,b} ,\psi_{N/m} \right ), \\
 N_{a,b}^{(2),\mu \nu} &= \sum \limits_{m=1}^{N} 
\left (i e_q \hat \epsilon_m \right )
\hat J^{(2),\mu \nu}  \left (q_{a,b} ,\psi_{N/m} \right ).
\end{split}
\ee
The two vectors  $q_{a,b}$ are 
given in Eqs~(\ref{eq4.4a},
\ref{eq4.4b}) 
and we identify $Q_N = P_X$.
\\

The formula in Eq.~\eqref{eq3.68}  requires
us to compute derivatives of  the Born matrix element squared.  Although one 
can calculate these derivatives for simple processes, where matrix elements squared are known, it becomes difficult to do so in complicated cases with a large number of particles. To facilitate computing derivatives also in such cases, we
relate them to the Green's functions that we have already introduced. In particular, we find 
\be
\begin{split}
D^{xa,b}_\mu N_a (p_b,x p_a, P_X) &= N_a^{(1),\mu} (p_b,x p_a, P_X), \\
D^{xa,b}_\nu D^{xa,b}_\mu N_a (p_b,x p_a, P_X) &= N_a^{(2),\mu \nu} (p_b,x p_a, P_X).
\end{split}
\ee
Given these relations, we can replace
\be
\begin{split}
D^{xa,b}_\mu \; |{\cal M}|^2(xp_a,p_b,Q_X) \rightarrow& {\rm Tr} \left [N_{a,\mu}^{(1)} x \hat p_a N_a^+ \hat p_b \right ] + {\rm Tr} \left [N_a x \hat p_a N_{a,\mu}^{(1),+} \hat p_b \right ] \\
& + {\rm Tr} \left [N_a \gamma_\mu N_a^+ \hat p_b \right ] - {\rm Tr} \left [N_a x \hat p_a N_a^+ \gamma_\mu \right ].
\end{split}
\ee
For the 
term with the second-order derivative in the function  $W_a(x)$ we find 
\be
\begin{split}
g^{\mu \nu}_{\perp} \;& D^{xa,b}_\nu D^{xa,b}_\mu \; |{\cal M}|^2(p_b,xp_a,Q_X) \rightarrow g^{\mu \nu}_{\perp} \bigg \{ {\rm Tr} \left [N_{a,\mu\nu}^{(2)} x \hat p_a N_a^+ \hat p_b \right ] + {\rm c.c.} \\
& + 2 {\rm Tr} \left [N_{a,\mu}^{(1)} x \hat p_a N_{a,\nu}^{(1),+} \hat p_b \right ] + 2 {\rm Tr} \left [N_{a,\mu}^{(1)} \gamma_\nu N_a^{+} \hat p_b \right ] + {\rm c.c.} \\
& - 2 {\rm Tr} \left [N_{a,\mu}^{(1)} x \hat p_a N_a^{+} \gamma_\nu \right ] + {\rm c.c.} - 2 {\rm Tr} \left [N_a \gamma_\mu N_a^+ \gamma_\nu \right ] \bigg \}.
\end{split}
\ee
We note that the above replacements are only valid if they are done simultaneously in {\it all} relevant terms.
\\

\begin{figure}
    \centering
    \includegraphics{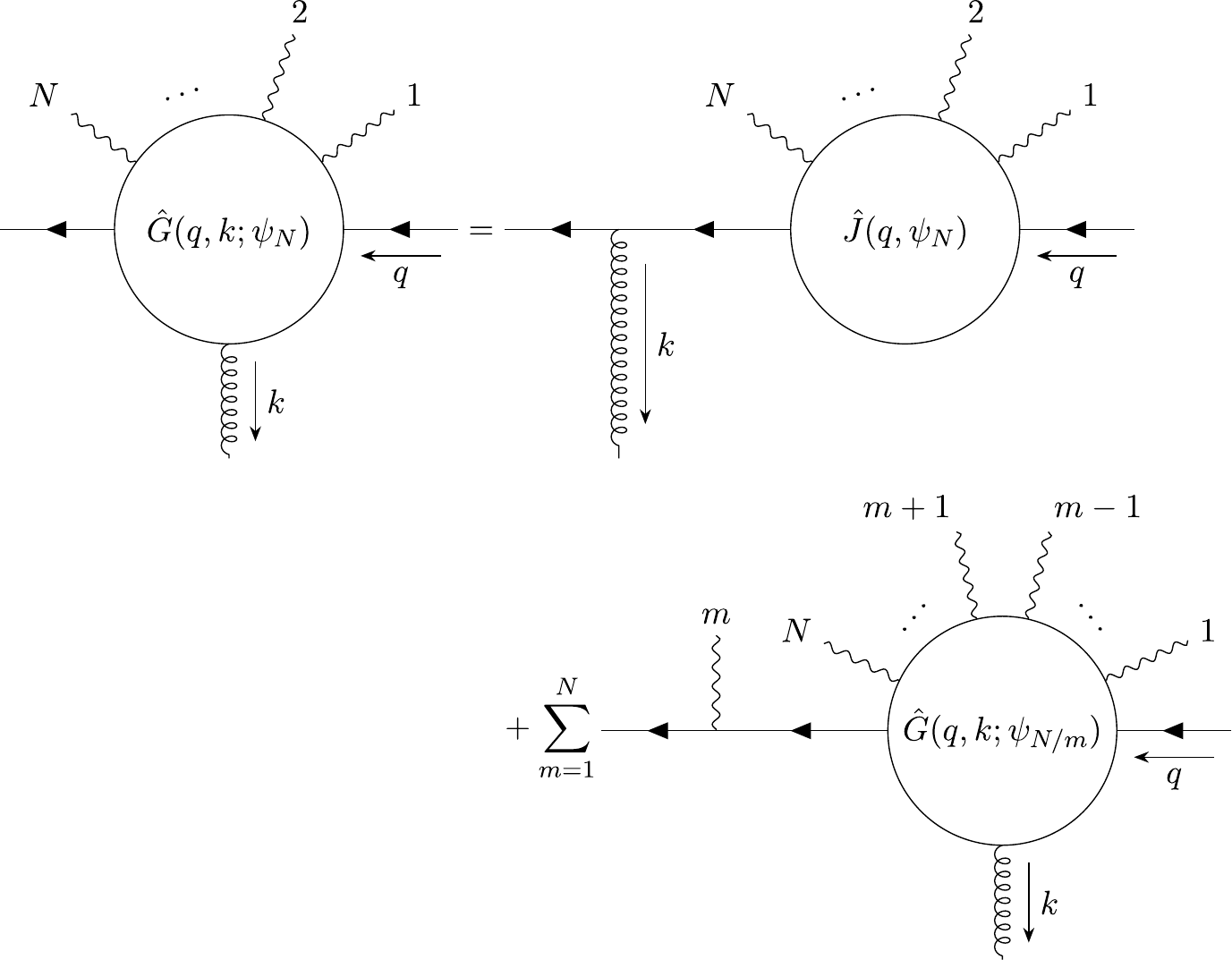}
    \caption{Pictorial representation of Eq.~\eqref{eq4.9}.}
    \label{fig4}
\end{figure}

The last  ingredients required for the 
final formula for subleading 
power corrections involve the Green's function 
$R_{\rm fin}^{\nu}$ and its expansion to first order in
the momentum $k_\perp$.
To compute these quantities, 
we introduce the current $\hat G^\nu$ 
that depends on the quark momentum $q$, 
the gluon momentum $k$ and the photon momenta and 
polarization vectors. 
This current  satisfies the following equation 
\be
\label{eq4.9}
\hat G^{\nu}(q,k;\psi_N) 
= \frac{i}{\hat q - \hat k - \hat Q_N}
\left [ 
i \gamma^\nu  \hat J(q,\psi_N)
+\sum \limits_{m=1}^{N} 
\left (i e_q \hat \epsilon_m \right )
\hat G^{\nu}  \left (q, k;\psi_{N/m} 
\right )
\right ],
\ee
where the first term on the right-hand side describes the gluon emission off the anti-quark leg,  and the second term refers to  a situation where the emission of one 
of the $N$ photons happens last, 
see  Fig.~\ref{fig4}.
The boundary condition for the 
recursion is 
\be
\hat G^\nu(q,k,\{\}) 
 = 0,
\ee
because  gluon  emissions off  the external quark line should not  be considered. 
For the same reason, 
the expression for $R^{\nu}_{\rm fin}$ reads
\be
R^{\nu}_{\rm fin}(q,k,\psi_N)
 = \sum \limits_{m=1}^{N} 
\left ( i e_q \hat \epsilon_m \right )
\hat G^{\nu}  \left (q , k;\psi_{N/m} 
\right ).
\ee
For the case $\vec k || \vec p_a$, we require 
$R^{\nu}_{\rm fin}$ in the strict collinear limit, in which 
case $q = p_a$ and $k = (1-x) p_a$. 
For the case $\vec k || \vec p_b$, 
$R^{\nu}_{\rm fin}$  should be evaluated for 
$q = p_a$ and $k = (1-x) p_b$.

We also require the expansion 
of $R^\nu_{\rm fin}$ to first 
order in $k_\perp$. 
Since the dependencies 
on $k_\perp$ arise after one of the two collinear boosts 
is applied to 
momenta 
$p_a$, $k$ and $p_b$, 
we will define 
the expansion of the current 
$G^\nu$ for   \emph{particular momentum deformations only}. We begin  with the $\vec k || \vec p_a$ 
case. Applying the 
$\Lambda_a$-boost, to $p_a$ 
and $k$ and expanding in $k_\perp$, we write 
\be
\begin{split}
\hat G^{\nu}\left (p_a + \frac{k_\perp}{2x},
(1-x) p_a + \frac{k_\perp(1+x)}{2x}, 
 \psi_N \right )
& = 
\hat G^{(0)\nu}\left (p_a ,
(1-x) p_a , 
\psi_N \right )
\\
 + &  
\hat G^{(1)\nu, \mu}\left (p_a, 
(1-x) p_a , 
\psi_N \right ) k_{\perp,\mu} + \cdots,
\end{split}
\ee
where $\hat G^{(0)\nu}$ 
is computed with the help of Eq.~(\ref{eq4.9}).
The  recurrence   relation for $\hat G^{(1)\nu, \mu}$ reads 
\be
\begin{split} 
& \hat G^{(1),\nu  \mu}(p_a,(1-x)p_a,\psi_N)
 = 
 \frac{1}{2} \frac{1}{x \hat p_a - \hat Q_N} \gamma^\mu \hat G^{(0),\nu}(p_a,(1-x)p_a,\psi_N)
 \\
 & - \frac{1}{x \hat p_a - \hat Q_N} 
 \left [ 
 \frac{1}{2x} \gamma^\nu \hat J^{(1),\mu}(p_a,\psi_N)
 +
 \sum \limits_{m=1}^{N}
(e_q \hat \epsilon_m) \hat G^{(1),\nu  \mu}
(p_a, (1-x)p_a, \psi_{N/m} )
\right ],
 \end{split}
 \ee
and the recursion 
starts with $G^{(1),\nu \mu} = 0$. Defining the expansion of 
$R^\nu_{\rm fin}$ as
\be
\begin{split} 
& R_{\rm fin}^{\nu}
\left (p_b + \frac{k_\perp}{2},
p_a + \frac{k_\perp}{2x},
(1-x) p_a + \frac{k_\perp(1+x)}{2x}, P_X \right )
\\
& \approx R_{\rm fin}^{(0),\nu}
(p_a,p_b,(1-x)p_a,P_X)
+ k_{\perp,\mu} R_{\rm fin}^{(1),\nu \mu}(p_a,p_b,(1-x)p_a,P_X) + {\cal O}(k_\perp^2),
\end{split} 
\ee
we find 
\be
R^{(1),\nu \mu}_{\rm fin}(p_b,p_a, (1-x) p_a,P_X)
 = \sum \limits_{m=1}^{N} 
\left ( i e_q \hat \epsilon_m \right )
\hat G^{(1),\nu \mu}  \left (p_a , (1-x) p_a,\psi_{N/m} 
\right ).
\ee

For  the $\vec k || \vec p_b$ 
case, we  apply the $\Lambda_b$ boost and write  
\be
\begin{split}
\hat G^{\nu}\left (p_a + \frac{k_\perp}{2},
(1-x) p_b + \frac{k_\perp(1+x)}{2x}, 
 \psi_N \right )
& = 
\hat G^{(0)\nu}\left (p_a ,
(1-x) p_b , 
\psi_N \right )
\\
 + &  
\hat G^{(1),\nu \mu}\left (p_a, 
(1-x) p_b , 
\psi_N \right ) k_{\perp,\mu} + \cdots.
\end{split}
\ee
We then derive an equation for $\hat G^{(1),\nu \mu}$. It reads 
\be
\begin{split} 
& \hat G^{(1),\nu \mu}(p_a,(1-x)p_b,\psi_N)
 = 
 \frac{1}{2x} \frac{1}{\hat P_{abN} } \gamma^\mu \hat G^{(0),\nu}(p_a,(1-x)p_b,\psi_N)
 \\
 & - \frac{1}{\hat  P_{abN}} 
 \left [ 
 \frac{1}{2} \gamma^\nu \hat J^{(1),\mu}(p_a,\psi_N)
 +
 \sum \limits_{m=1}^{N}
(e_q \hat \epsilon_m) \hat G^{(1), \nu \mu}
(p_a, (1-x)p_b, \psi_{N/m} )
\right ],
 \end{split}
 \ee
 where 
$P_{abN} =p_a -p_b(1-x)- Q_N
$.  Defining the expansion of 
$R^\nu_{\rm fin}$ as
\be
\begin{split} 
& R_{\rm fin}^{\nu}
\left (
p_b + \frac{k_\perp}{2x},
p_a + \frac{k_\perp}{2},
(1-x) p_b + \frac{k_\perp(1+x)}{2x}, P_X \right )
\\
& \approx R_{\rm fin}^{(0),\nu}
(p_a,p_b,(1-x)p_b,P_X)
+ k_{\perp,\mu} R_{\rm fin}^{(1),\nu \mu}(p_a,p_b,(1-x)p_b,P_X) + {\cal O}(k_\perp^2),
\end{split} 
\ee
we find for the $\vec k || \vec p_b$ case 
\be
R^{(1),\nu \mu}_{\rm fin}(p_a,p_b, (1-x) p_b,P_X)
 = \sum \limits_{m=1}^{N} 
\left ( i e_q \hat \epsilon_m \right )
\hat G^{(1),\nu \mu}  \left (p_a , (1-x) p_b,\psi_{N/m} 
\right ).
\ee

\section{Examples of application}
\label{sect:5}

In this section, we apply the master formula in Eq.~(\ref{eq3.68}) 
 to compute the next-to-leading power correction in the zero-jettiness variable  
to various processes. 
We start  with the Drell-Yan process $q \bar q \to l^+ l^-$ and  the 
two-photon production $q \bar q \to \gamma \gamma$.
These processes are sufficiently simple to allow 
an  analytic computation of the subleading contribution in the 
zero-jettiness expansion.  Then, we turn to the process 
$q \bar q \to 4 \gamma$. In this case, the matrix element and the required  Green's functions 
are  complicated, so that  we employ the generalized 
currents introduced in the 
previous section to perform the calculation.

\subsection{The Drell-Yan process}

 We consider the photon-mediated  production of a pair of leptons in the 
 annihilation of a quark 
 and an anti-quark
\be
q(p_a) + \bar q(p_b) \to \gamma^* \to l(p_1) + \bar l(p_2).
\label{eq5.1}
\ee
The calculation of  the next-to-leading power corrections involves several quantities that we need to specify. 
They include 
the leading order matrix element and the phase space  appearing  in Eq.~(\ref{eq3.68}). Since the  $1/\ep$ singularities have already been canceled, 
we can compute  the relevant quantities in four space-time dimensions. 

The analysis of the collinear 
$\vec k || \vec p_a$ contribution requires the (boosted) phase space ${\rm d} \Phi_2^{xa}$ that corresponds 
to the process in Eq.~(\ref{eq5.1}) where the quark 
momentum $p_a$ is replaced with $xp_a$. The phase space 
reads 
\be
{\rm d} \Phi_2^{xa} = \frac{1}{8 \pi } {\rm d} \beta \frac{{\rm d} \varphi}{(2 \pi)}.
\label{eq5.2}
\ee
In Eq.~(\ref{eq5.2})  $\varphi$ is the azimuthal angle of the
outgoing lepton in the reference frame where the
$z$-axis is aligned with the collision axis,  and 
the  parameter $\beta \in [0,1]$ is related to the polar  angle of the lepton.    With  this parametrization, the momenta
$p_{1,2}$ read 
\be
\begin{split} 
& p_1 = x (1-\beta) p_a + \beta p_b 
+ \sqrt{x s \beta (1-\beta)} n_\perp,
\\
& p_2 = x  \beta  p_a + (1-\beta) p_b 
- \sqrt{x s \beta (1-\beta)} n_\perp,
\end{split} 
\label{eq5.3}
\ee
where $s = 2 p_a \cdot  p_b$, $p_{a,b} \cdot n_\perp = 0$ 
and $n_\perp^2 = -1$.  We note that the phase space parametrization in Eq.~(\ref{eq5.2})  does not depend on the parameter $x$, so that if we set
$x =1$  also in Eq.~(\ref{eq5.3}), we obtain both the $x = 1$ 
Born phase space and the momenta parametrization. 

For the collinear region $\vec k || \vec p_b$, we require the phase space 
${\rm d} \Phi_2^{xb}$. We can use Eq.~(\ref{eq5.2}) to describe it provided that we use the following parametrization of the   momenta $p_{1,2}$
\be
\begin{split} 
& p_1 =  (1-\beta) p_a
+x \beta  p_b
+ \sqrt{x s \beta (1-\beta)} n_\perp,
\\
& p_2 =  \beta  p_a +x (1- \beta)  p_b 
- \sqrt{x s \beta (1-\beta)} n_\perp.
\end{split} 
\label{eq5.3a}
\ee
This parametrization ensures that in the soft $x=1$ limit
Eqs~(\ref{eq5.3}, \ref{eq5.3a}) coincide. 

The appropriately normalized Born matrix element squared summed over polarizations and colors reads
\be
\sum \limits_{\rm pol,col} \frac{|{\cal M}|^2(p_b,p_a;p_1,p_2)}{4 N_c 
(Q_q e^2)^2 }  
= 2  \frac{ 
s_{a1}^2 + s_{b1}^2}{s^2} 
 = 2 ( 1 - 2 \beta + 2 \beta^2),
\ee
where $s_{a1} = 2 p_a \cdot p_1$, $s_{b1} = 2 p_b \cdot p_1$, 
  $N_c$ is the number of colors, $e$ is the positron electric charge and $Q_q$ is the electric charge of the quark in units of $e$.  
The leading order cross section evaluates to 
\be
{\rm d} 
\sigma_0 = 16 \pi \; \bar \sigma_0 \;
{\rm d} \Phi^{ab}_2 \; (1 - 2 \beta + 2 \beta^2), 
\label{eq5.6}
\ee
where
\be
\bar \sigma_0 = \frac{\pi Q_q^2 \alpha_{\rm QED}^2}{s N_c},
\ee
and ${\rm d} \Phi^{ab}_2$
is given by Eq.~(\ref{eq5.2}).

To compute the next-to-leading power corrections 
from the master formula 
in Eq.~(\ref{eq3.68}), 
we have to calculate a significant number of terms.  We will perform the computation setting 
$N_c \to 1, Q_q \to 1$ 
and $e \to 1$ and restore the relevant factors at the end.  Then, we have to use  
\be
|{\cal M}|^2(p_b,p_a;p_1,p_2) \to 8 
\frac{s_{a1}^2 + s_{b1}^2}{s^2} = 8 ( 1- 2 \beta + 2 \beta^2),
\ee
With this normalization, the Green's functions $N_{a,b}$ 
read
\be
N_a = N_b =\frac{i}{s_{12}}  \; \gamma_\mu \; \left ( \bar u(p_1) \gamma^\mu v(p_2) \right ).
\label{eq5.8}
\ee

It follows from Eq.~(\ref{eq5.8}) that neither $N_a$ nor $N_b$ depends on 
$p_a$ and $p_b$, which means that it is not affected by the boost and 
that it does not  depend on the gluon momentum $k$. Hence, we find  
\be
N_{a,b}^{(1),\mu} = 0.
\ee
Furthermore, in case of the Drell-Yan process,    no gluon emissions from the 
internal lines can occur, which  implies that 
\be
R_{\rm fin}^\mu = 0.
\ee
Another  simplification is that for a $2 \to 2$ process  $\kappa_2 = 0$, which follows from the fact that Born amplitudes for such processes have  vanishing mass dimension. 
\\

With these preliminary remarks out of the way, 
 we proceed with the calculation of the  subleading power corrections, using the general formula  in Eq.~\eqref{eq3.66}. We will 
 start with the discussion of 
 the collinear $\vec k || \vec p_a$ contribution which 
 means that we employ the parametrization of momenta 
 $p_{1,2}$ given in Eq.~(\ref{eq5.3}) to write the corresponding expressions. 
 Several  ingredients 
 need to be discussed.
\begin{itemize} 

\item \emph{Traces that involve $p_a,p_b$ and some combinations of $N_a$ and $N_b$}. These are straightforward to compute given the expressions for these Green's functions.  We find e.g.
\be
\begin{split}
F_{\rm fin,a}
  = & - 8 \left (1-2 \beta + 2 \beta^2 \right) + 32 \; \frac{\beta (1-\beta)}{x^2} + 16 \; \frac{\beta (1-\beta)}{x} + 16 \; \frac{\beta (1-\beta)}{x^2} \\
 & -16 \; \frac{(1-x) \beta (1-\beta)}{ x}
 - 8 \; \frac{1+2\beta-2\beta^2}{x^2}
 -8 \;\frac{1-2\beta+2\beta^2}{ x}.
\end{split}
\ee

\item \emph{Terms that involve derivatives of the various quantities w.r.t. momenta of the incoming partons}. Such derivatives
appear in several terms in Eq.~(\ref{eq3.66}) and also in the definition of the function $W_a(x)$, c.f.  Eq.~(\ref{eq3.10}). 
We start by discussing derivatives of the matrix element squared. 
In principle, these derivatives may 
not be uniquely defined 
given the need to account 
for the momentum conservation, etc. However, in our formulas, the potentially ambiguous derivatives, always
involve contractions that
make them unique. For example,
we find 
\be
g_\perp^{\mu \nu} D_\nu^{ab} |M^2(p_b,p_a,p_1,p_2)|^2  
= \frac{16 p_{1,\perp}^\mu}{s_{12}^2}
\left (  
s_{a1} - s_{b1}
\right ),
\ee
where $D_\nu^{ab} = 
\partial/\partial p_a^\nu - \partial/\partial p_b^\nu$.
Furthermore, using Eq.~(\ref{eq3.75})
and Eq.~(\ref{eq3.74a}), it is easy to see that 
\be
\left [ 
\kappa_2 + 2 p_a^\mu \frac{\partial}{\partial p_{a}^{\mu}}
+ ( g^{\mu \nu} + \omega_{ab}^{\mu \nu}  )  L_{\mu \nu} 
\right ]|M^2(p_b,p_a,p_1,p_2)|
= 0.
\ee

\item Terms that involve derivatives and traces can be computed in a straightforward way using the above results. 
For example, we find a compact expression for the function $W_a$, 
\be
W_a (x) = 4 \frac{1-x}{x} (1-2\beta+2\beta^2).
\ee

We also find that in the Drell-Yan case
\be
g_\perp^{\rho \alpha} 
  \left( D^{xa,b}_\rho \; |{\cal M}|^2(p_b,xp_a, ...) - 2 {\rm Tr} \left [N_a \gamma_{\rho} 
N_a^+ \hat p_b \right ] \right) 
 {b_a}^{\mu \nu} _\alpha L_{\mu \nu}  = 0.
\ee

Another contribution with derivative operators and traces evaluates to 
\be
\begin{split}
s & (1-x) \frac{g_\perp^{\alpha \beta}}{4} \Bigg \{
- 2 {\rm Tr} \left [  N_a \gamma_\beta N_a^+ \hat p_b \right ] \\
& + {\rm Tr} 
\left [ N_a \gamma_\beta  \gamma_\rho \hat p_a \left( R_{{\rm fin}}^{\rho,+} + \frac{N_b^+ (\hat p_b - (1-x) \hat p_a) \gamma^\rho}{(1-x) s}\right) 
\hat p_b \right ] + {\rm c.c.}
\\
& +
\frac{2 x}{1-x}
{\rm Tr} \left [ 
N_a \hat p_a \left ( R_{\beta}^{{\rm fin},+} 
- \frac{N_b^+ \hat p_a  \gamma_\beta }{s}
\right ) \hat p_b
\right ] + {\rm c.c.} \Bigg \} {b_a}^{\mu \nu} _\alpha L_{\mu \nu}  \\
& = - 4 (1-x^2)  \frac{(1-2\beta)}{x^2} \bigg\{ \left[ p_1^\mu - (1-2 \beta) (1-\beta) \; x p_a^\mu + (1-2 \beta) \beta \; p_b^\mu \right ] \partial_{1 \mu} \\
& - \left[ p_2^\mu + (1-2 \beta) \beta \; x p_a^\mu - (1-2 \beta) (1-\beta) \; p_b^\mu \right ] \partial_{2 \mu} \bigg\},
\end{split}
\ee
where $\partial_{1(2),\mu}$
are derivatives 
$\partial/\partial p_{1,2}^\mu$. 

\end{itemize}

 Expressions for
 the case $\vec k || \vec p_b$
 can be obtained from the
 formulas for $\vec k || \vec p_a$  by  replacing 
\be
\beta \rightarrow 1-\beta, \qquad p_a \leftrightarrow p_b.
\label{eq5.17}
\ee

The total subleading contribution is obtained from  Eq.~\eqref{eq3.66}, using the partial results described above. We find 
\be
\begin{split}
\frac{{\rm d} \sigma^{\rm DY,{\rm NLP}} }{{\rm d} \tau} 
 & = \frac{4 [\alpha_s] C_F Q }{s} {\rm d} \sigma_0 \left[  \left( -1 + \frac{1}{2} {\cal D} \right) + \frac{1}{2}\log \left( \frac{\tau Q}{s}\right) {\cal D} \right] {\cal O}(p_1,p_2)
\\
  & + \frac{2 [\alpha_s] C_F Q }{s} \int \limits_{0}^{1} 
  {\rm d} \sigma_0 \; 
  {\rm d} x \; \bigg[ - \frac{1}{2 (1-x)_+} 
  \left ( {\cal D}\Big |_{ca}
  + {\cal D}\Big |_{cb}
  \right ) 
  \\
  & + \left( \frac{ \left [ \bar \beta  \; p_a^\mu - \beta \; p_b^\mu \right ] \partial_{1 \mu} + \left [  \beta \; p_a^\mu - \bar \beta  \; p_b^\mu \right ]
    \partial_{2 \mu} }{2 \; (1-x)_+} \right) \Bigg|_{ca} \\
  & + \left( \frac{\bar \beta  \; p_a^\mu \partial_{1\mu} + \beta \; p_a^\mu \partial_{2\mu}}{2} + \frac{{\cal P} (\beta, x, p_a, p_b; p_1,p_2,\partial_1,\partial_2)}{8(1 - 2 \beta + 2 \beta^2 )} \right) \Bigg |_{ca} \\
  & - \left( \frac{ \left [ \bar \beta  \; p_a^\mu - \beta \; p_b^\mu \right ] \partial_{1 \mu} + \left [  \beta \; p_a^\mu - \bar \beta  \; p_b^\mu \right ]
    \partial_{2 \mu} }{2 \; (1-x)_+} \right) \Bigg|_{cb} \\
  & + \left( \frac{\beta \; p_b^\mu \partial_{1\mu} + \bar \beta  \; p_b^\mu \partial_{2\mu}}{2} + \frac{{\cal P} (\bar \beta, x, p_b, p_a;p_1,p_2,\partial_1,\partial_2)}{8(1 - 2 \beta + 2 \beta^2 )} \right) \Bigg |_{cb}  \; \bigg] \; {\cal O}(p_1,p_2),
\end{split}
\label{eq5.20}
\ee
where $\bar \beta = 1- \beta$ and ${\rm d} \sigma_0$ is  given 
in Eq.~(\ref{eq5.6}).
We note that bars with a subscript $ca$ or  $cb$
indicate that after applying derivatives to the observable ${\cal O}(p_1,p_2)$, the ensuing scalar products must be evaluated in a particular collinear kinematics given in Eqs~(\ref{eq5.3}, \ref{eq5.3a}) 
 for the $ca$ and $cb$ cases, respectively. 
 The differential operator ${\cal D}$ reads
\be
{\cal D} = p_1^\mu \frac {\partial }{\partial p_1^\mu}  + p_2^\mu
\frac {\partial }{\partial p_2^\mu}.
\label{5.21D}
\ee
The other differential operator ${\cal P}(\beta,x,p_a,p_b;p_1,p_2,\partial_1,\partial_2)$ appearing in Eq.~\eqref{eq5.20} also
acts on the observable ${\cal O}(p_1,p_2)$. 
It  is given by the following expression 
\be
\begin{split} 
 &  {\cal P}(\beta,x,p_a,p_b;p_1,p_2,\partial_1,\partial_2) = 
  -2 \left(  \frac{1+x^2}{x^2}  \left(1- 6 \beta + 6 \beta^2\right) + \frac{2 f_0(\beta) }{x}
  \right) 
  \\
&  
+  g_2(x, \beta) \; p_1^\mu \partial_{1\mu}  + g_2(x,1-\beta)\; p_2^\mu \partial_{2\mu} 
+g_1(x,\beta)  \; p_a^\mu \partial_{1\mu}   + \frac{g_1\left (x_1,\beta \right ) }{x}  \;   p_b^\mu \partial_{2\mu} 
\\
& + g_1(x,1-\beta) p_a^\mu \partial_{2\mu}  + \frac{g_1\left (x_1,1-\beta \right ) }{x}  p_b^\mu \partial_{1\mu} \\
& + \frac{(1+x^2) f_0(\beta)}{2 x^2} \bigg\{ \Big[ -2 (1-\beta)^2 x^2 \ p_a^\mu p_a^\nu - 2 \beta^2  \ p_b^\mu p_b^\nu \\ 
& - x \ (p_a p_b) \ g^{\mu \nu} + 2 \left( x \ p_a^\mu p_1^\nu + p_b^\mu p_1^\nu \right)  + 4 x \beta (1-\beta) \ p_a^\mu p_b^\nu \Big]\partial_{1\nu} \partial_{1\mu} \\
& + \Big[ \left(f_0  (\beta) - 2 \right) \left( x^2 \ p_a^\mu p_a^\nu + p_b^\mu p_b^{\nu}  \right) + \left( x \ p_1^\mu p_a^\nu + p_1^\mu p_b^\nu \right) + \left( x \ p_a^\mu p_2^\nu + p_b^\mu p_2^\nu \right)  \\ 
&  - x \ ( p_a p_b) \ g^{\mu \nu} - x \left(1 - 2\beta^2 \right)  p_a^\mu p_b^\nu + x \left( 1- 4 \beta + 2 \beta^2 \right)  p_b^\mu p_a^\nu \Big] \partial_{2\nu} \partial_{1\mu} \\
& + \left( p_1 \leftrightarrow p_2, \ \beta \leftrightarrow 1-\beta \right) \bigg\}, 
\end{split} 
\ee
where $x_1 = 1/x$,
\be
\begin{split} 
 &  g_1(x,\beta) = -4 (1-\beta) f_0(\beta) +f_1(1-\beta)  x + \frac{f_2(1-\beta)}{x},
\\
& g_2(x,\beta) = f_3(1-\beta)  +  \frac{f_3(\beta) }{x^2},
\end{split} 
\ee
and 
\be
\begin{split} 
& f_0(\beta) = 1 - 2\beta + 2 \beta^2, \;\;\;
f_1(\beta) = -14 \beta^3+16 \beta^2-7 \beta+1,
\\
& f_2(\beta) = 1 +\beta - 8 \beta^2 + 10 \beta^3,
\;\;\;
f_3(\beta) = f_0(-\beta) - 2.
\end{split} 
\ee
We note that the complexity of the above formula is related to the fact that the observable ${\cal O}(p_1,p_2)$
is considered to be generic. 
If e.g. all derivatives applied to an observable, that appear in the final formula, are dropped, the expression for 
next-to-leading power corrections for the Drell-Yan process simplifies dramatically. 

\begin{figure}
    \centering
    \includegraphics[scale=0.6]{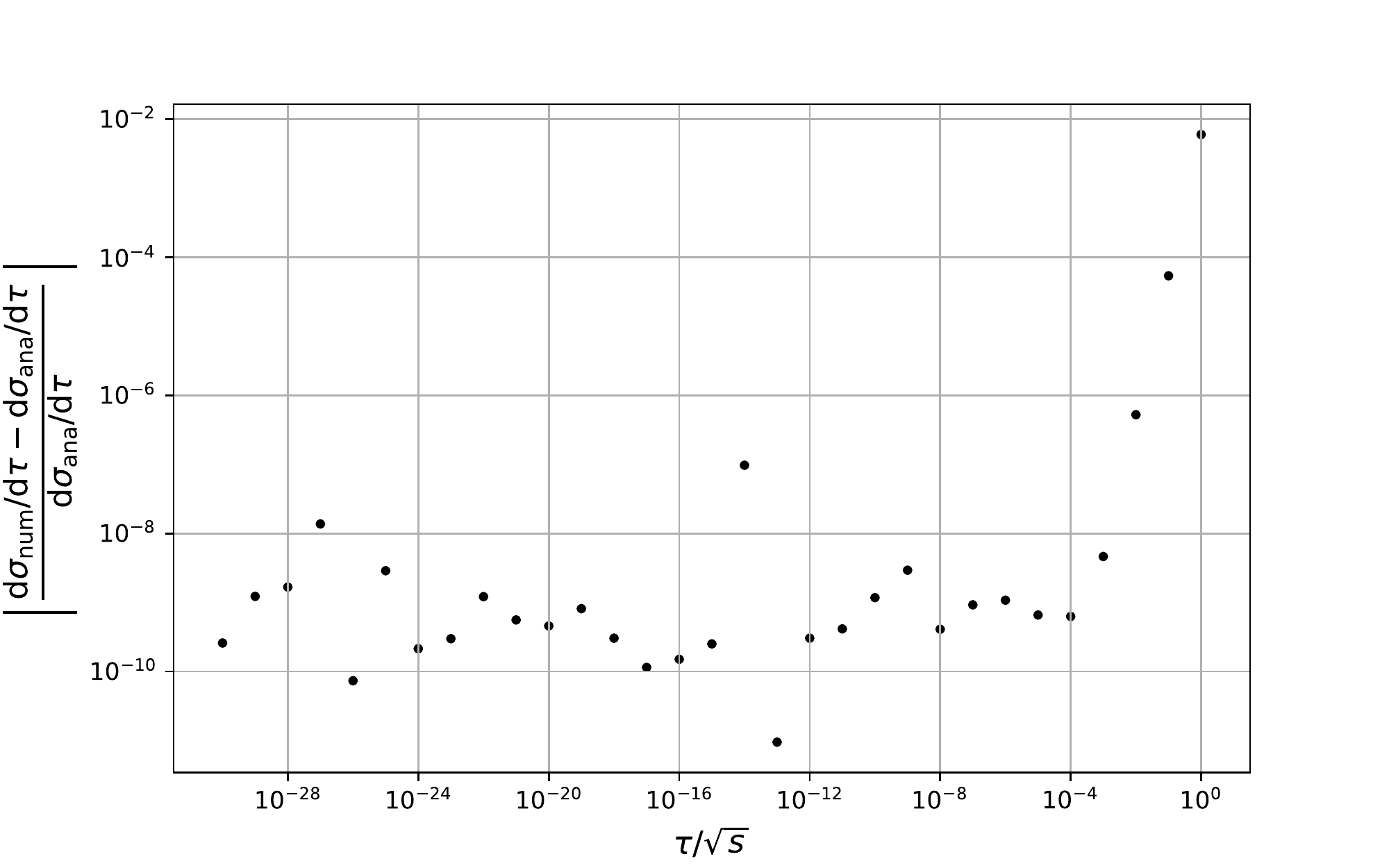}
    \caption{Relative difference between the small-jettiness expansions of the Drell-Yan cross section with an observable and its numerical integration.}
    \label{fig3}
\end{figure}

To ensure the correctness of the master formula, we perform the following checks.  First, we repeat the calculation of the zero-jettiness power corrections to the Drell-Yan process employing explicit 
expression for the matrix element squared  and using explicit parametrization of the phase space.  We find  a complete analytic agreement between the result of such  ``explicit'' calculation with the result that one obtains  when using the master formula to derive the power-suppressed term. 

We can also use  the master  formula to rederive results for the vector boson production discussed in Sec. \ref{sect2}.  To do so, we choose the observable that constrains  the invariant mass of two leptons
\be
{\cal O}( p_1,  p_2) = \delta(2 p_1 \cdot p_2 - m_V^2)  .
\ee
Since  in Sec.~\ref{sect2} we work with hadronic, rather than partonic cross sections, we need to do the same here.    Hence, we employ  Eq.~\eqref{eq2.7c} to calculate   power corrections to the hadronic cross section using the  master formula  for  ${{\rm d} \sigma (s) } / {{\rm d} \tau}$ adapted to the Drell-Yan case, and integrate over the leptonic final states to compare with Eq.~\eqref{eq2.34}. The derivatives of the observable will be related (via integration by parts) to derivatives of the luminosity function and other terms in the formula. We  find that ${\cal O}(\tau \log \tau)$ terms agree immediately, while other  
${\cal O}(\tau)$ terms agree after integrating  by parts over $z$, to transform terms with  ${\cal L}_1 \left ( m_V^2/z\right ) $ into ones with ${\cal L} \left ( m_V^2/z\right )$. The relevant equation reads  
\be
\int \limits_{0}^{1} {\rm d} z \;  {\cal L}_{1} \left ( m_V^2/z\right ) f(z) = - \frac{f(1)}{m_V^2} + \frac{1}{{\cal L} \left ( m_V^2 \right )}\int \limits_{0}^{1} {\rm d} z \; {\cal L} \left ( m_V^2/z\right ) \left(\frac{z^2}{m_V^2}f(z) \right)^{'} .
\ee

We  may use  Eq.~\eqref{eq5.20} to recover results for power corrections, obtained for the Drell-Yan process $q \bar q \to V$ in Ref.~\cite{Ebert:2018lzn}.  This calculation is more complex because in addition to the invariant mass, the rapidity of the vector boson (or, equivalently, the dilepton rapidity) is constrained.   Furthermore, the definition of the zero-jettiness variable used in  Ref.~\cite{Ebert:2018lzn} differs from what we employ here since  Born-projected momenta for the incoming partons are used there to compute zero-jettiness.  Nevertheless, choosing the observable 
%Nevertheless, by choosing the observable that adapts to their dilepton mass ($M$) and rapidity ($Y$) constraints
\be
\begin{split}
{\cal O}( p_1,  p_2) &= \delta \left( 2 p_1 \cdot p_2 - M^2 \right) \ \delta \left( \frac{1}{2} \ln \frac{P_b \cdot (p_1 + p_2)}{P_a \cdot (p_1 + p_2)} - Y \right),
%& =  \delta \left( 2 \; \xi_a \xi_b \; P_a \cdot P_b - M^2 \right) \ \delta \left( \frac{1}{2} \ln \frac{\xi_a}{\xi_b} - Y \right),
\end{split}
\ee
where $P_{a,b}$ are the hadronic initial-state momenta, and adding the appropriate convolution with parton distribution functions, it is possible to use our master formula to rederive results of Ref.~\cite{Ebert:2018lzn}. More information about this comparison can be found in Appendix~\ref{sect_compDY}.

Another check of the master formula originates  from an opportunity to perform another specialized computation for Drell-Yan process 
$q \bar q \to l^+ l^-$. Indeed, as shown in 
Appendix~\ref{sect3D}, it is straightforward 
to derive power corrections to the rapidity of \emph{one} of the charged leptons, using dedicated  phase-space parametrisation. 
Alternatively, we can derive the same result from the master formula. 
To this end, we choose the observable 
\be
{\cal O}( p_1,  p_2) = \delta(\tilde y - y_0) = \delta \left( \frac{1}{2} \ln \frac{p_b \cdot  p_1}{p_a \cdot  p_1} - y_0 \right),
\ee
that fixes the lepton rapidity. 
%The idea is then to compare the result we get by applying this delta in the lepton rapidity to the above result with the one in Appendix \ref{(and summing the contribution from both collinear sectors). 
For the comparison it is important to realize that the variable $x$ in this section and the variable $z$ in the appendix are, in fact, the same, and a both given by the following formula. 
\be
x = z = \frac{1}{s} ( p_1 +  p_2)^2.
\ee

Furthermore, the rapidity is expressed through variables $x, \beta$  differently in two collinear sectors, i.e. 
\be
y_{ca} = \frac{1}{2} \ln \frac{x (1-\beta)}{\beta}, \qquad y_{cb} = \frac{1}{2} \ln \frac{(1-\beta)}{x \beta}.
\ee
It follows 
\be
\beta = \begin{cases}
    \left ( 1+\frac{1}{x}  e^{2y} \right )^{-1}, &\quad \text{sector $ca$},\\
    \left ( 1+x e^{2y} \right )^{-1} , &\quad \text{sector $cb$}.
\end{cases}
\ee
After expressing everything in terms of the rapidity and evaluating the derivatives of the observable (that lead to  derivatives of the rapidity-constraining delta function that we need to get rid of by integrating by parts), we find perfect agreement in the distributions in $y_0$ obtained from both approaches.  %Interestingly, all $1/x$ poles appearing in the second approach cancel after summing over all the terms involving the derivatives of the observable. 

In addition, we have performed a numerical check, as we now describe. 
%For example,  choosing the simplest  observable $\mathcal O(\tilde p_1, \tilde p_2) = 1$ is not possible, because in this case the $1/x$ terms do not cancel at $x \rightarrow 0$. 
To keep things simple, we take  an observable that  constrains the invariant mass of the two leptons 
\begin{equation}
    \mathcal O(p_1, p_2) =   \theta\left( (p_1 + p_2)^2 - s_0 \right),
     \label{eq5.25}
\end{equation}
and compute the fiducial cross section by performing the phase-space integration in  Eq.~\eqref{eq4.2} at fixed values of $\tau$. To this end,  we remove 
the $\delta$-function responsible for the overall energy-momentum conservation by integrating  over the three-momentum of one of the leptons and the energy of another lepton, following the discussion in Appendix~\ref{sect3D}.  The zero-jettiness $\delta$-function 
is removed by integrating over the energy of the emitted gluon.  Integrations over 
the emission angles of one of the leptons and the gluon are performed numerically.
%We choose the beam axis as the reference for both polar angles and notice the symmetry of the problem with respect to rotations around it, which we exploit to replace the two azimuthal angle integrals by a single one over the relative azimuthal angle of the gluon and lepton 1. We then use these three angles as the integration variables for the numerical computation. 
For the numerical integration itself, we take  $s_0=\num{0.1}~{\rm GeV}^2$, $Q=\num{0.1}~{\rm GeV}$, $s=1~{\rm GeV}^2$ and set all couplings and charges to $1$. We perform the calculation of the differential cross section at finite $\tau$ for $\tau \in [\tau_{\rm min}, 1]$, where 
$\tau_{\rm min}$ is $10^{-30}$.

We wish to compare the results of the numerical and analytic computations. The latter (for the subleading power) is given in 
Eq.~\eqref{eq5.20}; they have to be supplemented with leading power results that are well 
known.  An important feature 
of the observable in 
Eq.~(\ref{eq5.25}) is that 
it is invariant under  Lorentz boosts applied to leptons.   This leads 
to significant simplifications in the final formula 
for next-to-leading power corrections to the Drell-Yan process. 
%vanish when taken as a whole, with the exception of the plus distribution terms not related to $\bar P_{qq}(x)$ in Eq.~\eqref{eq3.68}\footnote{In these terms the observable derivatives appear due to the transformation used to extract the soft singularity at $x=1$, which also involves a rescaling, and not due to the boost from the phase space manipulation.}. 
We find  
\be
\begin{split}
\frac{{\rm d} \sigma^{\rm DY,{\rm NLP}} }{{\rm d} \tau} 
 & = \frac{4 [\alpha_s] C_F Q }{s} {\rm d} \sigma_0 \bigg[ -1 - \frac{x_0}{1-x_0} -  \frac{1}{4(1 - 2 \beta + 2 \beta^2 )} \\
 & \times \int \limits_{0}^{1} {\rm d} x \left(  \frac{1+x^2}{x^2}  \left(1- 6 \beta + 6 \beta^2\right) + \frac{2 f_0(\beta) }{x}
  \right) \theta(s x - s_0) \bigg],
\end{split}
\ee
where  $x_0 = s_0/s$. We note that there is no  $\log \tau$ term in the subleading power corrections for the observable in Eq.~(\ref{eq5.25}), but  this  feature is certainly observable-dependent. 

In Fig.~\ref{fig3} we plot the relative difference between the numerical and analytic results, 
normalized to the analytic result.  At very small values 
of $\tau$, the precision 
of the numerical calculation 
is insufficient to constrain 
subleading power corrections, 
but it is good enough to check the leading power contributions. 
However, for values
$\tau \in [10^{-8}, 10^{-4}]$,
the precision becomes 
sufficient to enable  the check of the subleading 
power correction. 
%We point out that for small values of $\tau$ it becomes small enough that it plateaus due to the limit of the numerical precision of the calculation, whereas for larger values it increases due to the higher-order terms in the $\tau$-expansion which are not included in the analytic result.
 
We performed a  numerical fit for the 
$\tau$-independent coefficients  of the fiducial cross section defined with the 
observable in Eq.~(\ref{eq5.25}). 
Making an ansatz  
\begin{equation*}
    \frac{\mathrm d \sigma}{\mathrm d \tau} = \tau^{-1} \left ( \log \tau C_{\text{LP, LL}} +  C_{\text{LP, NLL}} \right )+ \log \tau C_{\text{NLP, LL}} + C_{\text{NLP, NLL}} + \tau C_{\text{NNLP}} + \tau^2 C_{\text{N3LP}} + \tau^3 C_{\text{N4LP}}~,
\end{equation*}
we compute ${\rm d} \sigma/{\rm d} \tau$ for 
different values of 
$\tau$ and perform a standard $\chi^2$ fit to determine the  
coefficients. The fit is done using the values within the range $\tau \in [10^{-30}, 10^{0}]$. %in order to stay away from the higher values of $\tau$ where more coefficients associated with higher powers would be required for a precise description of the cross section, and also because in that region the numerical integration becomes inaccurate. 
The result of the fit for the relevant terms is shown in Table \ref{tab1} together
with the results obtained from the analytic computation. 
Excellent agreement among the $\tau$-independent coefficients is observed.
\\

\begin{table}[t]
    \centering
    \caption{
    Comparison of the  expansion coefficients of 
       the fiducial cross section of a Drell-Yan process  in the zero-jettiness variable  through  next-to-leading power, obtained through a numerical fit and  an analytic computation,
       for the observable in Eq.~(\ref{eq5.25}).
 We  take
$s_0 = 0.1~{\rm GeV}^2, Q = 0.1~{\rm GeV}, s = 1~{\rm GeV}^2$ and set all couplings and charges to $1$. 
     }
    \vspace{6pt}
    \begin{tabular}{c S[table-format=3.10] S[table-format=3.10]}
        \toprule
        coefficient & fit & analytic \\
        \midrule
        $C_{\rm LP, LL}$ & -4.740740718 & -4.740740741 \\
        $C_{\rm LP, NLL}$ & 13.741118266 & 13.741118217 \\
        $C_{\rm NLP, LL}$ & 0.00017 & 0.00000 \\
        $C_{\rm NLP, NLL}$ & -1.0710 & -1.0725 \\
        %$C_{\rm NLP, LL}$ & \num{0.000179950} & \num{0.000000000} \\
        %$ C_{\rm NLP, NLL}$ & \num{-1.071083950} & \num{-1.072546919} \\
        \bottomrule
    \end{tabular}
    \label{tab1}
\end{table}

\subsection{Production 
of two photons in $q \bar q$ collisions}

Next we consider the production of two photons in
the annihilation of a quark and an anti-quark 
\be
q(p_a) + \bar q(p_b) \to \gamma(p_1) + \gamma(p_2).
\label{eq5.26}
\ee
Since this is also a $2 \to 2$  process, we can use 
the same phase space and momenta parametrization as  in the Drell-Yan case. However, the main difference between the two cases  is that  in the di-photon production  the quantities $R_{\rm fin}^\mu$ and $N_{a,b}^{(1),\mu}$ do not vanish. Because of this,  we can check all the entries in  the master formula for 
subleading power corrections given in  Eq.~(\ref{eq3.68}).

For the di-photon production 
Eq.~(\ref{eq5.26}), the leading order cross section reads
\be
{\rm d} 
\sigma_0^{2 \gamma} = 16 \pi \; \bar \sigma_0^{2 \gamma} \;
{\rm d} \Phi_2 \; \frac{(1 - 2 \beta + 2 \beta^2)}{\beta (1-\beta)}, 
\ee
where the phase space 
and the momenta parametrization can be found 
in Eqs~(\ref{eq5.2}, \ref{eq5.3}), 
and 
\be
\bar \sigma_0^{2 \gamma} = \frac{\pi Q_q^4 \alpha_{\rm QED}^2}{s N_c}.
\ee
Similar to the Drell-Yan case, we perform the computation setting 
$N_c \to 1, Q_q \to 1$ and $e \to 1$, and restore the relevant factors at the end. With this normalization, the required Green's functions can be computed either using formulas provided in Sec.~\ref{sec:currents} 
or simply collecting relevant 
Feynman diagrams which, for 
the process in Eq.~(\ref{eq5.26}) is quite straightforward. 
We find   
\be
N_{a,b} (p_{b},p_{a},P_X) = -i \left[ \frac{ \gamma_\nu (\hat p_{a} - \hat p_1) \gamma_\mu}{s_{a1}} + \frac{ \gamma_\mu (\hat p_{a} - \hat p_2) \gamma_\nu}{s_{a2}} \right] {\epsilon^\mu_1}^* {\epsilon^{\nu}_2}^*,
\ee
where $\epsilon_{i}$  are the polarization vectors of the photons and $s_{ai} = 2 p_a p_i$, $i=1,2$.
We note that, 
when constructing 
 $N_{a,b}$, we obtain the momentum of an anti-quark 
 ($p_b$) using  momentum conservation.  Similarly, 
 the function $R_{\rm fin}$ 
 is easy to construct; it reads
\be
\begin{split}
R_{\rm fin}^\rho (p_b,p_a,k,P_X) = i \bigg[ & \frac{ \gamma_\nu (\hat p_{a} - \hat k - \hat p_1) \gamma^\rho (\hat p_{a} - \hat p_1) \gamma_\mu}{(s_{a1}+s_{ak}-s_{k1}) \; s_{a1}} \\
& + \frac{ \gamma_\mu (\hat p_{a} - \hat p_2) \gamma^\rho (\hat p_{a} - \hat k - \hat p_2) \gamma_\nu}{(s_{a2}+s_{ak}-s_{k2}) \; s_{a2}} \bigg] {\epsilon^\mu_1}^*
{\epsilon^\nu_2}^*.
\end{split}
\label{eq5.30}
\ee
We note that the above expressions can be used for both $\vec k || \vec p_a$ and 
$\vec k || \vec p_b$ cases. 
In the first case, 
in Eq.~(\ref{eq5.30})
we have to take
$p_a \to x p_a$, $k \to (1-x) p_a$ in the strict collinear limit, and in the second case 
$p_a \to p_a $ and $k \to (1-x) p_b$. 

Given  the above expressions,  it is clear that the Green's functions  $N^{(1),\rho}_{a,b}$ and $R_{\rm fin}^{(1),\rho \sigma}$ do not vanish. They can be obtained by expanding the above formulas in the relevant small parameters, routing the  momentum perturbation in a particular way. We find 
\be
\begin{split}
N^{(1),\rho}_{a,b} (p_b,p_a,P_X) = -i \bigg[ & \frac{ \gamma_\nu \gamma^\rho \gamma_\mu}{s_{a1}} + 2 (p_a^\rho-p_1^\rho) \; \frac{ \gamma_\nu (\hat p_{a} - \hat p_1) \gamma_\mu}{s_{a1}^2} \\
& + \frac{ \gamma_\mu \gamma^\rho \gamma_\nu}{s_{a2}} + 2 (p_a^\rho-p_2^\rho) \; \frac{ \gamma_\mu (\hat p_{a} - \hat p_2) \gamma_\nu}{s_{a2}^2} \bigg] {\epsilon^\mu_1}^* {\epsilon^\nu_2}^*,
\end{split}
\ee
and 
\be
\begin{split}
& R_{\rm fin}^{(1),\rho \sigma} (p_b,p_a,k,P_X) = \frac{i}{2} {\epsilon^\mu_1}^* {\epsilon^\nu_2}^* \Bigg[ 
\frac{ \gamma_\nu (\hat p_{a} - \hat k - \hat p_1) \gamma^\rho \gamma^\sigma \gamma_\mu}{x \; (s_{a1}+s_{ak}-s_{k1}) \; s_{a1}} 
- \frac{ \gamma_\nu \gamma^\sigma \gamma^\rho (\hat p_{a} - \hat p_1) \gamma_\mu}{(s_{a1}+s_{ak}-s_{k1}) \; s_{a1}} \\
& + 2 \left( \frac{p_a^\sigma - p_1^\sigma}{x s_{a1}} - \frac{p_{a}^\sigma - k^\sigma - p_1^\sigma}{(s_{a1}+s_{ak}-s_{k1})} \right) \frac{ \gamma_\nu (\hat p_{a} - \hat k - \hat p_1) \gamma^\rho (\hat p_{a} - \hat p_1) \gamma_\mu}{(s_{a1}+s_{ak}-s_{k1}) \; s_{a1}} \Bigg] + \left\{ 1 \leftrightarrow 2 \right \}.
\end{split}
\ee
The comment about momenta assignments for the two collinear cases below Eq.~(\ref{eq5.30}) also applies  
for $N^{(1),\mu}$ and 
$R_{\rm fin}^{(1), \rho \sigma}$ in the above formulas. 
\\

We proceed with the  calculation of  the subleading power corrections, and discuss various contributions that appear in the general formula Eq.~(\ref{eq3.66}).
\begin{itemize} 

\item For the finite reminder
(c.f. Appendix~\ref{appB}) in the $\vec k || \vec p_a$ case, we find
\be
F_{\rm fin,a}^{2 \gamma}
  =  8 \left ( \frac{1-x}{\beta (1-\beta) x} \right)^2 - 8 \; \frac{1+2\beta-2\beta^2}{\beta (1-\beta) x}+32.
\ee

\item As we already mentioned, since we consider  massless particles, the following equation holds
\be
\left [ 
\kappa_m + 2 p_a^\mu \frac{\partial}{\partial p_{a}^{\mu}}
+ ( g^{\mu \nu} + \omega_{ab}^{\mu \nu}  )  L_{\mu \nu} 
\right ]|\mathcal M^2(p_b,p_a,p_1,p_2)|
= 0.
\ee

\item The function  $W_a$  evaluates to
\be
W_a^{2 \gamma} (x) = 2 \; \frac{1-x}{x} \; \frac{(1-2\beta+2\beta^2)}{(1-\beta)^2 \beta^2} + \frac{32}{x}.
\ee
It contains one term  that  does not vanish in the soft $x \to 1$ limit. 

\item We also find 
\be
\begin{split}
g_\perp^{\rho \alpha} &  
  \left( D^{xa,b}_\rho \; |{\cal M}|^2(p_b,xp_a, ...) - 2 {\rm Tr} \left [N_a \gamma_{\rho} 
N_a^+ \hat p_b \right ] \right) 
 {b_a}^{\mu \nu} _\alpha L_{\mu \nu} \\
& = 8 \frac{(1-2\beta+2 \beta^2)}{s x (1-\beta)\beta} \bigg\{ \left[ \frac{(1-2\beta)}{(1-\beta) \beta} \; p_1^\mu - \frac{(1-2\beta)^2}{\beta} \; x p_a^\mu + \frac{(1-2\beta)^2}{(1-\beta)} \; p_b^\mu \right ] \partial_{1 \mu} \\
& - \left[ \frac{(1-2\beta)}{(1-\beta) \beta} \; p_2^\mu + \frac{(1-2\beta)^2}{(1-\beta)} \; x p_a^\mu - \frac{(1-2\beta)^2}{\beta} \; p_b^\mu \right ] \partial_{2 \mu} \bigg\}.
\end{split}
\ee

\item Another contribution with derivatives and traces that involves multiple  Green's functions evaluates to 
\be
\begin{split}
s (1-x) & \frac{g_\perp^{\alpha \beta}}{4} \Bigg \{
- 2 {\rm Tr} \left [  N_a \gamma_\beta N_a^+ \hat p_b \right ] \\
& + {\rm Tr} 
\left [ N_a \gamma_\beta  \gamma_\rho \hat p_a \left( R_{{\rm fin}}^{\rho,+} + \frac{N_b^+ (\hat p_b - (1-x) \hat p_a) \gamma^\rho}{(1-x) s}\right) 
\hat p_b \right ] + {\rm c.c.}
\\
& +
\frac{2 x}{1-x}
{\rm Tr} \left [ 
N_a \hat p_a \left ( R_{\beta}^{{\rm fin},+} 
- \frac{N_b^+ \hat p_a  \gamma_\beta }{s}
\right ) \hat p_b
\right ] + {\rm c.c.} \Bigg \} {b_a}^{\mu \nu} _\alpha L_{\mu \nu}  \\
 = & - \frac{4 (1-2 \beta) (1-2 x^2 \beta + 2 x^2 \beta^2)}{x^2 (1-\beta)^2 \beta^2} \\
& \times \bigg\{ \left[ p_1^\mu - (1-2 \beta) (1-\beta) \; x p_a^\mu + (1-2 \beta) \beta \; p_b^\mu \right ] \partial_{1 \mu} \\
& \quad - \left[ p_2^\mu + (1-2 \beta) \beta \; x p_a^\mu - (1-2 \beta) (1-\beta) \; p_b^\mu \right ] \partial_{2 \mu} \bigg\}.
\end{split}
\ee
\end{itemize}
As in the previous section, to get ${\rm C}^{{\rm NLP},b}$ we should replace 
\be
\beta \rightarrow 1-\beta, \qquad p_a \leftrightarrow p_b, 
\ee
in the above formulas. 

With all the necessary ingredients, the total subleading contribution can be obtained using Eq.~\eqref{eq3.66}. In this case we get
\begin{align}
\frac{{\rm d} \sigma^{2 \gamma,{\rm NLP}} }{{\rm d} \tau} 
 & = \frac{4 [\alpha_s] C_F Q }{s} {\rm d} \sigma_0^{2 \gamma} \left[  \left( -1 + \frac{1}{2} {\cal D} \right) + \frac{1}{2} \log \left( \frac{\tau Q}{s}\right) {\cal D} \right] {\cal O}(p_1,p_2) \nonumber
\\
  & + \frac{2 [\alpha_s] C_F Q }{s} {\rm d} \sigma_0^{2 \gamma} \int \limits_{0}^{1} {\rm d} x \; \bigg\{ - \frac{1}{2 (1-x)_+} \; \left ( {\cal D}\Big |_{ca}
  + {\cal D}\Big |_{cb}
  \right )  \nonumber
  \\
  & + \left( \frac{ \left [ \bar \beta \; p_a^\mu - \beta \; p_b^\mu \right ] \partial_{1 \mu} + \left [  \beta \; p_a^\mu - \bar \beta  \; p_b^\mu \right ]
    \partial_{2 \mu} }{2 \; (1-x)_+} \right) \Bigg|_{ca} \\
  & + \left( 
  \frac{\bar \beta  \; p_a^\mu \partial_{1\mu} + \beta \; p_a^\mu \partial_{2\mu}}{2} + \frac{{\cal P}_{2 \gamma} (\beta, x, p_a, p_b;p_1,p_2,\partial_1,\partial_2)}{8(1 - 2 \beta + 2 \beta^2 )} 
  \right) \Bigg |_{ca} \nonumber \\
  & - \left( \frac{ \left [ \bar \beta  \; p_a^\mu - \beta \; p_b^\mu \right ] \partial_{1 \mu} + \left [  \beta \; p_a^\mu - \bar \beta  \; p_b^\mu \right ]
    \partial_{2 \mu} }{2 \; (1-x)_+} \right) \Bigg|_{cb} \nonumber \\
  & + \left( \frac{\beta \; p_b^\mu \partial_{1\mu} + \bar \beta \; p_b^\mu \partial_{2\mu}}{2} + \frac{{\cal P}_{2 \gamma} (\bar \beta, x, p_b, p_a; p_1,p_2,\partial_1,\partial_2)}{8(1 - 2 \beta + 2 \beta^2 )} \right) \Bigg |_{cb}  \; \bigg\} \; {\cal O}(p_1,p_2), \nonumber
\end{align}
where $\bar \beta = 1-\beta$ and vertical bars indicate that terms have to be evaluated in the appropriate collinear  kinematics. The differential operator ${\cal D}$ is defined in Eq.~(\ref{5.21D}) and ${\cal P}_{2 \gamma}(\beta,x,p_a,p_b;p_1,p_2,\partial_1,\partial_2)$  is given by
\be
\begin{split} 
 & {\cal P}_{2 \gamma} (\beta,x,p_a,p_b;p_1,p_2,\partial_1,\partial_2) =
   {\cal P}(\beta,x,p_a,p_b;p_1,p_2,\partial_1,\partial_2) 
   \\
   & + \left(  \frac{1+x^2}{x^2} \; \frac{5+4 (\bar \beta \beta )^2}{\bar \beta \; \beta} - \frac{8}{x \: \bar \beta \; \beta}
  \right)
  +g_\gamma(x,\beta)  \; p_a^\mu \partial_{1\mu}   + \frac{g_\gamma \left (x_1,\beta \right ) }{x}  \;   p_b^\mu \partial_{2\mu}
  \\
  & + g_\gamma (x,\bar \beta ) p_a^\mu \partial_{2\mu}  + \frac{g_\gamma \left (x_1,\bar \beta  \right ) }{x}  p_b^\mu \partial_{1\mu} - \frac{g_\gamma(x,\beta)}{x \bar \beta(1-2\beta)}  p_1^\mu \partial_{1\mu} + \frac{g_\gamma(x, \bar \beta)}{x \beta (1-2\beta)}  p_2^\mu \partial_{2\mu} ,
\end{split}
\ee
where
\be
g_\gamma(x,\beta) = \frac{(1-2\beta)^2}{\beta} \; \frac{(1-x^2)}{x}.
\ee
Similar to the Drell-Yan case, the results shown above were checked against the direct expansion of the NLO matrix element squared of the process 
$q(p_a) + \bar q(p_b) 
\to \gamma(p_1) + \gamma(p_2)
+g(k)$ through next-to-leading power in the gluon momentum in the soft and collinear limits and then integrating over the unresolved phase space.  
Full agreement with the above formulas has been found. This completes the check of all the entries present in the master formula given in  Eq.~(\ref{eq3.68}).
\\

\subsection{Production 
of four photons in $q \bar q$ collisions}

In this subsection, we apply  the  master formula presented at the end of Sec.~\ref{sect3a}, to calculate the subleading power corrections in the zero-jettiness to the production of a high-multiplicity colorless final state. For this purpose, we developed a \texttt{FORTRAN} code capable of computing the subleading power corrections to the production of an \emph{arbitrary} number of photons in $q \bar q$ collisions.   

The central element of the code is the computation of the generalized currents 
described in Sec. \ref{sec:currents} 
which can be done 
using recursive functions in \texttt{FORTRAN~90} for an arbitrary number of final-state particles $N$.  
The use of such functions makes coding straightforward. However,  it also requires careful optimization since the 
calculation of matrix currents is, in fact, quite  expensive. 
In addition,  phase space routines 
for an arbitrary number of final-state particles are available (see e.g. \cite{Kleiss:1985gy} and
\cite{Platzer:2013esa}), making it straightforward to write a program  to compute the subleading power correction in the zero-jettiness variable to a process 
$q \bar q \to N \gamma$.\footnote{In practice, we have employed  the multi-particle   
phase-space generator written by K.~Asteriadis.}

An important limitation  of  the current code is that it works for one observable at a time. 
 This observable should be such that it keeps all  photons hard (i.e., not collinear to the incoming quarks and not soft) or, at the very least, it should regulate the cross section in potentially singular  regions of the phase space. A possible choice is the product of the squared transverse momenta of all photons, i.e. 
\be
\mathcal{O}(P_{N\gamma}) = \frac{p^2_{1,\perp} \: p^2_{2, \perp} \; \cdots \; p^2_{N, \perp}}{s^{N}},
\label{eq5.43}
\ee
where $s = 2 p_a \cdot p_b$. The 
  transverse momentum squared of the $i$-th  photon is given by
\be
p^2_{i,\perp} = 2\frac{(p_a \cdot p_i) (p_b \cdot p_i)}{p_a \cdot p_b}.
\ee

We have checked the numerical  code by using it to calculate the subleading power corrections for the production of two photons (using the observable given in Eq.~\eqref{eq5.43}), and comparing the result with  the integration of the  analytic expression 
for subleading corrections to the $q \bar q \to \gamma \gamma$ process
presented in the previous subsection. We found excellent agreement between the results of the two calculations. 

We then used the numerical code to compute the subleading power zero-jettiness 
correction   to 
$q \bar q \to 4 \gamma$
for the observable 
in Eq.~(\ref{eq5.43}). 
We found that computation of the 
  subleading power correction for four-photon production with a percent precision  required ${\cal  O}(10~000)$ CPU hours. This is to be contrasted with $\mathcal O(5)$ CPU hours needed to compute the fiducial leading order cross section for the four-photon production. This increase is related to the complexity and the number of the many different currents that are required at subleading power but,  probably, with further optimization, significant improvements in efficiency  can be achieved. 
  
  In order to validate our numerical results for the zero-jettiness 
power correction   to 
$q \bar q \to 4 \gamma$, we used the same \texttt{FORTRAN} code to compute bin-integrated  cross section for  $q \bar q \to 4 \gamma + g$ with the  observable 
in Eq.~(\ref{eq5.43}), i.e. 
\be
\int \limits_{\tau_{\rm min}}^{\tau_{\rm max} } {\rm d} {\tau} 
\frac{{\rm d} \sigma_{4 \gamma}}{{\rm d} \tau} \; {\cal O}(P_{4 \gamma} ),
\ee
for several bins $[\tau_{\rm min}, \tau_{\rm max}]$ drawn from the interval  $\tau \in [10^{-4}, 1]$.
We then write 
\be
\frac { {\rm d} \sigma_{4 \gamma}
}{{\rm d} \tau}
= \frac { {\rm d} \sigma^{\rm LP}_{4 \gamma}
}{{\rm d} \tau}
+ \frac{{\rm d} \sigma_{4 \gamma}^{\rm NLP}}{{\rm d} \tau}, 
\ee
and use the well-known result for the 
leading-power cross section $ {\rm d} \sigma^{\rm LP}_{4 \gamma}/{\rm d} \tau$,  and  the following 
ansatz 
\begin{equation*}
    \frac{\mathrm d \sigma^{\text{NLP}}_{4 \gamma}}{\mathrm d \tau} = \ \log \tau (C_{0,\text{NLP}}+ \tau C_{0,\text{NNLP}} +\tau^2 C_{0,\text{N3LP}}) + C_{1,\text{NLP}} + \tau C_{1,\text{NNLP}} + \tau^2 C_{\text{1,N3LP}} + \tau^3 C_{\text{1,N4LP}}~,
    \label{eq6.52}
\end{equation*}
 for the subleading one. 
 We then perform a standard $\chi^2$ fit to determine  coefficients in the above equation 
 by integrating the ansatz for each $\tau$-bin. %The logarithmic $\tau$-bins used for the fit are within the range $[10^{-4}, 10^{0}]$%. 
 The fitted coefficients $C_{0,\text{NLP}}$ and $C_{1,\text{NLP}}$ are then  used to determine  the subleading soft and collinear coefficients using Eq.~(\ref{eq3.66}).
  
  The results of the numerical evaluation of  the subleading power corrections to  four-photon production  and their  comparison with the fitted results is shown  
   %Finally,  for the numerical  evaluation, we set $s=4\times 10^4$ GeV and set all the couplings and charges to $1$.
  %The results of the coefficients as defined in Eq. (\ref{eq3.66}) are presented in 
 in  Table~\ref{tab2}. We find an agreement between numerically-calculated  and fitted coefficients within the error of the fit.  We note that further reduction  of the fit error is possible, but would require
 a more 
 significant  computational effort.
\\

\begin{table}[t]
    \centering
    \caption{Next-to-leading power  coefficients as defined in Eq.~(\ref{eq3.66}) for $4\gamma$ production at $\sqrt{s}=200$ GeV with the observable defined in Eq.~\eqref{eq5.43}. In this case, the collinear coefficient ${\rm C}^{{\rm NLP},b}$ is equal to ${\rm C}^{{\rm NLP},a}$. To compute these coefficients, we have set quark electric charges to 
    one, $e e_q \to 1$. Fitted results 
    for  the next-to-leading power 
    coefficients are compared 
    with the results obtained by a numerical integration of the derived analytic formula.}
    \vspace{6pt}
    \label{tab2}
    \begin{tabular}{ccc}
        \toprule
        coefficient &  numeric &  fitted \tabularnewline
        \midrule
       $ {\rm C}^{{\rm NLP},s}$ & \hfill $2.61598(7) \times 10^{-7}$  & \hfill $2.5(1) \times 10^{-7}$\\
        ${\rm C}^{{\rm NLP},a}$ & \hfill $8.61(8) \times 10^{-7}$ & \hfill $8.9(5) \times 10^{-7}$ \\
       % ${\rm C}^{{\rm NLP},b}$ & $3.49(3) \times 10^{-4} $ \\
        \bottomrule
    \end{tabular}
\end{table}

%\begin{table}[t]
%    \centering
%    \caption{Coefficients of the $4\gamma$ cross section expanded in the zero-jettiness $\tau$ to next-to-leading power.}
%    \label{tab3}
%    \begin{tabular}{cc}
%        \toprule
%        coefficient & result(pb) \\
%        \midrule
%        NLP, LL & $6.98(3) \times 10^{-5} $\\
%        NLP, NLL & $2.06(6) \times 10^{-4} $ \\
%        \bottomrule
%    \end{tabular}
%\end{table}

\section{Conclusions}
\label{sect:conc}

We discussed the computation of  next-to-leading power corrections in the zero-jettiness variable to the production of \emph{arbitrary} colorless final states at hadron colliders 
at next-to-leading order in perturbative QCD. Our goal was to investigate whether a similar degree of universality that exists for leading power corrections can be achieved for the subleading ones.   We have relied on the powerful tools 
developed to study infra-red and collinear limits of QCD which employ momenta redefinition and Lorentz boosts, and we have shown how to use these methods to construct an expansion of the generic phase space and matrix elements squared at next-to-leading power, restricted to the production of colorless finals states.  

The most challenging aspect of these expansions comes from the collinear limit where the universality of the limit is lost at next-to-leading power in the  sense that the result depends on the radiative process albeit in the simplified  kinematics.  
We have argued that complicated Green's functions that arise from these expansions can be calculated 
recursively using analogs of Berends-Giele currents \cite{Berends:1988zn} which should enable applications 
of the derived formulas to processes with high multiplicity final states. 
We have
provided an example by computing the next-to-leading power correction in the zero-jettiness variable  to the fiducial cross section for the  production of four hard photons in $q \bar q$ collisions, 
and we have constructed a numerical code which 
can be used to compute such power  corrections to $q \bar q \to N \gamma$
process for any $N$.

We note that we only considered the $q \bar q$ annihilation channel in this paper whereas also $q g \to X+q $  and similar channels are needed for a complete next-to-leading order computation.  The most important difference between $qg \to X+q$ and $q \bar q \to X +g$ channels is that in the former, a soft final-state quark only contributes to subleading power so that the analysis of the soft  limit is significantly simpler than in $q \bar q \to X+g$ case.  On the other hand,  we do not anticipate any significant differences between $qg$ and $q \bar q$ channels in the collinear limits. Thus, we believe that the methodology developed in this paper can be applied to all partonic channels in a straightforward way. 

An important shortcoming for the numerical implementation of our method  is its explicit dependence on  observables.  This is in strong contrast to calculations at leading power where one obtains all observables in a single Monte-Carlo run  by computing many of them  for each generated kinematic point, and storing them in histogram bins. 
It is important to find a way to do this for power corrections as well, since it  will make such computations observable-independent and significantly  more efficient.  

Eventually, one would like to extend the current understanding of the next-to-leading power corrections in the context of existing slicing schemes to arbitrary collider processes,  similar to what has been achieved at leading 
power. This is a highly non-trivial task,  and there are lessons that 
one can take from the computation 
described in this paper. For example, 
at next-to-leading order,  next-to-leading power contributions to arbitrary processes originate exclusively from soft and collinear limits that can be treated independently. 
Similar to the leading power case,
at next-to-leading power the soft contributions  can be treated universally  and the collinear contributions  -- which appear to be the major bottleneck --  are localized on the external legs.   At the same time, extension to QCD final states will require understanding of jet algorithms and their interplay with power corrections, and, as we already see, 
observables introduce a significant degree of complexity into the analysis 
of subleading power corrections even for colorless final states. 
All in all, it  remains to be seen to what extent the approach introduced  in this paper 
can be used to extend slicing schemes  to next-to-leading power 
for \emph{arbitrary} processes at NLO QCD and beyond.

\section*{Acknowledgments}
We have benefited from conversations with I.~Novikov.   We are grateful to K.~Asteriadis for 
providing the phase-space generator that was used  in the 
numerical code.  
This  research  was supported by the German Research Foundation (DFG, Deutsche For\-schungs\-ge\-mein\-schaft) under grant 396021762-TRR 257.

\appendix

\section{Explicit formulas for boosts}
\label{appA}

For the analysis of collinear contributions, four boosts are required. 
In the main text, they are denoted as $\Lambda_{a}$, $\Lambda_{b}$, 
$\Lambda_{ax}$, $\Lambda_{bx}$.  In this appendix we present these 
quantities  explicitly. 

A general formula that describes a Lorentz boost that transforms a four-vector $Q_i$ to a four-vector $Q_f$
\be
Q_f^\mu =  \left [ \Lambda_{\rm gen}(Q_f,Q_i) \right ]\indices{^\mu_\nu} Q_i^\nu,
\ee
reads
 \be
\left [ \Lambda_{\rm gen}(Q_f, Q_i) \right ]\indices{^\mu_\nu} = g\indices{^\mu_\nu} -
\frac{2(  Q_f +  Q_i)^\mu ( Q_f + Q_i)_{\nu}}{(Q_f+Q_i)^2} + \frac{ 2 Q_f^\mu Q_{i,\nu} }{Q_f^2}.
\label{eqa.2}
\ee
The above equation is only valid if  $Q_f^2 = Q_i^2$.
We use this formula to compute  expressions for the Lorentz transformations in the various limits. 

\subsection{Case $\vec k || \vec p_a$}

We begin  with the discussion of the collinear boosts in the case when the gluon is emitted along 
the direction of the incoming quark with the momentum $p_a$, $\vec k || \vec p_a$.  Then, 
\be
Q_f = Q_a = x p_a + p_b,
\;\;\; Q_i = p_a + p_b - k, 
\;\;\; k = (1-x) p_a + \tka,
\ee
and we need to expand 
the Lorentz boost in 
Eq.~(\ref{eqa.2}) to 
second order in $\tka$. We find 
\be
\Lambda^{\mu \nu}_{a}(Q_f, Q_i) = g^{\mu \nu}
+ \frac{\tka^\mu  Q_a^\nu - \tka^\nu Q_a^\mu}{\tilde Q_a^2}
- \frac{1}{2} \frac{Q_a^\mu  Q_a^\nu}{Q_a^4} \tka^2
-\frac{1}{2} \frac{ \tka^\mu \tka^\nu}{Q_a^2}
+{\cal O}(\tka^3).
\label{eqa.4}
\ee
We will also need the inverse of  $\Lambda_a$. It is easy to see that, to the required order, 
$\Lambda_a^{-1}$ is obtained from 
$\Lambda_a$ by replacing $\tka \to -\tka$. Then
\be
\left [  \Lambda_a^{-1} \right ] ^{\mu \nu}(Q_i, Q_f) = g^{\mu \nu}
- \frac{\tka^\mu  Q_a^\nu - \tka^\nu Q_a^\mu}{ Q_a^2}
- \frac{1}{2} \frac{ Q_a^\mu Q_a^\nu}{Q_a^4} \tka^2
-\frac{1}{2} \frac{ \tka^\mu \tka^\nu}{Q_a^2}.
\label{eqa.5}
\ee

This transformation needs to be applied to $p_a$, $p_b$ and $k$.
The calculation of 
$\Lambda_a p_{a,b}$ and $\Lambda_a k$ requires us to compute 
scalar products $Q_a \cdot p_{a,b}$ and $\tka \cdot p_{a,b}$.  Since $Q_a =  x p_a + p_b$, we find   
\be
Q_a^2 = xs,\;\;\; Q_a \cdot p_a =  s/2, \;\;\;
Q_a \cdot p_b = x s/2,
\ee
where $s = 2 p_a \cdot p_b$. 
Furthermore,
\be
\frac{Q_a \cdot  p_a}{Q_a^2}
 = \frac{1}{2x},
 \;\;\;
 \frac{Q_a \cdot p_b}{Q_a^2}
 = \frac{1}{2}.
\ee

Since $k = (1-x) p_a + \tka$, 
we find 
\be
k \cdot p_a = \tka \cdot p_a,
\ee
and, using  $k^2 = 0$, we obtain 
\be
\tka^2 = -2 (1-x) \tka \cdot p_a
= -2(1-x) k \cdot p_a.
\ee
It follows from Eq.~(\ref{eq4.46}) that 
$\tka \cdot p_b = - \tka \cdot p_a$.

Combining these formulas, we find the following expressions for the boosted momenta
\be
\begin{split} 
& \Lambda_a p_a = p_a + \frac{1}{2x} \; \tka + Q_a \; \frac{k p_a}{Q_a^2} 
\left ( \frac{1-3x}{2x} \right ), 
\\
& \Lambda_a p_b = p_b + \frac{1}{2} \; \tka + Q_a \; \frac{k p_a}{Q_a^2} 
\left ( \frac{3-x}{2} \right ), 
\\
& \Lambda_a k = (1-x) p_a + \frac{1+x}{2x} \; \tka+  Q_a \; \frac{k p_a}{Q_a^2} 
\left ( \frac{1-x^2}{2x} \right ).
\end{split}
\label{eq4.103}
\ee
Additionally, we write the formula for the Lorentz transformation of $\tka$ and of $p_a - k$. We obtain
\be
\begin{split}
& \Lambda_a(p_a - k) = x p_a - \frac{1}{2} \tka - Q_a \; \frac{k \cdot p_a}{Q_a^2} \frac{3-x}{2},
\;\;\; \Lambda_a \tka
 = \tka  -  \frac{\tka^2}{Q_a^2}
 \; Q_a.
\end{split}
\label{eq4.104}
\ee
\\

To extract soft singularities from the collinear case $\vec k || \vec p_a$, 
we need a Lorentz boost $\Lambda_{ax}^{-1}$. It reads
\be
\left [ \Lambda_{ax}^{-1}
\right ]^{\mu \nu} 
 = g^{\mu \nu} 
 - \frac{2(xp_a + p_b + \sqrt{x} P_{ab})^\mu
 (xp_a + p_b + \sqrt{x} P_{ab})^\nu
 }{s(2 x+
 \sqrt{x}(x+1))}
 + \frac{2(xp_a + p_b)^\mu P_{ab}^\nu }{s \sqrt{x}}.
 \label{eqa.13}
 \ee
The expansion of 
$\Lambda_{ax}^{-1}$ around $x=1$ is given 
by the following formula
\be
[\Lambda_{ax}^{-1}]^{\mu \nu} = g^{\mu \nu} 
- \frac{1-x}{2} \omega_{ab}^{\mu \nu} 
+{\cal O}((1-x)^2),
\label{eqa.14}
\ee
where 
\be
\omega_{ab}^{\mu \nu} 
 = \frac{p_a^\mu p_b^\nu - p_a^\nu p_b^\mu}{p_a \cdot p_b}.
 \label{eqa.15}
\ee

\subsection{Case $\vec k || \vec p_b$}

We continue with the discussion of the collinear boosts in case the gluon is emitted along 
the direction of the incoming anti-quark with momentum $p_b$, $\vec k || \vec p_b$.  Then, 
\be
Q_f = Q_b = p_a + x p_b,
\;\;\; Q_i = p_a + p_b - k, 
\;\;\; k = (1-x) p_b + \tkb,
\ee
and we need to expand 
the Lorentz boost in 
Eq.~(\ref{eqa.2}) to 
second order in $\tkb$. We find 
\be
\Lambda^{\mu \nu}_{b}(Q_f, Q_i) = g^{\mu \nu}
+ \frac{\tkb^\mu  Q_b^\nu - \tkb^\nu Q_b^\mu}{\tilde Q_b^2}
- \frac{1}{2} \frac{Q_b^\mu  Q_b^\nu}{Q_b^4} \tkb^2
-\frac{1}{2} \frac{ \tkb^\mu \tkb^\nu}{Q_b^2}
+{\cal O}(\tkb^3).
\ee
To the required order, 
the inverse $\Lambda_b^{-1}$ is obtained from 
$\Lambda_b$ by replacing $\tkb \to -\tkb$. Then
\be
\left [  \Lambda_b^{-1} \right ] ^{\mu \nu}(Q_f, Q_i) = g^{\mu \nu}
- \frac{\tkb^\mu  Q_b^\nu - \tkb^\nu Q_b^\mu}{ Q_b^2}
- \frac{1}{2} \frac{ Q_b^\mu Q_b^\nu}{Q_b^4} \tkb^2
-\frac{1}{2} \frac{ \tkb^\mu \tkb^\nu}{Q_b^2}.
\ee

The calculation of 
$\Lambda_b p_{a,b}$ and $\Lambda_b k$ requires the
scalar products $Q_b \cdot p_{a,b}$ and $\tkb \cdot p_{a,b}$.  Since $Q_b =  x p_b + p_a$, we find   
\be
Q_b^2 = xs,\;\;\; Q_b \cdot p_a = x s/2, \;\;\;
Q_b \cdot p_b = s/2,
\ee
so that 
\be
\frac{Q_b \cdot  p_a}{Q_b^2}
 = \frac{1}{2},
 \;\;\;
 \frac{Q_b \cdot p_b}{Q_b^2}
 = \frac{1}{2x}.
\ee

Since $k = (1-x) p_b + \tkb$, 
we find 
\be
k \cdot p_b = \tkb \cdot p_b,
\ee
and because  $k^2 = 0$, we obtain 
\be
\tkb^2 = -2 (1-x) \tkb \cdot p_a
= -2(1-x) k \cdot p_b.
\ee
It follows from Eq.~(\ref{eq4.46})  that 
$\tkb \cdot p_a = - \tkb \cdot p_b$.

Using the above  results, we find the following expressions for the boosted momenta
\be
\begin{split} 
& \Lambda_b p_b = p_b + \frac{1}{2x} \; \tkb + Q_b \; \frac{k p_b}{Q_b^2} 
\left ( \frac{1-3x}{2x} \right ), 
\\
& \Lambda_b p_a = p_a + \frac{1}{2} \; \tkb + Q_b \; \frac{k p_b}{Q_b^2} 
\left ( \frac{3-x}{2} \right ), 
\\
& \Lambda_b k = (1-x) p_b + \frac{1+x}{2x} \; \tkb+  Q_b \; \frac{k p_b}{Q_b^2} 
\left ( \frac{1-x^2}{2x} \right ).
\end{split}
\label{eqa.24}
\ee
We also find
\be
\begin{split}
& \Lambda_b(p_b - k) = x p_b - \frac{1}{2} \tkb - Q_b \; \frac{k p_b}{Q_b^2} \frac{3-x}{2},
\;\;\; \Lambda_b \tkb
 = \tkb  -  \frac{\tkb^2}{Q_b^2}
 \; Q_b.
\end{split}
\label{eqa.25}
\ee
\\

The  Lorentz boost $\Lambda_{bx}^{-1}$ reads 
\be
\left [ \Lambda_{bx}^{-1}
\right ]^{\mu \nu} 
 = g^{\mu \nu} 
 - \frac{2(p_a +x  p_b + \sqrt{x} P_{ab})^\mu
 (p_a + xp_b + \sqrt{x} P_{ab})^\nu
 }{s(2 x+
 \sqrt{x}(x+1))}
 + \frac{2(p_a + x p_b)^\mu P_{ab}^\nu }{s \sqrt{x}}.
 \label{eqa.26}
 \ee
The expansion of 
$\Lambda_{bx}^{-1}$ around $x=1$  is given by 
\be
[\Lambda_{bx}^{-1}]^{\mu \nu} = g^{\mu \nu} 
+ \frac{1-x}{2} \omega_{ab}^{\mu \nu} 
+{\cal O}((1-x)^2),
\label{eqa.27}
\ee
where the tensor 
$\omega_{ab}^{\mu \nu}$ 
can be found  in Eq.~(\ref{eqa.15}).

\section{Formulas for remainders}
\label{appB}

In Eq.~(\ref{eq3.68}), we have defined a  remainder for the $\vec k || \vec p_a$ 
case 
\be
F_{{\rm fin},a} 
= F_{{\rm rem},a} + \frac{s}{2} F_{rr,a}
+ \frac{s}{2} \left (
C_{1a}^k + C_{2a}^k + C_{3a}^k 
\right ),
\ee
where
\be
\begin{split}
   F_{{\rm rem},a} = & - |{\cal M}|^2(p_b,xp_a,...) + \frac{1}{2sx} 
{\rm Tr} 
\left [ 
 N_a \hat p_b \gamma^\nu \hat p_a  N_b^+ \hat p_a \gamma_\nu \hat p_b
\right ] + {\rm c.c.}
\\
& 
- \frac{1}{2x}
{\rm Tr} 
\left [
 N_a \hat p_b \gamma_\nu \hat p_a  R^{\nu,+}_{\rm fin} \hat p_b
\right ] + {\rm c.c.}  
+\frac{1}{x} 
{\rm Tr}
\left [ 
 N_a  \hat p_b  N_a^+ \hat p_b
\right ],
\end{split}
\ee
and
\be
F_{rr,a}
 = \frac{1}{s}
{\rm Tr}
\left [ 
 R_{\rm fin}^\mu \hat p_a  N_b^+ 
\hat p_a \gamma_\mu \hat p_b 
\right ] +{\rm c.c.}+
  \frac{2}{s} 
{\rm Tr} 
\left [ N_b \hat p_a 
 N_b^+ \hat p_a 
\right ]
-{\rm Tr}
\left [R_{\rm fin}^\nu \hat p_a R_{\rm fin}^{\mu,+} 
\hat p_b\right ] g_{\perp,\mu \nu}.
\ee

When computing the collinear 
expansion of the matrix element 
squared in the $\vec k || \vec p_a$
limit,  we pointed out that three terms need to be expanded to second order in the transverse momentum $k_\perp$, after the Lorentz boost is applied. They are 
\be
\begin{split}
& C_{1a} = \frac{2}{2 p_a \cdot  k} 
 {\rm Tr} 
\left [  N_a
\hat { \kappa}_a  
 N_a^+ \hat p_b
\right ]_{\Lambda_a},
 \\
& C_{2a} =   {\rm Tr} 
\left [ 
\frac{N_a \hat { \kappa}_a  \gamma_\nu \hat p_a  N_{{\rm fin},a}^{+,\nu} 
\hat p_b }{(-2 p_a \cdot k) }
\right ]_{\Lambda_a} + {\rm c.c.},
\\
& 
C_{3a} = \frac{2 \kappa_{a,\nu} }{(1-x)(-2 p_a \cdot k)}
{\rm Tr} 
\left [ 
N_a x \hat p_a 
\left ( R_{\rm fin}^{+,\nu} 
- N_b^+ \frac{\hat p_a  \gamma^\nu }{s}
\right ) \hat p_b
\right ]_{\Lambda_a} + {\rm c.c.} \; .
\end{split}
\ee
Since the Lorentz boost of 
$\kappa_a$ gives the 
transverse momentum $k_\perp$ 
and since two  powers of 
$k_\perp$ are needed to obtain 
the non-vanishing contribution we define quantities that contribute 
to the cross section in the collinear limit
\be
C_{ia}^{k} = \lim_{k_\perp^2 \to 0} \; \langle 
C_{ik}
\rangle_{k_\perp}.
\ee
In the above formula the two brackets indicate that one averages over $k_\perp$ directions.  To compute $C_{ia}^k$, 
$i=1,2,3$, we require the expansion of $N_a, N_b$ and $R_{\rm fin}^\nu$ after the boost $\Lambda_a$ to first order in $k_\perp$. Performing the boost, 
and expanding in $k_\perp$, 
we find 
\begin{equation}
\begin{split}
& N_a\left (
p_b + \frac{k_\perp}{2},x p_a -\frac{k_\perp}{2}, P_X \right ) 
 = N_a^{(0)} - \frac{k_{\perp,\mu}}{2} N_a^{(1),\mu} + \dots,
\\
& N_b\left (
p_b -(1-x) p_a  -\frac{k_\perp}{2x}, 
p_a +\frac{k_\perp}{2x}, P_X\right  )
= N_b^{(0)} + \frac{k_{\perp,\mu}}{2 x} N_b^{(1),\mu} + 
\dots, 
\\
& R_{\rm fin}^{\nu}\left ( 
p_b + \frac{k_\perp}{2},
p_a + \frac{k_\perp}{2 x} ,  p_a(1-x) + \frac{k_\perp(1+x)}{2x},P_X
\right )
= 
R_{\rm fin}^{(0),\nu}
+ R_{\rm fin}^{(1),\nu \mu} k_{\perp,\mu} 
+ \dots,
\\
\end{split}
\end{equation}
where ellipses stand for ${\cal O}(k_\perp^2) $ 
contributions. The 
quantities $N_a^{(1),\mu}, N_b^{(1),\mu}$ and $R^{(1),\nu \mu}_{\rm fin}$ are particular 
Green's functions that can be computed following the discussion 
in Sec.~\ref{sec:currents}. For our purposes here, we assume that they 
are known. We find 
\be
C_{1a}^{k}= \frac{1-x}{2} g_{\perp}^{\mu \nu}
D^{xa,b}_\nu \;
{\rm Tr} 
\left [ N_a(x p_a, \dots )
\gamma_\mu
N_a^+(x p_a, \dots)  \hat p_b
\right ],
\ee
where  $D_\nu^{xa,b} = x^{-1} \partial/\partial p_{a}^{\nu} - \partial/\partial p_{b}^{\nu}$.

The second limit is more complex. 
To write it in the compact  form, 
we introduce two  matrix functions
\be
\begin{split}
   & X^{(0),\alpha,+} 
= R_{\rm fin}^{(0),\alpha,+}
(1-x) + N_b^{(0),+}
\frac{\hat q_{ba} \gamma^\alpha}{s}, \\
&X^{(1),\alpha \mu,+} 
= R_{\rm fin}^{(1),\alpha \mu,+}
(1-x) + N_b^{(1),  \mu ,+}
\frac{\hat q_{ba} \gamma^\alpha}{2 x s},
\end{split}
\ee
where  $\hat q_{ba} = \hat p_b - (1-x) \hat p_a$, 
and write 
\be
\begin{split}
 C_{2a}^{k} = & 
 -\frac{g_\perp^{\mu \nu}}{4}
{\rm Tr}
\left [N_a^{(0)} \gamma_\nu \gamma^\alpha \hat p_a  
\left ( \frac{N_b^{(0),+} \gamma_\mu \gamma_\alpha \hat p_b}{xs}  -
X_\alpha^{(0),+} \gamma_\mu
- 2 X_{\alpha \mu}^{(1),+} \hat p_b
\right )
\right ] 
\\ 
& + \frac{g_\perp^{\mu \nu}}{4}
{\rm Tr} 
\left [ \left ( 
-N_{a,\mu}^{(1)} \gamma_\nu \gamma_\alpha \hat p_a 
+ \frac{1}{x}N_a^{(0)} \gamma_\nu \gamma_\alpha 
\gamma_\mu
\right )
X^{(0),\alpha,+}
\hat p_b
\right ] + {\rm c.c.}.
\end{split} 
\ee
Finally, the third term reads 
\be
\begin{split}
 C_{3a}^{k} = &
\frac{g_{\perp,\mu \nu}}{2}
{\rm Tr} 
\left [ 
\left ( - N^{(1),\mu}_a x \hat p_a
+ N_a^{(0)} \gamma^\mu 
\right )
\left ( R_{\rm fin}^{(0),\nu,+} 
- N_b^{(0),+} \frac{\hat p_a  \gamma^\nu }{s}
\right ) \hat p_b
\right ] 
\\
& + 
g_{\perp, \mu \nu}
{\rm Tr} 
\left [ 
N^{(0)}_a x \hat p_a 
\left ( R^{(1),\nu \mu,+}_{\rm fin} 
- N_b^{(1),\mu,+} \frac{\hat p_a  \gamma^\nu }{2 x s}
\right ) \hat p_b
\right ] 
\\
& + \frac{ g_{\perp,\mu \nu} }{2} 
{\rm Tr} 
\left [ N_a^{(0)} x \hat p_a R^{(0),\nu,+}_{\rm fin} \gamma^\mu \right ]
 - \frac{1}{s} {\rm Tr} 
\left [ N_a^{(0)} \hat p_a  N_b^{(0),+} 
(x \hat p_a + \hat p_b) 
\right ] + {\rm c.c.}.
\end{split}
\ee
\\

\section{Derivation of Eq.~(\ref{eq3.40})}
\label{app:eq340}

When simplifying  
Eq.~(\ref{eq3.40}), 
we wrote  the  integral of the function $W_3^{(a)}$ defined 
in Eq.~(\ref{eq6.33}) 
in the following way
\be
\begin{split}
 &   
\int {\rm d} x \; 
{\rm d} \Phi_m^{ab} 
\frac{W^{(a)}_3(x,p_a,p_b,P_X,{\cal O}(P_X) )}{(1-x)^2} 
=
\\
& - \frac{1}{2}
\left ( \kappa_m + 2 p_a^\mu \frac{ 
\partial }{\partial p_a^\mu }
+\left ( g^{\rho \sigma } 
+ \omega_{ab}^{\rho \sigma}
\right ) L_{\rho \sigma} \right ) {\cal O}(P_X) \; |{\cal M}|^2(p_b,x p_a, P_X)
\\
& - \frac{1}{2}
\int {\rm d} x \; 
{\rm d} \Phi(x p_a,p_b,P_X) \;
\frac{1}{(1-x)_+} \\
& \times \left ( \kappa_m + 2 p_a^\mu \frac{ 
\partial }{\partial p_a^\mu }
+\left ( g^{\rho \sigma } 
+ \omega_{ab}^{\rho \sigma}
\right ) L_{\rho \sigma} \right ) {\cal O}(P_X) \; |{\cal M}|^2(p_b,x p_a, P_X).
\end{split}
\label{eqc.1}
\ee
 In this appendix, we explain how to derive Eq.~(\ref{eqc.1}). To this end, we note 
that the function 
$W_3^{(a)}$ (c.f. Eq.~\eqref{eq6.33}) is written
as a difference of three terms, i.e. 
\be
W_3^{(a)}(x) 
= F(x) - F(1) + (1-x) 
F'(1),
\label{eqc.2}
\ee
where $F'(1)={\rm d} F(x)/{\rm d} x$ at $x = 1$. Then, using integration by parts,  it is easy to see that the following equation holds 
\be
\int \limits_{0}^{1}
{\rm d} x 
\frac{W^{(a)}_3(x)}{(1-x)^2}
= -F'(1) - 
\int \limits_{0}^{1} 
\frac{{\rm d} x}{1-x}
\left ( x F'(x) - F'(1) \right ).
\ee
Hence, we have 
\be
\int \limits_{0}^{1} 
{\rm d} x \; 
{\rm d} \Phi_m^{ab} \; \frac{W_3^{(a)}(x)}{(1-x)^2}
= - F'(1) \; {\rm d} 
\Phi_m^{ab} 
- \int \limits_{0}^{1} 
{\rm d} x \;
\frac{{\rm d} \Phi_m^{ab}}{1-x}
\left ( x F'(x) - F'(1) 
\right ). 
\label{eqc.5}
\ee
Comparing Eq.~(\ref{eqc.2}) and 
Eq.~(\ref{eq3.40}), we find 
\be
\begin{split}
& F(x) =  \lambda^{\kappa_m} 
{\cal O}(\lambda \Lambda_{ax}^{-1} P_X) 
|{\cal M}|^2(x p_a, p_b,\lambda \Lambda_{ax}^{-1} P_X),
\\
& F'(1) =  \frac{1}{2} 
\left ( 
\kappa_m + 
2 p_a^\mu \frac{ 
\partial }{\partial p_a^\mu }
 +\left ( g^{\rho \sigma } 
+ \omega_{ab}^{\rho \sigma}
\right ) L_{\rho \sigma}
\right ) 
 |{\cal M}|^2 \left ( p_a, 
p_b, 
P_X \right ) 
{\cal O}(P_X).
\end{split}
 \ee

To simplify Eq.~(\ref{eqc.5}), we need to compute $F'(x)$. To this end, 
we introduce $x_1 = x+\Delta x$
and note that 
\be
F'(x) = 
\lim_{\Delta x \to 0} \frac{F(x_1) - F(x) }{\Delta x}.
\ee
We also note that because of the nature of the Lorentz boosts 
with $x_1$ and $x$, the  following relation holds 
\be
\lambda_{x_1}
\Lambda_{ax_1}^{-1}
 = (I + \delta K) \lambda_x \Lambda_{ax}^{-1},
\ee
where $I^{\mu \nu} = g^{\mu \nu}$
and 
\be
\delta K^{\mu \nu} = \frac{\Delta x}{2x}
\left ( g^{\mu \nu} 
+\omega_{ab}^{\mu \nu} 
\right ).
\ee
Hence, 
\be
\begin{split} 
F'(x) 
& = \lim_{\Delta x \to 0} 
\frac{1}{\Delta x}
\Bigg  [ 
\lambda_1^{\kappa_m} 
{\cal O}\left( (I + \delta K) \lambda \Lambda_{ax}^{-1} P_X \right) 
|{\cal M}|^2 \left( x_1 p_a, p_b, (I + \delta K)
\lambda \Lambda_{ax}^{-1} P_X \right)
\\
& - 
\lambda^{\kappa_m} 
{\cal O}( \lambda \Lambda_{ax}^{-1} P_X) 
|{\cal M}|^2(x p_a, p_b, 
\lambda \Lambda_{ax}^{-1} P_X)
\Bigg  ] 
\\
&= 
\frac{\lambda^{\kappa_m}}{2x}
\left [ 
\kappa_m + 
2 p_a^\mu \frac{ 
\partial }{\partial p_a^\mu }
 +\left ( g^{\rho \sigma } 
+ \omega_{ab}^{\rho \sigma}
\right ) L_{\rho \sigma}
\right ]
 {\cal O}( Q_X) 
|{\cal M}|^2(x p_a, p_b, 
Q_X), 
\end{split}
\ee
where derivatives that appear in 
$L_{\rho \sigma}$ are computed 
with respect to  momenta $Q_X  \lambda =\Lambda_{ax}^{-1} P_X$.
Finally, we change the momentum of the colorless system $ P_X \to 
\lambda^{-1} \Lambda_{ax} P_X$ in the phase space, and obtain the result shown in Eq.~(\ref{eqc.1}).
\\

\section{ Comparison to results in the literature for the Drell-Yan process}
\label{sect_compDY}

In this appendix, we show how to derive results for zero-jettiness power corrections  to the process $q \bar{q} \rightarrow V g$ obtained  
in Ref.~\cite{Ebert:2018lzn}, from our Drell-Yan master formula in Eq.~\eqref{eq5.20}. At first glance the two results look very different, since we work with partonic cross sections for $q \bar q \to l^+ l^-$, where as  authors of Ref.~\cite{Ebert:2018lzn} work with hadronic production cross section of a   
fixed-mass vector boson, and study its  rapidity 
distribution. 
%and they have imposed constraints in the vector boson's (or dilepton system) invariant mass and rapidity. 
Furthermore, the definition of the zero-jettiness variable itself is different, as Born-projected momenta 
 are employed in Ref.~\cite{Ebert:2018lzn} to calculate it. Nevertheless, as discussed below, it is possible  to start with  Eq.~\eqref{eq5.20},  and derive Eq.~(5.36) in  Ref.~\cite{Ebert:2018lzn}, by taking into account two important points. 

First, we incorporate  the constraint on the   dilepton rapidity $Y$ (which is equivalent to the rapidity of the vector boson) and the dilepton  invariant mass $M$  as the observable function in study. %This observable fixes  momenta fractions of the incoming partons when the hadronic cross section is computed by integrating the partonic one with parton distribution functions. 
It reads
%The observable function reads 
\be
\begin{split}
{\cal O}^{\text{DY}}( p_1,  p_2) &= \delta \left( 2 p_1 \cdot p_2 - M^2 \right) \ \delta \left( \frac{1}{2} \ln \frac{P_b \cdot (p_1 + p_2)}{P_a \cdot (p_1 + p_2)} - Y \right), 
%\\
%& =  \delta \left( 2 \; \xi_a \xi_b \; P_a \cdot %P_b - M^2 \right) \ \delta \left( \frac{1}{2} %\ln \frac{\xi_a}{\xi_b} - Y \right),
\end{split}
\ee
where $P_{a,b}$ are the momenta of incoming hadrons.  Denoting the momenta fractions of $q$ and $\bar q$ as $\xi_{a,b}$, one may expect that the subleading power correction  to the hadronic cross section computed in 
Ref.~\cite{Ebert:2018lzn}
can be obtained by integrating the subleading partonic cross section, obtained with the help of our master formula, with parton distribution functions $f_{a,b}$
\be
\frac{{\rm d} \sigma^{\rm DY,{\rm NLP}}_{\rm had} }{{\rm d} \tau \; {\rm d} Y \; {\rm d} M^2}
\Bigg |_{[16]} = \int^1_0 {\rm d} \xi_a \; {\rm d} \xi_b \; f_a(\xi_a) f_b(\xi_b) \frac{{\rm d} \sigma^{\rm DY,{\rm NLP}} }{{\rm d} \tau} \bigg|_{{\cal O} = {\cal O}^{\text{DY}}}, 
%{\cal O}^{\text{DY}}(p_1,p_2).
\label{eqE.2}
\ee
%where $f_{a,b}(\xi_{a,b})$ are  parton %distribution functions. 
where we also need to write  the partonic center-of-mass energy squared $s$ as $\xi_a \xi_b S$, 
with $S = 2 P_a \cdot P_b$.

We note that, depending on the collinear sector,  our variable $x$  corresponds to  variables $z_{a,b}$ in Ref.~\cite{Ebert:2018lzn}. 
We also note that ${\rm d} \sigma^{\rm DY,{\rm NLP}}$ is described by our master formula 
and
includes derivatives with respect to lepton momenta that act on the observable ${\cal O}^{\rm DY}$.
These derivatives are re-written using  integration by parts, resulting in derivatives of parton distribution functions.

Upon doing that, we find that  the result that follows from Eq.~\eqref{eqE.2}  \emph{does not agree} with the result for subleading power corrections presented in  Ref.~\cite{Ebert:2018lzn}. The reason for this is the  difference in the zero-jettiness definition, as we now explain.  Indeed, 
the  definition employed in this paper uses  partonic momenta of the process, while 
in Ref.~\cite{Ebert:2018lzn}
Born-projected initial-state momenta are used. 
 These are obtained using the Born-like momentum fractions constructed from  vector boson mass and its rapidity, as opposed to $\xi_{a,b}$ 
 momenta fractions that we would use to define 
 zero-jettiness. Hence,  
to match the result of Ref.~\cite{Ebert:2018lzn},
  we need to account for this fact. 
   
  To this end, we compute the difference between to definitions. Suppose we find a gluon 
  momentum $k$ that satisfies the 
  zero-jettiness constraint $T_0(k) = \tau$, where $T_0(k)$ is the definition of zero-jettiness adopted in this paper. Then, when the same momentum is used to evaluate zero-jettiness as defined 
  in Ref.~\cite{Ebert:2018lzn} yields
\be
T_0(k) \Bigg |_{[16]} = \begin{cases}
    \begin{aligned}
    \tau & \left[ 1-(2-\psi) \; \frac{\omega_k}{\sqrt{s}} + \mathcal{O}(w
    _k^2) \right],  & \text{soft},\\
    \tau & \; z_a \left[ 1 + \frac{1+z_a}{z_a s} p_ak + \mathcal{O}((p_ak)^2) \right],  & \text{collinear $a$}, \\
    \tau & \; z_b \left[ 1 + \frac{1+z_b}{z_b s} p_bk + \mathcal{O}((p_bk)^2) \right],  & \text{collinear $b$.} \\
    \end{aligned}
\end{cases}
\label{eqE.4}
\ee
We observe that  in addition to subleading terms, there is also a re-scaling at leading power  in the collinear contributions. Given the relations of Eq.~\eqref{eqE.4}, we calculate how each of the three 
contributions (soft, $ca$ and $cb$) changes. 
Although this is rather straightforward, 
one should exercise significant care  when switching from one definition of the zero-jettiness to the other. For example,  in  our formula the term from Eq.~\eqref{eq3.40}, along with the manipulations  presented in Appendix~\ref{app:eq340},   are  affected by the presence of the extra factor $1/(z_{a,b})$ and have to be carefully re-analyzed. 
%\footnote{In that calculation, since $\lambda = \sqrt{x}$, one can simply make the change $\kappa_m \rightarrow \kappa_m - 2$ in order to get this additional term.}. This creates a difference proportional to the Born multiplied by a plus distribution in $1/(1-z_{a,b})_+$ and cancels the $\epsilon$-poles stemming from the Leading-Power collinear contributions.

Finally, putting everything together, we can express the result from Ref.~\cite{Ebert:2018lzn} for Drell-Yan in terms of our master  formula in the following manner
\be
\frac{{\rm d} \sigma^{\rm DY,{\rm NLP}}_{\rm had} }{{\rm d} \tau  \; {\rm d} Y \; {\rm d} M^2}
\Bigg |_{[16]} = \int^1_0 {\rm d} \xi_a \; {\rm d} \xi_b \; f_a(\xi_a) f_b(\xi_b) \frac{{\rm d} \sigma^{\rm DY,{\rm NLP}} }{{\rm d} \tau} \bigg|_{{\cal O} = {\cal O}^{\text{DY}}} 
%{\cal O}^{\text{DY}}(p_1,p_2),
\label{eqE.7}
\ee
where
\be
\begin{split}
\frac{{\rm d} \sigma^{\rm DY,{\rm NLP}} }{{\rm d} \tau} 
=& \left[ \frac{{\rm d} \sigma^{\rm DY,{\rm NLP}} }{{\rm d} \tau} \Big |_{\text{Eq.}~(6.19)}
- \frac{Q \tau}{2 s} \left( \frac{z_a + z_b}{z_a z_b} \right) \frac{{\rm d} \sigma^{\rm DY,{\rm LP}} }{{\rm d} \tau} \right] \left( \frac{1}{z_a z_b} \right) \\
&+\frac{4 [\alpha_s] C_F Q }{s} \int \limits_{0}^{1} \int \limits_{0}^{1} 
  \frac{{\rm d} \sigma_0}{2 z_a z_b} \left[ \frac{\delta(1-z_b)}{(1-z_a)_+} + \frac{\delta(1-z_a)}{(1-z_b)_+} \right] {\cal O}(p_1,p_2) \; {\rm d} z_a \ {\rm d} z_b,
\end{split}
\label{eqd.5}
\ee
and
%and the leading-power cross section reads 
\be
\begin{split}
\frac{{\rm d} \sigma^{\rm DY,{\rm LP}} }{{\rm d} \tau} = \frac{ [\alpha_s] C_F}{\tau} \int \limits_{0}^{1} \int \limits_{0}^{1} {\rm d} \sigma_0 \bigg[ & 4 \delta(1-z_a)\delta(1-z_b) + \frac{1+z_a^2}{z_a(1-z_a)_+} \delta(1-z_b) \\
& + \frac{1+z_b^2}{z_b (1-z_b)_+} \delta(1-z_a)\bigg] {\cal O}(p_1,p_2) \; {\rm d} z_a \; {\rm d} z_b,
\end{split}
\ee
is the finite contribution arising from the leading power terms. The Drell-Yan leading order cross section ${\rm d} \sigma_0$ is given in Eq.~\eqref{eq5.6}.
In Eq.~\eqref{eqE.7} we should further integrate over the leptonic angle $\beta$, and take $Q=M$ to account for the definition of  the zero-jettiness in Ref.~\cite{Ebert:2018lzn}. The result obtained with the help 
of Eq.~(\ref{eqd.5}) is  
in complete agreement with  Eq.~(5.36) of Ref.~\cite{Ebert:2018lzn}.

\section{ The Drell-Yan process with  dilepton final states}
\label{sect3D}

In this appendix, we consider a photon-mediated dilepton production in quark-antiquark annihilation
\be
q(p_a) + \bar q(p_b) \to l(p_1) + \bar l(p_2),
\label{eq3.1}
\ee
and compute zero-jettiness power corrections by directly expanding the phase space and the matrix element squared of the process. As we will see, this procedure allows us to compute  power corrections to the rapidity distribution 
of a single charged lepton.

The Born cross section can be written as 
\be
{\rm d} \sigma_0  = \bar \sigma_0  \; \left ( 1 - 2\ep  + \cos \theta_1^2 \right )
 [{\rm d} \Omega^{(d-1)}_1],
\ee
where $\theta_1$ is the angle between the lepton with the momentum $p_1$ and the initial quark with the momentum $p_a$,
\be
\bar \sigma_0 = \frac{\pi Q_q^2 \alpha_{\rm QED}^2 N_\ep}{2 N_c \; s^{1+\ep}},
\ee
and $[\Omega^{(d-1)}_1]$ is the solid angle of $p_1$ defined in the center-of-mass frame
of the colliding quark and anti-quark,  normalized to the solid angle in $d-2$ dimensions.
%\be
%[\Omega^{(d-1)}_3] = \frac{{\rm %d} \Omega_3^{(d-1)}}{\Omega^{(d-%2)}},
%\ee
Furthermore, $Q_q$ is the quark charge in units of the positron charge,
$\alpha_{\rm QED}$ is the fine structure constant, $N_c=3$ is the number of colors, and 
\be
N_\ep = 2^{2\ep} \frac{(4\pi)^\ep}{\Gamma(1-\ep)},
\ee
is the normalization factor.

To compute power correction in the zero-jettiness, we consider the real-emission process 
\be
q(p_a) + \bar q(p_b) \to l(p_1) + \bar l(p_2)  + g(k), 
\ee
and impose the appropriate constraint on the final-state gluon. We write 
\be
\begin{split} 
\frac{ {\rm d}  \sigma}{{\rm d} \tau}
& = \frac{1}{8s N_c^2}
\int [{\rm d} p_1][{\rm d} p_2] [{\rm d} k]
(2\pi)^d \delta( p_a + p_b - k - p_1 - p_2 )
\\
& \times \delta(\tau - T_0(p_a,p_b,k) ) \sum \limits_{\rm col, pol}^{} |{\cal M}|^2(p_a,p_b,p_1,p_2,k).
\end{split}
\label{eq2.7}
\ee
We then integrate over the momentum of the anti-lepton $p_2$, and find
\be
\begin{split} 
& \frac{ {\rm d}  \sigma}{{\rm d} \tau} = \frac{2 \pi }{8s N_c^2}
\int [{\rm d} p_2] [{\rm d} k]
\delta( ( p_a + p_b - k - p_1)^2 )
\\
& \times 
\delta(\tau - T_0(p_a,p_b,k) )
\sum \limits_{\rm col, pol}^{} |{\cal M}|^2(p_a,p_b,p_1,p_2,k).
\end{split} 
\ee
Since 
\be
\delta( ( p_a + p_b - k - p_1)^2 )
 = \delta( s - 2\sqrt{s} (\omega_k + E_1) + 2 \omega_k E_1 \rho_{1k} ),
\ee
we can remove this $\delta$-function by integrating over the lepton energy $E_1$ and then remove the zero-jettiness $\delta$-function by integrating over the gluon energy $\omega_k$. We  find
\be
\begin{split} 
\frac{ {\rm d}  \sigma}{{\rm d} \tau} & = \bar \sigma_0 \; \frac{  [\alpha_s] C_F  }{\tau^{1+2\ep} }
\int [{\rm d} \Omega^{(d-1)}_1] \;  [ {\rm d} \Omega^{(d-1)}_k]
\; \frac{ \left (1- 2 \frac{\omega^*_k}{\sqrt{s}} \right )^{1-2\ep}}{\left ( 1 -  \frac{ \omega^*_k}{\sqrt{s}}  \rho_{1k} \right )^{2-2\ep}}
\left ( \frac{Q}{ \sqrt{s} \psi_k}  \right )^{2-2\ep}
\\
& \times \sum \limits_{\rm col, pol}^{} \frac{  \tau^2 \; |{\cal M}|^2(p_a,p_b,p^*_1,p_2,k^*)}{ 4g_s^2 N_c C_F Q_q^2 e^4},
\end{split} 
\label{eq2.10}
\ee
where
\be
\omega_k = \frac{\tau Q}{\sqrt{s} \psi_k},\;\;\; E_1 = \frac{\sqrt{s}}{2}
     \frac{ 1 - \frac{2 \omega_k}{\sqrt{s}} } { \left ( 1 -  \frac{ \omega_k}{\sqrt{s}}  \rho_{1k} \right ) }.
\label{eq2.11a}
\ee
Since $E_1 \geq 0$ is required, we derive the following constraint
\be
\omega_k \leq \frac{ \sqrt{s}}{2} \; \Rightarrow   \; \psi_k \geq \frac{ 2 \tau Q}{s}.
\label{eq2.11}
\ee

Although the constraints appear to be  more complex than what has been discussed in Sec.~\ref{sect2}, the situation is rather similar in that there are two distinct cases for the gluon emission angle: $\theta_k \sim 1,\;  \omega_k \sim \tau$ and $\theta_k \sim 
\sqrt{\tau/Q},\;\;\; \omega_k \sim \sqrt{s}$.  The first case  describes the soft emission and the second case -- the collinear one.

We then  expand  the integrand in the soft region and in the  two collinear regions. We begin
with the discussion of the soft expansion. Computing the matrix element squared, expanding in $\omega_k \sim \tau$ and using the fact
that $\rho_{bk} = 2- \rho_{ak}$,  we find that two 
angular integrals,  given in 
Eq.~(\ref{eq3.31}), 
are required to calculate  the soft contribution to the
cross section through next-to-leading power.
Keeping $\tau$-suppressed terms through next-to-leading power and using the explicit  expression for the matrix element squared 
$|{\cal M}|^2(p_1,p_2,p_3,p_4,k)$
in Eq.~(\ref{eq2.10}), 
we find 
\be
\frac{{\rm d } \sigma^{(s)}}{{\rm d} \tau} = 
\frac{4 [\alpha_s] C_F}{\tau^{1+2\ep}} \left ( \frac{Q^2}{s} \right )^{-\ep}
\; {\rm d} \sigma_0  \;  \left[ \frac{1}{\ep} + \frac{1-2 \ep}{1 - \ep} \frac{Q \tau}{s}
   \right ].
   \label{eq3.14}
\ee
An interesting feature of this result is that ${\cal O}(\tau)$ power corrections to the cross section are  $\ep$-finite in this case.

We continue with computing  the collinear 
contribution, 
and focus on the case   when the
gluon is emitted along the incoming quark with momentum $p_a$. Then 
\be
\psi_k = \rho_{ak} = \rho.
\ee
We write
\be
\rho = \frac{2 \tau Q}{s (1-z)}, 
\ee
with $ 0 < z < 1$ so that
\be
\omega_k = \frac{\tau Q}{\sqrt{s} \psi_k} = \frac{\sqrt{s}}{2} (1-z).
\ee
It follows that  
\be
E_1 = \frac{\sqrt{s} z}{ 2 - (1-z) \rho_{1k} },
\ee
and, furthermore, 
\be
\frac{ \tau Q}{s \rho} = \frac{ (1-z)}{2}.
\ee

The expansion in the collinear case is the expansion in $\rho \sim \tau$ at fixed $z$.  The scalar products are
\be
2 p_a \cdot k = \sqrt{s} \omega_k \rho = \frac{s}{2} (1-z) \rho,
\;\;\;
2 p_b \cdot k = \sqrt{s} \omega_k (2 - \rho )
 =  \frac{s (1-z)  }{2}  \left (2 - \frac{ 2 \tau Q}{1-z} \right ).
\ee

We will also need to account for the scalar product 
between the momenta  of the emitted gluon and the outgoing lepton, and write it in such a way that its expansion in the 
collinear limit becomes possible.  To do this, we consider the unit vector 
$\vec k/\omega_k = \vec n_k$
and write it as 
\be
\vec n_k = (1 - \rho) \vec n_1 + \sqrt{\rho(2-\rho)} \vec n_{k,\perp}, 
\ee
where $\vec n_1$ is the unit vector in the direction of the 
momentum $\vec p_1$, 
and  $\vec n_{k \perp}$ is a unit  vector which is orthogonal to $\vec n_1$.
With this, we find 
\be
\rho_{1k} = 1 - \vec n_1 \cdot \vec n_k = \rho_{1a} + \rho \; (1 - \rho_{1a}) - \sqrt{\rho(2-\rho)} ( \vec n_{k,\perp} \cdot \vec n_{1} ).
\label{eq2.23}
\ee
Hence, expansion of this quantity in $\tau$ is straightforward but, because 
of the last term in Eq.~(\ref{eq2.23}) 
the expansion proceeds in 
powers of $\sqrt{\rho} \sim \sqrt{\tau}$.

Upon the expansion in $\tau$ around collinear limit, we will  average over  directions
of $\vec n_{k,\perp}$. This removes all contributions that 
contain odd powers of $\vec n_{k,\perp}$. The angular integration 
that is needed to compute the 
contribution at next-to-leading power is given by the standard formula
\be
( \vec n_{k,\perp} \cdot \vec n_{1} )^2  \rightarrow  \frac{\rho_{a1}(2-\rho_{a1}) }{d-2}.
\ee

 We find it to be convenient to combine  both  collinear contributions. We then write 
\be
\frac{{\rm d } \sigma^{(ca+cb)}}{{\rm d} \tau} = 
\bar \sigma_0\;
\frac{[\alpha_s] C_F}{\tau^{1+\ep}} Q^{-\ep}
\int \limits_{0}^{1} {\rm d} z \;
\left ( g^{(1)}_{c}(z,c_1) + \frac{\tau Q}{s} g^{(2)}_{c}(z,c_1)
\right ).
\label{eq3.25}
\ee
The function $g_c^{(1)}$ has the following structure 
\be
g_c^{(1)}(z,c_1) = (1-z)^{-\ep} \tilde P_{qq}(z) \ \tilde g^{(1)}_{c}(z,c_1),
\ee
where $\tilde P_{qq}(z)$ is related to $q \to q^* g$ collinear 
splitting.  The function $g_c^{(2)}$ describes 
the power correction. We expose its structure by  writing it as 
\be
g^{(2)}_{c}(z,c_1) =
4 (1+\ep)(1-2\ep+c_1^2) \left ( \frac{1}{(1-z)^{2+\ep}} 
+ \frac{\ep}{(1-z)^{1+\ep}} 
\right ) 
+ g^{(2)}_{c,\rm fin}(z,c_1), 
\label{eq3.26}
\ee
which separates divergent $z \to 1$ contributions from terms that do not contain non-integrable singularities  on the  interval $0 < z < 1$; 
such contributions are described by the function 
$g^{(2)}_{c,\rm fin}(z)$.  
%The 
%parametrization as in 
%Eq.~(\ref{eq3.26}) is convenient if numerical %integration over $z$ is attempted or if %constraints on lepton kinematics are introduced. %  We will discuss such examples in what follows. 
We  note that  
a direct integration of the 
function $g_c^{(2)}$ over $z$
at fixed $c_1$ is possible. 
Integrating
over  $z$, and  discarding all terms that vanish in the  $\ep \to 0$ limit, we find 
\be
\begin{split} 
\int \limits_{0}^{1} \; {\rm d} z \; g^{(2)}_{c}(z,c_1)
& =
\frac{2 \left(  3 + 37 
c_1^2 - 11 c_1^4 + 3 c_1^6 \right ) }{3 (1 - c_1^2)^2 }
- 4 \; c_1^3   \ln \frac{1+c_1}{1-c_1}.
\end{split} 
\ee
We observe that, similar to the soft contribution, the next-to-leading power  collinear contribution is  $\ep$-finite as well.

We are now in position to write the next-to-leading power zero-jettiness correction for the process in Eq.~(\ref{eq3.1}) at fixed $c_1$.   Combining the 
soft and collinear contributions using 
Eqs~(\ref{eq3.14}, \ref{eq3.25},
\ref{eq3.26}), we obtain 
\be
\begin{split} 
\frac{ {\rm d} \sigma^{\rm NLP} }{{\rm d} \tau {\rm d} c_{1}}
& = \frac{\alpha_s C_F }{2 \pi}  \bar \sigma_0  \frac{Q}{s}
\Bigg  [
  \frac{2 \left( 9 + 31 c_{1}^2 - 17 c_{1}^4 + 9 c_{1}^6 \right ) }{3 (1 - c_{1}^2)^2} - 4 c_{1}^3 \ln \frac{1+c_{1}}{1-c_{1}}
\Bigg  ].
\end{split} 
\ee

Finally, we note that we can write this  result using the  lepton rapidity in the partonic center-of-mass frame as a variable.  The relation between
rapidity and the scattering angle is
\be
y = \frac{1}{2} \ln \frac{p_b p_1}{p_a p_1} 
= \frac{1}{2} \ln \frac{1+ \cos \theta_1}{1-\cos \theta_1}.
\ee
It follows that 
\be
\; c_1 = \cos \theta_{1} = \frac{ \sh (y)}{\ch(y)}.
\ee
Changing the variables, we obtain
\be
\frac{ {\rm d} \sigma^{\rm NLP} }{{\rm d} \tau {\rm d} y}
 = \frac{\alpha_s C_F }{2 \pi}  \bar \sigma_0  \frac{Q}{s}
\Bigg  [
  \frac{32}{3} \ch(2 y) - \frac{16}{3} + \frac{20}{3 \ch^2(y)} - \frac{6}{\ch^4(y)}
  - 8 y \frac{\sh^3(y)}{\ch^5(y)}
\Bigg  ].
\ee
We observe the known fact \cite{Moult:2016fqy, Ebert:2018lzn} that zero-jettiness power corrections exhibit an  exponential growth at 
large lepton rapidities. Ways to address this problem were discussed in Ref.~\cite{Ebert:2018lzn}.
%The problem can be taken care of by choosing the %normalization factor $Q$ appropriately, e.g. $ Q %\sim 1/\ch(2y)$.

\bibliographystyle{JHEP}
\bibliography{rref}

\end{document}